\documentclass[11pt]{article}
\pdfoutput=1

\newcommand\beq{\begin{eqnarray}}
\newcommand\eeq{\end{eqnarray}}

\usepackage{epsfig, epstopdf}
\usepackage{amsmath}
\usepackage{tikz}
\usetikzlibrary{positioning,arrows}
\usetikzlibrary{decorations.pathmorphing}
\usetikzlibrary{decorations.markings}

\def\beq{\begin{equation}}
\def\eeq{\end{equation}}
\def\barr{\begin{array}}
\def\earr{\end{array}}

\def\gev{\, {\rm GeV}}

\def\lsim{\mathrel{\rlap{\lower4pt\hbox{$\sim$}}
    \raise1pt\hbox{$<$}}}                
\def\gsim{\mathrel{\rlap{\lower4pt\hbox{$\sim$}}
    \raise1pt\hbox{$>$}}}            

\usepackage{graphicx}
\usepackage{setspace, relsize}
\usepackage{placeins}
\usepackage{multirow}
\usepackage{xcolor}
\usepackage{dcolumn}
\usepackage{bm}
\usepackage{jheppub}
\usepackage{latexsym}
\usepackage{multirow}
\usepackage{color}
\usepackage{graphicx}
\usepackage{epsfig}  
\usepackage{dcolumn}
\usepackage{bm}
\usepackage{dcolumn}
\usepackage{textcomp}
\usepackage{float}
\usepackage{hypcap}
\usepackage[]{hyperref}
\usepackage{makecell}
\usepackage{multirow}
\usepackage{color}
\usepackage{pifont}
\usepackage{subfigure}
\usepackage{amssymb}
 \usepackage{amsmath}
 \usepackage{caption}
\usepackage{array}
\newcolumntype{L}[1]{>{\raggedright\let\newline\\\arraybackslash\hspace{0pt}}m{#1}}
\newcolumntype{C}[1]{>{\centering\let\newline\\\arraybackslash\hspace{0pt}}m{#1}}
\newcolumntype{R}[1]{>{\raggedleft\let\newline\\\arraybackslash\hspace{0pt}}m{#1}}

 \usepackage[normalem]{ulem}
\usepackage{color}

\usepackage{blindtext}
 
\makeatletter
\renewcommand{\paragraph}{\@startsection{paragraph}{4}{0ex}%
    {-3.25ex plus -1ex minus -0.2ex}%
    {1.5ex plus 0.2ex}%
    {\normalfont\normalsize\bfseries}}
\makeatother
 
\stepcounter{secnumdepth}
\stepcounter{tocdepth}

\allowdisplaybreaks
  \newcommand{\capdef}{}
  \newcommand{\mycaption}[2][\capdef]{\renewcommand{\capdef}{#2}
       \caption[#1]{{\footnotesize #2}}}
  \newcommand{\be}{\begin{equation}}
   \newcommand{\ee}{\end{equation}}

\hypersetup{
  bookmarks=true,         
  unicode=false,          
  pdftoolbar=true,        
 pdfmenubar=true,        
 pdffitwindow=true,     
 pdfstartview={FitH},    
 pdfsubject={Neutrino Oscillations Phenomenology},   
 pdfnewwindow=true,      
 pdfcreator={RevTeX},
 colorlinks=true,       
 linkcolor=red,          
 citecolor=blue,        
 filecolor=black,      
 urlcolor=blue,           
  }

\title{Multi-charged TeV scale scalars and fermions in the framework of a radiative seesaw model }

\author[a,b]{Avnish,}
\author[a,b]{Kirtiman Ghosh}

\affiliation[a]{Institute of Physics, Sachivalaya Marg, Sainik School Post, Bhubaneswar 751005, India}
\affiliation[b]{Homi Bhabha National Institute, Training School Complex, Anushakti Nagar, Mumbai 400085, India}

\emailAdd{avnish@iopb.res.in}
\emailAdd{kirti.gh@gmail.com}

 \abstract{Explaining the tiny neutrino masses and non-zero mixings have been one of the key motivations for going beyond the framework of the Standard Model (SM). We discuss a collider testable model for generating neutrino masses and mixings via radiative seesaw mechanism. That the model does not require any additional symmetry to forbid tree-level seesaws makes its collider phenomenology interesting. The model includes multi-charged fermions/scalars at the TeV scale to realize the Weinberg operator at 1-loop level. After deriving the constraints on the model parameters resulting from the neutrino oscillation data as well as from the upper bound on the absolute neutrino mass scale, we discuss the production, decay and resulting collider signatures of these TeV scale fermions/scalars at the Large Hadron Collider (LHC). We consider both Drell-Yan and photoproduction. The bounds from the neutrino data indicate the possible presence of a long-lived multi-charged particle (MCP) in this model. We obtain bounds on these long-lived MCP masses from the ATLAS search for abnormally large ionization signature. When the TeV scale fermions/scalars undergo prompt decay, we focus on the 4-lepton final states and obtain bounds from different ATLAS 4-lepton searches. We also propose a 4-lepton event selection criteria designed to enhance the signal to background ratio in the context of this model.
}


\begin{document}
\tikzset{
  vector/.style={decorate, decoration={snake,amplitude=.4mm,segment length=2mm,post length=1mm}, draw},
  tes/.style={draw=black,postaction={decorate},
    decoration={snake,markings,mark=at position .55 with {\arrow[draw=black]{>}}}},
  provector/.style={decorate, decoration={snake,amplitude=2.5pt}, draw},
  antivector/.style={decorate, decoration={snake,amplitude=-2.5pt}, draw},
  fermion/.style={draw=black, postaction={decorate},decoration={markings,mark=at position .55 with {\arrow[draw=blue]{>}}}},
  fermionbar/.style={draw=black, postaction={decorate},
    decoration={markings,mark=at position .55 with {\arrow[draw=black]{<}}}},
  fermionnoarrow/.style={draw=black},
  scalar/.style={dashed,draw=black, postaction={decorate},decoration={markings,mark=at position .55 with {\arrow[draw=blue]{>}}}},
  scalarbar/.style={dashed,draw=black, postaction={decorate},decoration={marking,mark=at position .55 with {\arrow[draw=black]{<}}}},
  scalarnoarrow/.style={dashed,draw=black},
  electron/.style={draw=black, postaction={decorate},decoration={markings,mark=at position .55 with {\arrow[draw=black]{>}}}},
  bigvector/.style={decorate, decoration={snake,amplitude=4pt}, draw},
  particle/.style={thick,draw=blue, postaction={decorate},
    decoration={markings,mark=at position .5 with {\arrow[blue]{triangle 45}}}},
  gluon/.style={decorate, draw=black,
    decoration={coil,aspect=0.3,segment length=3pt,amplitude=3pt}}
}

\maketitle
\flushbottom

\section{Introduction}

The Standard Model (SM) based on the gauge symmetry $SU(3)_C\times SU(2)_L\times U(1)_Y$ is a successful description of fundamental particles and interactions. Almost all the predictions of the SM have been verified experimentally. However, there are two important experimental observations (among many others) that compel us to think of the SM as a low energy theory requiring new physics at a high scale. These are the existence of the dark matter in the universe, and the tiny non-zero masses of the neutrinos and their mixings. The SM has no candidate for the dark matter. While the simplest way to generate neutrino masses is to add right-handed neutrino fields to the SM particle content, it is hard to explain their extreme smallness. These issues have led to a plethora of new dynamics beyond the SM.  The Large Hadron Collider (LHC) at CERN is aiming to uncover any such dynamics that may be operative at the scale of a few TeVs.

Neutrino masses are at least six orders of magnitude smaller than the next lightest standard model fermion. Such a small mass could be understood if neutrinos are Majorana particles\footnote{It is important to note that the current experimental data (from neutrino oscillation as well as scattering experiments) is inconclusive in determining the Dirac or Majorana nature of neutrinos as well as the mechanism of neutrino mass generation.} and Majorana masses for neutrinos are generated from higher dimensional operators which violate lepton number by two units. The most studied example of such an operator is the dimension-5 Weinberg operator \cite{Weinberg:1979sa}:
\begin{equation}
  {\cal O}_5~=~\frac{c_{\alpha\beta}}{\Lambda}\left(\bar{L^C_{\alpha L}}\tilde H^*\right)\left(\tilde H^\dagger L_{\beta L}\right)~+~{\rm h.c.},\nonumber
\end{equation}
where, $\alpha,~\beta$ are the generation indices, $L_L~=~(\nu_L, l_L)^T$ is the left-handed lepton doublet of the SM, $H~=~(h^+, \frac{h+i\eta}{\sqrt 2})^T$ is the Higgs doublet and $\tilde H=i\sigma_2 H^*$. $\Lambda$ is the scale of new physics and $c_{\alpha\beta}$ is a model-dependent coefficient. Weinberg operator gives rise to Majorana masses (suppressed by $\Lambda$) for the neutrinos after electroweak symmetry breaking (EWSB). At tree level, there are only three ways to generate the  Weinberg operator, namely, the type-I \cite{Klein:2019iws,Minkowski:1977sc,Yanagida:1979as,GellMann:1980vs,Mohapatra:1979ia}, the type-II \cite{Magg:1980ut,Schechter:1980gr,Wetterich:1981bx,Lazarides:1980nt,Mohapatra:1980yp,Cheng:1980qt,Perez:2008ha} and the type-III \cite{Foot:1988aq}  seesaw mechanisms. In the framework of tree-level seesaw models, the smallness of neutrino masses ($m_\nu$s) is explained via new physics at a very high scale of $\Lambda$. For instance, assuming $c_{\alpha \beta} \sim {\cal O}(1)$, $m_\nu \sim 0.1$ eV requires new physics at a scale $\Lambda\sim 10^{14}-10^{15}$ GeV which is impossible to probe in the ongoing as well as proposed collider experiments. However, there are two alternative classes of models in which ${\cal O}_5$ is forbidden at tree level and neutrino masses are generated radiatively \cite{Babu:2013pma,Sierra:2014rxa,Okada:2015vwh,Nomura:2016run,Nomura:2016ask,Nomura:2016dnf,Nomura:2017vzp,Cepedello:2017lyo,Cepedello:2018rfh,Cai:2017jrq,Babu:2019mfe,Gargalionis:2019drk,Cepedello:2020lul,Arbelaez:2020xcg,Ma:2009gu,Ma:2012xj,Kanemura:2010bq,Krauss:2002px,Branco:1988ex,Aoki:2009vf,Aoki:2008av,Ma:2007yx,Ma:2006km,Jana:2019mgj,Cheung:2004xm,Cheung:2017kxb,Cheung:2018itc,Nomura:2018lsx,FileviezPerez:2009ud,Babu:2002uu,Babu:1989pz,Ma:1998dn,Babu:1988ki,Bonnet:2012kz} or from a tree-level effective operator with $d>5$ \cite{Babu:2009aq,Anamiati:2018cuq,Gu:2009hu,Babu:1999me,Giudice:2008uua,Gogoladze:2008wz,Chen:2006hn,Bonnet:2009ej}. The additional suppression to the neutrino masses, arises from the loop integrals (in case of former) or higher powers of $\Lambda$ in the denominator (in case of later), brings down the new physics scale $\Lambda$ to TeV scale and hence, makes these models testable at the LHC. Radiative neutrino mass generation scenarios often require a $Z_2$ symmetry to forbid the tree-level contribution(s) to the Weinberg operator. Apart from forbidding the generation of neutrino masses at the tree-level, the $Z_2$ symmetry also ensures the stability of the lightest $Z_2$-odd particle which (if weakly interacting) could be a cosmologically viable candidate for cold dark matter (CDM) \cite{Kajiyama:2013rla,Farzan:2014aca,Restrepo:2015sjs,Kashiwase:2015pra,Ahriche:2016rgf,Nomura:2016vxr,Sierra:2016qfa,Ahriche:2016ixu,Guo:2016dzl,Simoes:2017kqb,Nomura:2017emk,Yao:2017vtm,Esch:2018ccs,CentellesChulia:2019xky,Loualidi:2020jlj,Restrepo:2013aga}. However, in collider experiments, a stable weakly interacting particle remains invisible and hence, contributes to the missing transverse energy ($E_T\!\!\!\!\!\!/~$) signature. Presence of $E_T\!\!\!\!\!\!/~$ in the final state poses serious problems in the reconstruction of the new particles masses as well as in the discovery of the new physics (NP) signatures over the SM backgrounds. On the other hand, the models, which do not require a $Z_2$ symmetry to forbid ${\cal O}_5$ at tree-level, lack the motivation of a candidate for CDM. Such models, however, give rise to smoking-gun signatures at the collider experiments and hence, are easily testable at the LHC.  

In this article, we have studied the detailed phenomenology of a model that generates neutrino masses at the 1-loop level. In the framework of the SM gauge symmetry, the model includes two new scalar $SU(2)_L$-doublets ($\Phi_{\frac{5}{2}}$ and $\Phi_\frac{3}{2}$), one scalar singlet ($k$) and at least two\footnote{Neutrino oscillation data indicates towards non-zero masses for atleast two neutrinos.} copies vector-like fermion singlets ($E$). The aim is to generate Weinberg operator (${\cal O}_5^{\rm 1 \text - loop}$) at 1-loop level\footnote{Weinberg operator at the 1-loop level has already been studied in details in the literature. In Ref.~\cite{Bonnet:2012kz}, 12 topologies, which contribute to the Weinberg operator at 1-loop level, have been identified. For each topology, there are several alternatives (models) for assigning different types (scalar or fermion) of fields running in the loop. A complete list of all these models leading to ${\cal O}_5^{\rm 1 \text - loop}$ can also be found in Ref.~\cite{Bonnet:2012kz}.} via T1-i topology \cite{Bonnet:2012kz}. To ensure the loop diagram(s) as the leading contribution to the neutrino masses, one needs to forbid the couplings which lead to ${\cal O}_5$ at tree-level. For instance, in this model, the Yukawa couplings involving the newly introduced singlet fermions and SM lepton and Higgs doublets give rise to ${\cal O}_5^{\rm tree}$ via Type-I seesaw mechanism. Charging matter fields under a $Z_2$ symmetry is enough to forbid such couplings. Alternatively, one can carefully choose the hypercharges\footnote{The electric charge ($Q$) is given by $Q=I_3+Y$, where $I_3$ is the third component of isospin.} of the singlet fermions ($Y_E$) to forbid such couplings. In the context of this particular scenario, $Y_E \pm Y_L \pm Y_H  \neq 0$ or equivalently, $Y_E \neq 0~{\rm or}~\pm 1$, where $Y_{L,H}$ is the hypercharge of the SM lepton(Higgs) doublet,  is enough to forbid Type I seesaw mechanism. Our choice $Y_E=2$ results into doubly charged singlet fermions ($E^{++}$) in the model. The generation of non-zero neutrino masses at 1-loop via T1-i topology requires the following hypercharge assignments  for the doublet and singlet scalars: $Y_{\Phi_{\frac{3}{2}\left(\frac{5}{2}\right)}}=\frac{3}{2}\left(\frac{5}{2}\right)$ and $Y_k=2$, respectively (see Ref.~\cite{Bonnet:2012kz} for details). These particular hypercharge assignments for the newly introduced doublets and singlets result in TeV scale multi-charged scalars (triply, doubly and singly charged) and fermions (doubly charged) in the model. In this article, we have studied collider signatures of these multi-charged scalars and fermions at the LHC with 13 TeV center-of-mass energy.

Single production of these multi-charged scalars and fermions at the LHC are suppressed. However, they can be pair produced via quark anti-quark fusion (Drell-Yan production) or photon-photon fusion (photoproduction). The parton densities of quarks and anti-quarks are significantly larger than the photon density\footnote{It is important to note that $\alpha_{EM}$ is of the same order of magnitude as $\alpha_{strong}^2$. Therefore, in the era of precision phenomenology at the LHC when the PDFs are already determined up to NNLO in QCD, consistency of calculations require PDFs which are corrected at least up to NLO QED. Next-to-leading order (NLO) QED corrections require photon as a parton inside the proton, with an associated parton distribution function (PDF).} and thus, photoproductions of charged scalar/fermions pairs are neglected in the phenomenological studies \cite{delAguila:2013mia,Kanemura:2013vxa,Du:2018eaw,Primulando:2019evb,Dev:2018kpa,Dev:2018sel,Alloul:2013raa} as well as by the experimental groups \cite{ATLAS:2012hi,ATLAS:2014kca,Aaboud:2018qcu,Chatrchyan:2012ya,CMS:2017pet,CMS:2016cpz,Aad:2019pfm,Aaboud:2018kbe}. However, it is important to note that at the Born level, photoproduction cross-section of a multi-charged (with charge $Q$) particle is enhanced by a factor of $Q^4$. Moreover, being $t(u)$-channel process, photoproduction falls relatively slowly with the parton center-of-mass energy compared to the $s$-channel Drell-Yan (DY) process. The importance of the photoproduction has already been pointed out in the recent literature \cite{PhysRevD.50.2335,PhysRevD.76.075013,Babu:2016rcr,Agarwalla:2018xpc,Ghosh:2017jbw,Baines:2018ltl,Kurochkin:2006jr,Acharya:2019vtb,_nan_2012,Danielsson:2016nyy} and it has been shown that in the larger mass region, photoproduction of $Q>1$ fermions/scalars could be significant/dominant compared to the conventional DY production. Therefore, in this work, we have considered both DY and photoproduction of the (multi-)charged scalars/fermions.

After being produced at the LHC, the multi-charged scalars/fermions decay into the SM leptons and/or bosons ($W^\pm,~Z$, and Higgs). The resulting signature is characterized by  multiple leptons (including 3,4,5,6 leptons, same-sign dilepton e.t.c.) final state. The decays of the (multi-)charged scalars/fermions usually proceed through the Yukawa couplings involving these new scalars/fermions and the SM leptons. It is important to note that such Yukawa couplings also contribute to the loop diagram(s) generating the neutrino masses and hence, required to be small. Therefore, depending on the choice of parameters, the total decay width of these multi-charged particles (MCPs) could be small enough ($\Gamma_{TOT}<10^{-16}$) to ensure the decay of these particles outside the detector. The energy loss of the charged particles inside the detector increases quadratically with its charge \cite{Bethe:1930ku} and hence, long-lived MCPs are expected to leave a very characteristic signature of high ionization in the detector (especially, in the pixel detector, transition radiation tracker and muon chamber) \cite{Aaboud:2018kbe,Aad:2013pqd,Khachatryan:2016sfv,Veeraraghavan:2013rqa}. Recently, in Ref.~\cite{Aaboud:2018kbe}, the ATLAS collaboration has presented a search for abnormally large ionization signature to constrain scenarios with long-lived MCPs at the LHC with $\sqrt s~=~13$ TeV and 36.1 fb$^{-1}$ integrated luminosity. Such constraints may also be applicable in our model for MCPs with $\Gamma_{TOT}<10^{-16}$ GeV. On the other hand, if the total decay widths are large enough $\Gamma_{TOT}>10^{-13}$ to ensure prompt decays of the scalars/fermions, the model gives rise to multi-lepton signatures at the LHC. We have studied 4-leptons final state in detail. Different other new physics scenarios also give rise to 4-leptons final state at the LHC. This final state is particularly interesting due to negligible background contributions from the SM and hence, have already been studied by the ATLAS \cite{Aaboud:2018zeb,Aad:2014iza,ATLAS:2012kr,Aaboud:2017qph,Aad:2015dha} and CMS \cite{Sirunyan:2017lae,Khachatryan:2016iqn,Chatrchyan:2014aea,Chatrchyan:2013xsw,Chatrchyan:2012mea} collaborations in the context of different new physics scenarios. We have used existing ATLAS searches \cite{Aaboud:2018zeb,Aaboud:2017qph} for 4-lepton final states to constraint the parameter space of this model. To enhance the reach at the LHC, we have also proposed a set of event selection criteria optimized for our model.        

This paper is organized as follows. The next section is devoted to the introduction of the model and the discussion about the generation of the neutrino masses at the 1-loop level. In section \ref{E_phenomenology}, we studied the collider phenomenology of the doubly charged fermions at the LHC with 13 TeV center-of-mass energy. The production, decay, and collider signatures of the scalars are presented in section \ref{scalar_phenomenology}.  We finally conclude in section \ref{conclusion}.

\section{The Model}
To realize Weinberg operator at 1-loop level, the model incorporates new SM $SU(2)_L$ singlet fermions ($E^{++}_\alpha$, where $\alpha~=~$1, 2 and 3) as well as a singlet scalar ($k^{++}$) and two doublet scalars ($\Phi_\frac{3}{2}$ and $\Phi_\frac{5}{2}$) in the framework of the SM gauge symmetry. The field content along with their gauge quantum numbers are summarized in the following:\\
\begin{table}[h!]
  \centering
  \begin{tabular}{c c}
    \hline\hline
    \multicolumn{2} {c}{$G_{\rm SM}~=~SU(3)_C \times SU(2)_L \times U(1)_Y$}\\
    \hline\hline
        {\bf Fermions:} & $Q_{\alpha L}~=~\begin{pmatrix} u_\alpha \\ d_\alpha \end{pmatrix}_L \sim \left(3,2,\frac{1}{6}\right)$,~~ $L_{\alpha L}~=~\begin{pmatrix} \nu_\alpha \\ e_\alpha \end{pmatrix}_L \sim \left(1,2,-\frac{1}{2}\right)$\\
        & \\
        & $u_{\alpha R}\sim \left(3,1,\frac{2}{3}\right)$, $d_{\alpha R}\sim \left(3,1,-\frac{1}{3}\right)$, $e_{\alpha R}\sim \left(1,1,-1\right)$\\
        & \\
        & $E^{++}_{\alpha L(R)}\sim \left(1,1,2\right)$ \\
        & \\
        {\bf Scalars:} & $H~=~\begin{pmatrix} h^{+} \\ \frac{h + i\eta}{\sqrt 2} \end{pmatrix} \sim \left(1,2,\frac{1}{2}\right)$\\
        & \\
        & $\Phi_{\frac{3}{2}}~=~\begin{pmatrix} \phi^{++}_{\frac{3}{2}} \\ \phi^+_{\frac{3}{2}} \end{pmatrix} \sim \left(1,2,\frac{3}{2}\right)$, ~~$\Phi_{\frac{5}{2}}~=~\begin{pmatrix} \phi^{+++}_{\frac{5}{2}} \\ \phi^{++}_{\frac{5}{2}} \end{pmatrix} \sim \left(1,2,\frac{5}{2}\right)$\\
        & \\
        & $k^{++}\sim\left(1,1,2\right)$\\
        \hline\hline
  \end{tabular}
  \caption{Field content of the model along with their gauge quantum numbers: $\left(SU(3)_C,SU(2)_L,U(1)_Y\right)$ is presented where $\alpha=1,~2,~{\rm and}~3$ is the generation index and the electric charges are determined by $Q=T_3+Y.$}
  \label{table:1}
\end{table}

The gauge interactions of the newly introduced scalars/fermions with the SM gauge bosons ($W^\pm,~Z$ and photon) can be obtained from the kinetic part of the Lagrangian,
\begin{eqnarray}
  {\cal L}_{\rm kin} ~\supset && \left(D_\mu \Phi_{\frac{3}{2}}\right)^\dagger\left(D^\mu \Phi_{\frac{3}{2}}\right)~+~\left(D_\mu \Phi_{\frac{5}{2}}\right)^\dagger\left(D^\mu \Phi_{\frac{5}{2}}\right)~+~\left(D_\mu k^{++}\right)^\dagger\left(D^\mu k^{++}\right)\nonumber\\
  &&+~\overline{E^{++}_\alpha}i\gamma^\mu D_\mu E^{++}_\alpha,
  \label{lag_kin}
\end{eqnarray}
where, the gauge covariant derivative $D_\mu$ is given by, $D_\mu=\partial_\mu-ig\tau^a W_\mu^a-ig^\prime Y B_\mu$ for $SU(2)_L$ doublets and $D_\mu=\partial_\mu-ig^\prime Y B_\mu$ singlets with $W^a_\mu$ and $B_\mu$ being the gauge bosons of $SU(2)_L$ and $U(1)_Y$, respectively. Here, $g~{\rm and}~g^\prime$ are the gauge couplings corresponding to $SU(2)_L$ and $U(1)_Y$, respectively, $Y$ is the hypercharge and $\tau^a$s are the generators of $SU(2)_L$ doublet representation. Assignment of the gauge quantum numbers (see Table~\ref{table:1}) allows Yukawa interactions involving the doubly charged singlet fermions, the SM lepton doublets and $Y=\frac{3}{2}\left(\frac{5}{2}\right)$ scalar doublet. The couplings involving the doubly charged singlet scalar ($k^{\pm\pm}$) and a pair of SM singlet leptons are also allowed. The relevant parts of the Yukawa Lagrangian are as follows:
\begin{eqnarray}
  {\cal L}_{\rm Yukawa} ~\supset && m_E^{\alpha\beta}\overline{E^{++}_\alpha}E^{++}_\beta~+~y_{\frac{3}{2}}^{\alpha\beta}\overline{L_{\alpha L}}{\Phi}_{\frac{3}{2}}\left(E^{++}_{\beta L}\right)^C ~+~y_{\frac{5}{2}}^{\alpha\beta}\overline{L_{\alpha L}}i\sigma_2\Phi_{\frac{5}{2}}^*E^{++}_{\beta R} \nonumber\\
  &&+~y_k^{\alpha\beta} \overline{e_{\alpha R}}k^{--}\left(e_{\beta R}\right)^C+{\rm h.c.},
  \label{lag_yuk}
\end{eqnarray}
where, $C$ stands for charge conjugation, $\alpha~{\rm and}~\beta$ are the generation indices, $y_{\frac{3}{2}\left(\frac{5}{2}\right)}$ and $y_k$ are Yukawa matrices and $m_E$ is the mass matrix for the vector-like doubly charged fermions. All these matrices are, in general, complex $3\times 3$ matrices. However, one has the freedom to choose a basis for $E^{++}_\alpha$ in which, $m_E$ is diagonal with real positive diagonal elements. It is important to note that $y_{\frac{3}{2}\left(\frac{5}{2}\right)}$ contributes to the neutrino masses at the 1-loop level. On the other hand, $y_{k}$ allows the decay of doubly charged scalars into a pair of SM charged leptons and hence, determines the phenomenology at the LHC. Apart from the usual quadratic (mass terms) and quartic terms involving the new doublet ($\Phi_{\frac{3}{2}\left(\frac{5}{2}\right)}$) and singlet ($k^{++}$) scalars, the most general renormalizable gauge invariant scalar potential also includes following phenomenologically important terms:
\begin{eqnarray}
  {\rm V}(H,\Phi_{\frac{3}{2}},\Phi_{\frac{5}{2}},k^{++})~\supset && \mu\left(H^Ti\sigma_2{\Phi}_{\frac{3}{2}}\right)k^{--}~+~\mu^\prime\left(H^\dagger{\Phi}_{\frac{5}{2}}\right)k^{--} \nonumber\\
  &&+~\lambda\left(H^Ti\sigma_2{\Phi}_{\frac{3}{2}}\right)\left(H^T\Phi^*_{\frac{5}{2}}\right) +c.c.
  \label{lag_scalar}
\end{eqnarray}
The cubic and quartic terms in Eq.~\ref{lag_scalar} are not only important for generating the neutrino masses at the 1-loop level but also determines the collider signatures of this model via controling the mixings among the doubly charged scalars ($\phi_{\frac{5}{2}}^{++},~\phi_{\frac{3}{2}}^{++}~{\rm and}~k^{++}$). After the electroweak symmetry breaking (EWSB), the mass Lagrangian for the multi-charged scalars (MCS) can be written as,
\begin{eqnarray}
  {\cal L}_{\rm MASS}^{\rm MCS}~= && m_{\frac{3}{2}}^2\left(\phi^{+}_{\frac{3}{2}}\right)^\dagger\phi^{+}_{\frac{3}{2}}~+~m_{\frac{5}{2}}^2\left(\phi^{3+}_{\frac{5}{2}}\right)^\dagger\phi^{3+}_{\frac{5}{2}} \nonumber\\
  &&+~\left(\begin{array}{ccc}  \phi^{++}_\frac{5}{2} &\phi^{++}_\frac{3}{2}  & k^{++}  \end{array}\right)^*
  \left(\begin{array}{ccc}  m_{\frac{5}{2}}^2  &  -\frac{\lambda v^2}{2} & \frac{\mu^{\prime}v}{\sqrt{2}} \\ -\frac{\lambda v^2}{2} &  m_{\frac{3}{2}}^2 & -\frac{\mu v}{\sqrt{2}} \\\frac{\mu^{\prime}v}{\sqrt{2}}  &  -\frac{\mu v}{\sqrt{2}} & m_k^2\end{array}\right)
    \left(\begin{array}{c} \phi^{++}_\frac{5}{2}\\ \phi^{++}_\frac{3}{2} \\ k^{++}  \end{array}\right)\,\, ,
    \label{lag_mass}
\end{eqnarray}
where, $v$ is the vacuum expectation value (VEV) of the SM Higgs doublet and $m_{\frac{3}{2}\left(\frac{5}{2}\right)}^2$ and $m_k^2$ are the coefficients of the quadratic terms involving hypercharge $\frac{3}{2}\left(\frac{5}{2}\right)$ doublet and doubly charged singlet, respectively. The tree-level masses for the singly ($\phi^{+}_{\frac{3}{2}}$) and triply ($\phi^{3+}_{\frac{5}{2}}$) charged scalars are given by $m_{\frac{3}{2}}$ and $m_{\frac{5}{2}}$, respectively. Whereas, the mass squared of the physical doubly charged scalars (denoted as $H_1^{++},~H_2^{++}~{\rm and}~H_3^{++}$) are given by the eigen values ($m_{H_1}^2,~m_{H_2}^2~{\rm and}~m_{H_3}^2$) of the doubly charged scalar mass matrix (denoted by $M_{\rm DCS}$ and given by the matrix in the last term of ${\cal L}_{\rm MASS}^{\rm MCS}$) which is a $3\times 3$ real symmetric matrix and hence, can be diagonalized by a orthogonal matrix $O$: $O M_{DCS} O^T~=~{\rm diag}\left(m_{H_1}^2,~m_{H_2}^2,~m_{H_3}^2\right)$. The physical doubly charged scalars are given by,
\begin{equation}
  H_a^{++}~=~O_{a1} \phi_{\frac{5}{2}}^{++}+O_{a2} \phi_{\frac{3}{2}}^{++}+O_{a3} k^{++},
  \label{mixing}
\end{equation}
where, $a \ni 1,~2,~3$. A list of Feynman rules which are relevant for rest of the article, are presented in Appendix A.

\subsection{Neutrino Masses at 1-loop level}
\label{nu_mass}

\begin{figure}[!t]
  \centering
\begin{minipage}{0.35\textwidth}
    \begin{tikzpicture}[line width=1.4 pt, scale=1.65,every     node/.style={scale=1.0}]
      \draw[fermion,black] (0.0,0.0)  --(-0.7,-0.7);
      \draw[fermion,black] (2,0) --(0,0);
      \draw[fermion,black] (2,0) --(2.7,-0.7);

      \draw[scalar,black] (0,0)  --(0.5,2.0);
      \draw[scalar,black] (0.5,2.0)  --(1.5,2.0);
      \draw[scalar,black] (1.5,2.0)  --(2.5,2.3);
      \draw[scalar,black] (1.5,2.0)  --(2,0);
      \draw[scalar,black] (0.5,2)  --(-0.5,2.3);

      \node at (1,-0.2) {$E^{--}$};
      \node at (1,1.80) {$k^{--}$};
      \node at (-0.1,1.0) {$\phi^{--}_{\frac{3}{2}}$};
      \node at (2.2,1.0) {$\phi^{--}_{\frac{5}{2}}$};
      \node at (0.5,2.2) {$\mu$};
      \node at (1.5,2.2) {$\mu^\prime$};
      \node at (-0.5,2.55) {$<H>$};
      \node at (2.5,2.55) {$<H>$};
      \node at (2.55,-0.3) {$\nu_L$};
      \node at (-0.55,-0.3) {$\nu_L$};
      \node at (0.1,-0.23) {$ y_{\frac{3}{2}} $};
      \node at (1.9,-0.23) {$ y_{\frac{5}{2}} $};
    \end{tikzpicture}
  \end{minipage}
  \hspace{0.2\textwidth}
  \begin{minipage}{0.35\textwidth}
    \begin{tikzpicture}[line width=1.4 pt, scale=1.65,every node/.style={scale=1.0}]
      \draw[fermion,black] (0.0,0.0)  --(-0.7,-0.7);
      \draw[fermion,black] (2,0) --(0,0);
      \draw[fermion,black] (2,0) --(2.7,-0.7);

      \draw[scalar,black] (0,0)  --(1,2.0);
      \draw[scalar,black] (1,2)  --(2,0);
      \draw[scalar,black] (1,2)  --(0.1,2.4);
      \draw[scalar,black] (1,2)  --(1.9,2.4);

      \node at (1,-0.2) {$E^{--}$};
      \node at (0.1,1.0) {$\phi^{--}_{\frac{3}{2}}$};
      \node at (1.9,1.0) {$\phi^{--}_{\frac{5}{2}}$};
      \node at (1,2.2) {$\lambda$};
      \node at (0.1,2.55) {$<H>$};
      \node at (1.9,2.55) {$<H>$};
      \node at (2.55,-0.3) {$\nu_L$};
      \node at (-0.55,-0.3) {$\nu_L$};
      \node at (0.1,-0.23) {$ y_{\frac{3}{2}} $};
      \node at (1.9,-0.23) {$ y_{\frac{5}{2}} $};
    \end{tikzpicture}
  \end{minipage}
\mycaption{Feynman diagrams generating neutrino masses at 1-loop level.}
\label{fd_nu_mass}
\end{figure}

In the framework of this model, Weinberg operator is generated at 1-loop level via the Feynman diagrams depicted in Fig.~\ref{fd_nu_mass}. Neutrinos get Majorana masses after the EWSB. Simultaneous presence of Yukawa couplings $y_{\frac{5}{2}}$ and $y_{\frac{5}{2}}$ violates lepton number conservation in this model.
The contributions of the box (Fig.~\ref{fd_nu_mass} left panel) and triangle (Fig.~\ref{fd_nu_mass} right panel) diagrams to the neutrino mass matrix $m_{\nu}$ can be computed as,
\begin{equation}
  \frac{ {m_{\nu}}^{\square}}{{\langle H \rangle}^2}~=~ \frac{\mu \mu^\prime}{16\,\pi^2\,v^2} \left(y_{\frac{5}{2}}^T M_\square^{-1} y_{\frac{3}{2}}+y_{\frac{3}{2}}^T M_\square^{-1} y_{\frac{5}{2}}\right)~{\rm and}~ \frac{{m_{\nu}}^{\Delta}}{{\langle H \rangle}^2}~=~\frac{\lambda}{16\,\pi^2} \left(y_{\frac{5}{2}}^T M_\Delta^{-1} y_{\frac{3}{2}}+y_{\frac{3}{2}}^T M_\Delta^{-1} y_{\frac{5}{2}}\right),
  \label{nu_mass}
\end{equation}
respectively, where, $M_\square^{-1}~{\rm and}~M_\Delta^{-1}$ are $3 \times 3$ matrices given by,
\begin{eqnarray}
  \left(M_\square^{-1}\right)^{\alpha \beta} & = & \sum_{a,b,c=1}^3\left(O_{a1}O_{b2}O_{c3}\right)^2  m_E^{\alpha \beta} I_4\left({m^{\alpha\beta}_E},m_{H_a},m_{H_b},m_{H_c}\right),\nonumber\\
  \left(M_\Delta^{-1}\right)^{\alpha \beta} & = & \sum_{a,b=1}^3 \left(O_{a1}O_{b2}\right)^2 m_E^{\alpha \beta} I_3\left(m^{\alpha\beta}_E,m_{H_a},m_{H_b}\right).
\end{eqnarray}
$m_E^{\alpha\beta}$ is the element of vector-like doubly charged fermion mass matrix (defined in ${\cal L}_{Yukawa}$) and $I_3(m^{\alpha\beta}_E,m_{H_a},m_{H_b})$ and $I_4(m^{\alpha\beta}_E,m_{H_a},m_{H_b},m_{H_c})$ are the loop integral factors given by,
\begin{eqnarray}
  I_3(m_A,m_B,m_C) &=& \frac{m^2_Aln\left(\frac{m^2_C}{m^2_A}\right)}{\left(m^2_A-m^2_B\right)\left(m^2_A -m^2_C\right)}+\frac{m^2_Bln\left(\frac{m^2_C}{m^2_B}\right)}{\left(m^2_B-m^2_A\right)\left(m^2_B -m^2_C\right)},\nonumber\\
  \frac{1}{v^2}\,I_4(m_A,m_B,m_C,m_D)&=&\frac{m^2_Aln\left(\frac{m^2_D}{m^2_A}\right)}{\left(m^2_A-m^2_B\right)\left(m^2_A -m^2_C\right)\left(m^2_A -m^2_D\right)}\nonumber\\
    &&+\frac{m^2_Bln\left(\frac{m^2_D}{m^2_B}\right)}{\left(m^2_B-m^2_A\right)\left(m^2_B -m^2_C\right)\left(m^2_B -m^2_D\right)}\nonumber\\
    &&+\frac{m^2_Cln\left(\frac{m^2_D}{m^2_C}\right)}{\left(m^2_C-m^2_A\right)\left(m^2_C -m^2_B\right)\left(m^2_C -m^2_D\right)}.\nonumber
\end{eqnarray}
It is important to note that one can always go to a particular $E^{++}_\alpha$ basis where $m_E$ is diagonal with positive entries. Therefore, in this basis,  $M_\square^{-1}~{\rm and}~M_\Delta^{-1}$ are also diagonal. Defining $M_{D}~=~\frac{\langle H \rangle^2}{16\,\pi^2} \left(\frac{\mu \mu^\prime}{v^2}M_\square^{-1}+\lambda M_\Delta^{-1}\right)~=~{\rm diag}\left(M_1,M_2,M_3\right)$, the loop induced neutrino mass matrix is given by,
\begin{equation}
  m_\nu~=~ \left(y_{\frac{5}{2}}^T M_D y_{\frac{3}{2}}+y_{\frac{3}{2}}^T M_D y_{\frac{5}{2}}\right)~=~U^*_{MNS}D_\nu U_{MNS}^\dagger,
  \label{lowhigh}
\end{equation}
where, $D_\nu~=~{\rm diag}\left(m_1,m_2,m_3\right)$ with $m_1,~m_2~{\rm and}~m_3$ being the masses of the SM neutrinos and $U_{MNS}$ is the Pontecorvo--Maki--Nakagawa--Sakata matrix \cite{Maki:1962mu,Pontecorvo:1967fh} determined by 3-angles and 3-phases (one Dirac phase and two Majorana phases). Note that in the low energy effective theory, neutrino mass matrix is determined by 9-parameters (3 neutrino masses, 3 mixing angles and 3 phases). Whereas, the number of parameters appearing from the high-energy theory are much larger.  In the basis where the SM charged lepton mass matrix and $m_E$ are diagonal, $y_{\frac{3}{2}}$ and $y_{\frac{5}{2}}$ contain 33 independent real parameters\footnote{$y_{\frac{3}{2}}$ and $y_{\frac{5}{2}}$ are $3\times 3$ complex matrices and hence, contain 18 real parameters each. However, 3 phases are unphysical and can be eliminated by a phase redefinition of the SM left handed lepton doublet ($L_{\alpha L}$).} and $m_E$ is determined by 3 additional parameters. However, not all the 36 parameters are independent in the context of the particular structure of the neutrino mass matrix in Eq.~\ref{lowhigh}. In particular, 3 more parameters can be eliminated by resealing the columns of the Yukawa matrices $y_{\frac{3}{2}}$ and $y_{\frac{5}{2}}$. Therefore, the light neutrino mass matrix is determined by 33 effective parameters at the high-scale. Since the number of parameters in the high-energy theory is much larger than the number of parameters describing low-energy neutrino phenomenology, proper parameterization \cite{Casas:2006hf,Casas:2001sr,Cordero-Carrion:2019qtu} of Yukawa matrices ($y_{\frac{5}{2}}$ and $y_{\frac{3}{2}}$) is required to ensure consistency of the model with the  available results from the neutrino oscillation experiments. In this regard, Eq.~\ref{lowhigh} can be re-written as,
\begin{equation}
  \left(D_{\sqrt \nu}^{-1}\right)^T U^T_{MNS}y_{\frac{5}{2}}^T M_{\sqrt D}^T M_{\sqrt D} y_{\frac{3}{2}} U_{MNS} D_{\sqrt \nu}^{-1}+\left(D_{\sqrt \nu}^{-1}\right)^T U^T_{MNS} y_{\frac{3}{2}}^T M_{\sqrt D}^T M_{\sqrt D} y_{\frac{5}{2}}U_{MNS} D_{\sqrt \nu}^{-1}~=~{\rm \bf I}_{3\times 3},\nonumber
\end{equation}
where, $M_{\sqrt D}~=~{\rm diag}\left(\sqrt{M_1},\sqrt{M_2},\sqrt{M_3}\right)$ and $D_{\sqrt \nu}^{-1}~=~{\rm diag}\left(m_1^{-\frac{1}{2}},m_2^{-\frac{1}{2}},m_3^{-\frac{1}{2}}\right)$. The most general form of matrices $Y_{\frac{3}{2}}$ and $Y_{\frac{5}{2}}$ which are consistent with the physical, low-energy neutrino parameters like, the three light neutrino masses ($m_1$, $m_2$ and $m_3$) and mixing angles as well as phases (contained in $U_{MNS}$), are given by,
\begin{eqnarray}
  M_{\sqrt D} y_{\frac{5}{2}} U_{MNS} D_{\sqrt \nu}^{-1}~=~A~~~&\Rightarrow&~~~y_{\frac{5}{2}}~=~M_{\sqrt D}^{-1}A D_{\sqrt \nu} U_{MNS}^{\dagger},\nonumber\\
  M_{\sqrt D} y_{\frac{3}{2}} U_{MNS} D_{\sqrt \nu}^{-1} ~=~B~~~&\Rightarrow&~~~y_{\frac{3}{2}}~=~ M_{\sqrt D}^{-1}B D_{\sqrt \nu} U_{MNS}^{\dagger},
  \label{param}
\end{eqnarray}
where, $A$ and $B$ are arbitrary $3 \times 3$ complex matrices subjected to the condition $A^TB~+~B^T A~=~{\rm \bf I}_{3\times 3}$. In Eq.~\ref{param}, the 33 parameters (contained in $y_{\frac{3}{2}}$, $y_{\frac{5}{2}}$ and $M^{-1}_{\sqrt D}$) introduced at the high-scale can be counted as 9 parameters in the SM neutrino sector (contained in $D_{\sqrt \nu}$ and $U_{MNS}$) plus 24 additional parameters contained in the matrices $A~{\rm and}~B$. It is important to note that in the limit $y_{\frac{3}{2}}~=~y_{\frac{5}{2}}$, the parameterization in Eq.~\ref{param} reduce to Casas-Ibarra parameterization proposed in Ref.~\cite{Casas:2006hf,Casas:2001sr}. After introducing the model and discussing the phenomenology in the context of neutrino masses and mixings, we are now equiped enough to discuss the collider phenomenology of this model. We have studied the collider phenomenology in the context of a simplified version of this model which will be discussed in the next section.

\subsection{Simplified version of the model for collider phenomenology}
In the previous section, we have considered 3-generations of doubly charged singlet fermions which are required to explain the observed data from neutrino oscillation experiments. However, in the context of collider phenomenology, we can easily restrict ourselves to the case of only one  doubly charged singlet fermion (denoted by $E^{++}$). Note that in the presence of more doubly charged fermion, it will be the lightest one which contributes dominantly to the collider signatures. With only one generation of doubly charged fermion in the scenario, the $3 \times 3$ Yukawa matrices $y_{\frac{3}{2}}$ and  $y_{\frac{5}{2}}$ in Eq.~\ref{lag_yuk} reduce to  $1\times 3$ vectors,
\begin{equation}
  y_{\frac{5}{2}}~=~\left(y_{\frac{5}{2}}^{eE},y_{\frac{5}{2}}^{\mu E},y_{\frac{5}{2}}^{\tau E}\right),~~~{\rm and}~~~y_{\frac{3}{2}}~=~\left(y_{\frac{3}{2}}^{eE},y_{\frac{3}{2}}^{\mu E},y_{\frac{3}{2}}^{\tau E}\right), \nonumber
\end{equation}
and the doubly charged fermion mass matrix $m_E$ is now a scalar. We further assumed real\footnote{We do not consider the phases of the Yukawa couplings ($f_{\frac{5}{2}\left(\frac{3}{2}\right)}$ and $y_{k}$) nor the ones of the PMNS matrix, $U_{MNS}$. Note that the phases of the Yukawa couplings do not play any significant role in the context of collider phenomenology.}
\begin{figure}[!t]
\centering
\includegraphics[width=0.7\linewidth]{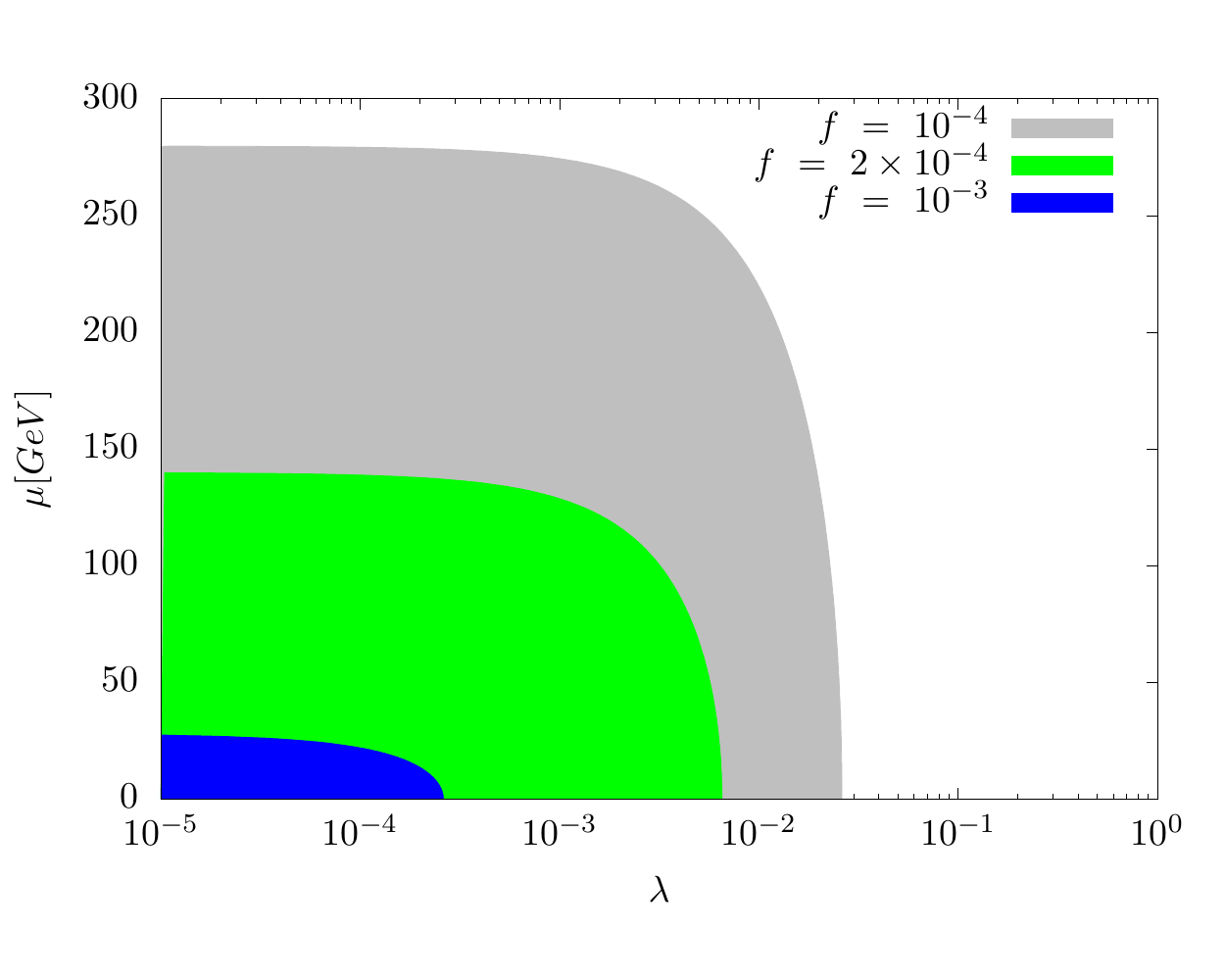}
\mycaption{The regions of parameter-space ($\mu$--$\lambda$ plane) which is consistent with  upper bound  ($~0.2$ eV) on the  absolute neutrino mass scale are depicted for three different values of the Yukawa couplings. For simplicity, we have assumed $\mu~=~\mu^\prime$  and $f~=~f_{\frac{5}{2}}~=~f_{\frac{3}{2}}$.}
\label{fig:nu_bound}
\end{figure}
$y_{\frac{5}{2}}$, $y_{\frac{3}{2}}$ and $y_{k}$ which couple to all three generations of the SM leptons with equal coupling strength (lepton flavour universality) {\em i.e.,}
\begin{equation}
y_{\frac{5}{2}\left(\frac{3}{2}\right)}^{eE}~=~y_{\frac{5}{2}\left(\frac{3}{2}\right)}^{\mu E}~=~y_{\frac{5}{2}\left(\frac{3}{2}\right)}^{\tau E}~=~f_{\frac{5}{2}\left(\frac{3}{2}\right)}~~~{\rm and}~~~y_{k}~=~f_{k}\,{\bf \rm I}_{3\times 3}, \nonumber
\end{equation}
where $f_{\frac{5}{2}\left(\frac{3}{2}\right)}$ and $f_{k}$ are real numbers. The loop induced neutrino masses (see Eq.~\ref{nu_mass}) are determined in terms of the Yukawa couplings $f_{\frac{5}{2}\left(\frac{3}{2}\right)}$, the tri-linear ($\mu$ and $\mu^\prime$) and the quartic ($\lambda$) scalar couplings (introduced in the scalar potential in Eq.~\ref{lag_scalar}) as well as the masses of the heavy fermions/scalars. Therefore, the allowed values of $f_{\frac{5}{2}\left(\frac{3}{2}\right)}$, $\mu$, $\mu^\prime$ and $\lambda$ are constrained from the upper limit on the  absolute neutrino mass scale ($\sum m_{\nu_{i}}~=~m_1+m_2+m_3$), defined as the sum of the masses of the neutrino mass eigenstates. Cosmological observations provide the strongest upper bound of about 0.2 eV \cite{Lesgourgues:2014zoa,Capozzi:2017ipn} on $\sum m_{\nu_{i}}$.  For TeV scale masses of the heavy fermion ($E^{++}$) and scalars ($\phi_{\frac{5}{2}}^{3+}$, $\phi_{\frac{3}{2}}^{+}$ and $H_i^{++}$), $M_{\square}^{-1}$ and $M_{\Delta}^{-1}$ in Eq.~\ref{nu_mass} are of the order of TeV$^{-1}$ and hence, $m_\nu^{\square}$ and $m_\nu^\Delta$ lighter than 0.1 eV require both $\frac{\mu \mu^\prime}{16\,\pi^2\,v^2}f_{\frac{5}{2}}f_{\frac{3}{2}}$ and $\frac{\lambda}{16\,\pi^2} f_{\frac{5}{2}}f_{\frac{3}{2}}~<~10^{-12}$, respectively. Assuming $\mu~=~\mu^{\prime}$ and $f_{\frac{5}{2}}~=~f_{\frac{3}{2}}$, in Fig.~\ref{fig:nu_bound}, we have depicted the region of parameter-space ($\mu$--$\lambda$ plane) which is consistent with  the upper bound  on the  absolute neutrino mass scale. In Fig.~\ref{fig:nu_bound}, we have assumed three different values of $f_{\frac{5}{2}\left(\frac{3}{2}\right)}$ of the same order of magnitude as the SM charged lepton Yukawa couplings. It can be seen from Fig.~\ref{fig:nu_bound} that only smaller values of $\mu$ and $\lambda$ are allowed. This has important consequences on the scalar mass spectrum and hence, on the collider phenomenology of this model.

\begin{figure}
  \begin{center}
    \begin{minipage}{.49\textwidth}
      \centering
  \begin{tikzpicture}[line width=1.4 pt, scale=1.65,every node/.style={scale=1.0}]
    \draw[fermion,black] (-1.0,1.0)  --(0.0,0.0);
    \draw[fermion,black] (-1.0,1.0) --(0.0,0.0);
    \draw[fermion,black] (-1.0,-1.0) --(0.0,0.0);
    \draw[vector,black] (0,0.0) --(1.0,0.0);
    \draw[fermion,black] (1,0.0) --(1.5,1.0);
    \draw[fermion,black] (1,0.0) --(1.5,-1.0);

    \draw[fermion,black] (1.5,1.0)  --(2.0, 1.0);
    \draw[fermion,black] (1.5,1.0) --(2.0,1.5);
    \draw[fermion,black] (1.5,1.0)  --(2.0,0.5);

    \draw[fermion,black] (1.5,-1.0)  --(2.0,-1.0);
    \draw[fermion,black] (1.5,-1.0)  --(2.0,-1.5);
    \draw[fermion,black] (1.5,-1.0)  --(2.0,-0.5);

    \node at (-0.5,0.8) {q};
    \node at (-0.5,-0.8) {$\bar{q}$};
    \node at (0.5,0.25) {$ \gamma / Z$};
    \node at (1.1,0.6) {$E^{++}$};
    \node at (1.1,-0.6) {$E^{--}$};
    \node at (1.7,1.4) {$e^{+}$};
    \node at (1.7,0.6) {$e^{+}$};
    \node at (2.1,0.9) {$\nu$};

    \node at (1.7,-1.4) {$e^{-}$};
    \node at (1.7,-0.6) {$e^{-}$};
    \node at (2.1,-1.1) {$\nu$};
  \end{tikzpicture}
    \end{minipage}
    \begin{minipage}{.49\textwidth}
      \centering
      \begin{tikzpicture}[line width=1.4 pt, scale=1.65,every node/.style={scale=0.9}]
    \draw[vector,black] (-1.0,0.0) --(0.0,0.0);
    \draw[vector,black] (-1.0,2.0) --(0.0,2.0);
    \draw[fermion,black] (0.0,0.0) --(0.0,2.0);
    \draw[fermion,black] (0.0,2.0) --(1.0,2.0);
    \draw[fermion,black] (0,0.0) --(1.0,0.0);

    \draw[fermion,black] (1.0,2.0) --(1.5,2.5);
    \draw[fermion,black] (1.0,2.0) --(1.5,1.5);
    \draw[fermion,black] (1.0,2.0) --(1.5,2.0);

    \draw[fermion,black] (1.0,0.0) --(1.5,-0.5);
    \draw[fermion,black] (1.0,0.0) --(1.5,0.5);
    \draw[fermion,black] (1.0,0.0) --(1.5,0.0);

    \node at (-0.5,2.2) {$\gamma$};
    \node at (-0.5,0.2) {$\gamma$};
    \node at (0.3,1.0) {$E^{++}$};
    \node at (0.5,2.2) {$E^{++}$};
    \node at (0.5,0.2) {$E^{--}$};
    \node at (1.2,0.4) {$e^{-}$};
    \node at (1.2,-0.4) {$e^{-}$};
    \node at (1.55,0.1) {$\nu$};
    \node at (1.2,2.4) {$e^{+}$};
    \node at (1.2,1.6) {$e^{+}$};
    \node at (1.55,1.9) {$\nu$};
  \end{tikzpicture}

    \end{minipage}

    \end{center}
  \caption{Feynman diagrams showing the Drell-Yan (left panel) and photo-fusion (right panel) pair production of doubly charged fermion and their subsequent decay into the SM leptons.}
  \label{prod}
\end{figure}
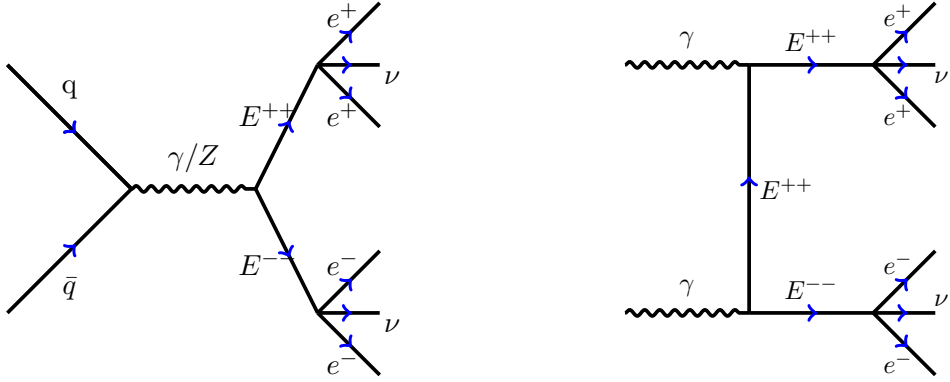

At tree level, the mass of the doubly charged fermion ($E^{++}$) is given by the parameter $m_E$. Being charged under the $U(1)_Y$, $m_E$ receives radiative corrections from the loops involving photon and $Z$-boson. The Yukawa interactions in Eq.~\ref{lag_yuk} also contribute to the radiative corrections via loops involving a heavy scalar and  a SM lepton. However, these corrections are suppressed by the Yukawa couplings. The corrections involving the SM gauge bosons in the loop are also estimated to be small (of the order of few hundred MeVs) \cite{Cirelli:2005uq}. Therefore, one can safely neglect the radiative corrections to $m_E$ in the context of the collider analysis. The tree level masses for the singly ($\phi_{\frac{3}{2}}^\pm$) and triply ($\phi_{\frac{5}{2}}^{3\pm}$) charged scalar are given by $m_{\phi^\pm}~=~m_{\frac{3}{2}}$ and  $m_{\phi^{3\pm}}~=~m_{\frac{5}{2}}$, respectively, whereas, the masses of the physical doubly charged ($H_i^{\pm\pm}$) scalars are given by the eigen values of the mass matrix in Eq.~\ref{lag_mass}. Since the allowed values of $\lambda$ and $\mu$ (see Fig.~\ref{fig:nu_bound}) are constrained from the upper bound on the absolute neutrino mass scale, the mixing between the doubly charged scalars ($\phi_{\frac{5}{2}\left(\frac{3}{2}\right)}^{\pm\pm}$ and $k^{\pm\pm}$) are small and hence, $H_1^{\pm\pm}$, $H_{2}^{\pm\pm}$ and $H_{3}^{\pm\pm}$ are dominantly $\phi_{\frac{5}{2}}^{\pm\pm}$, $\phi_{\frac{3}{2}}^{\pm\pm}$ and $k^{\pm\pm}$, respectively. Mixing between the doubly charged scalars gives rise to small mass splitting between triply(singly) charged scalar $\phi_{\frac{5}{2}}^{3\pm}(\phi_{\frac{3}{2}}^{\pm})$ and the corresponding doubly charged scalar $H_1^{++}(H_2^{++})$. At the leading order in $\lambda$ and $\mu$, the mass splitting can be calculated as,  
\begin{equation}
  m_{\phi^{3\pm(\pm)}}-m_{H_{1(2)}^{\pm\pm}}~\approx~ \frac{v^2}{8\,m_{\frac{5}{2}\left(\frac{3}{2}\right)}}\,\,\left(\frac{4\mu^2}{m_{k}^2-m_{\frac{5}{2}\left(\frac{3}{2}\right)}^2}\,\, + \,\,\frac{\lambda^2v^2}{m_{\frac{3}{2}\left(\frac{5}{2}\right)}^2-m_{\frac{5}{2}\left(\frac{3}{2}\right)}^2}\right)\,\, ,
  \label{splitting}
\end{equation}
for $m_{\frac{5}{2}} ~\neq~m_{\frac{3}{2}}~<~m_k$.  For TeV scale masses of the heavy fermion/scalars and $\lambda\,v~\sim~\mu ~\sim~10^2$ GeV (the largest possible values consistent with the bound on absolute neutrino mass scale), the mass splittings (due to the mixing between doubly charged scalars) between $\phi^{3\pm}$ and $H_{1}^{\pm\pm}$ as well as $\phi^{\pm}$ and $H_{2}^{\pm\pm}$ are estimated to be of the order of few tens of MeVs. Radiative corrections also contribute to mass splittings between different components of the scalar doublets. In the context of present scenario, the mass splittings due to radiative corrections are estimated to be of the order of GeV~\cite{Cirelli:2005uq}. In particular, the radiative splitting between the triply- and doubly-charged component of the $Y=\frac{5}{2}$ doublet  is estimated to be $m_{\phi^{3\pm}}-m_{H_1^{\pm\pm}}~\sim~\frac{5}{2}\alpha_{EM}\, m_Z~\sim~1.8$ GeV. Whereas, the loop induced splitting between doubly- and singly-charged component of the $Y=\frac{3}{2}$ doublet is given by $m_{H_2^{\pm\pm}}-m_{\phi^{\pm}}~\sim~\frac{3}{2}\alpha_{EM}\, m_Z~\sim~1.1$ GeV. These splittings are small, though they play a crucial role in determining the decays of the (multi-)charged scalars which will be discussed in details in section~\ref{scalar_phenomenology}. To summarize, the scalar spectrum of this model contains a degenerate pair of triply ($\phi^{3\pm}$) and doubly ($H_{1}^{\pm\pm}$) charged scalars ($m_{\phi^{3\pm}}~\approx~m_{H_{1}^{\pm\pm}}~\approx~m_{\frac{5}{2}}$), another degenerate pair of singly ($\phi^{\pm}$) and doubly ($H_{2}^{\pm\pm}$) charged scalars ($m_{\phi^{\pm}}~\approx~m_{H_{2}^{\pm\pm}}~\approx~m_{\frac{3}{2}}$) and a doubly charged scalar, $H_{3}^{\pm\pm}$ ($m_{H_{3}^{\pm\pm}}~\approx~m_{k}$).

After introducing the phenomenological model, we are now equipped enough to discuss it's signatures at the LHC. However, before going into the discussion of collider signatures, it is important to discuss about the low-energy constraints on the parameter space of this model resulting from the observables like, lepton-flavor violations, muon $g-2$, oblique parameters {\em e.t.c.} It has been shown in Ref.~\cite{Cheung:2017kxb} that the discrepancy between the experimental measurement and the SM prediction of muon magnetic moment $(g-2)_\mu$ \cite{Davier:2010nc,Aoyama:2012wk,Keshavarzi:2018mgv,Hagiwara:2011af} can be explained for  a substantial part of the parameter space while satisfying the experimental upper bounds \cite{Adam:2013mnn,TheMEG:2016wtm,Lindner:2016bgg,Meucci:2019jog} on the lepton-flavor violating decays like, $\mu~\to~e \gamma$, $\tau~\to~e \gamma$, $\tau~\to~\mu \gamma$ {\em e.t.c.} Whereas, the contributions to the oblique parameters \cite{Cheung:2017kxb} are automatically suppressed due to the degeneracy between $\phi^{3\pm}_{\frac{5}{2}}(\phi^{\pm}_{\frac{3}{2}})$ and $H_{1}^{\pm\pm}(H_{2}^{\pm\pm})$.

\section{Phenomenology of doubly charged fermion}
\label{E_phenomenology}

\begin{figure}[!t]
  \centering
  \includegraphics[width=0.48 \linewidth]{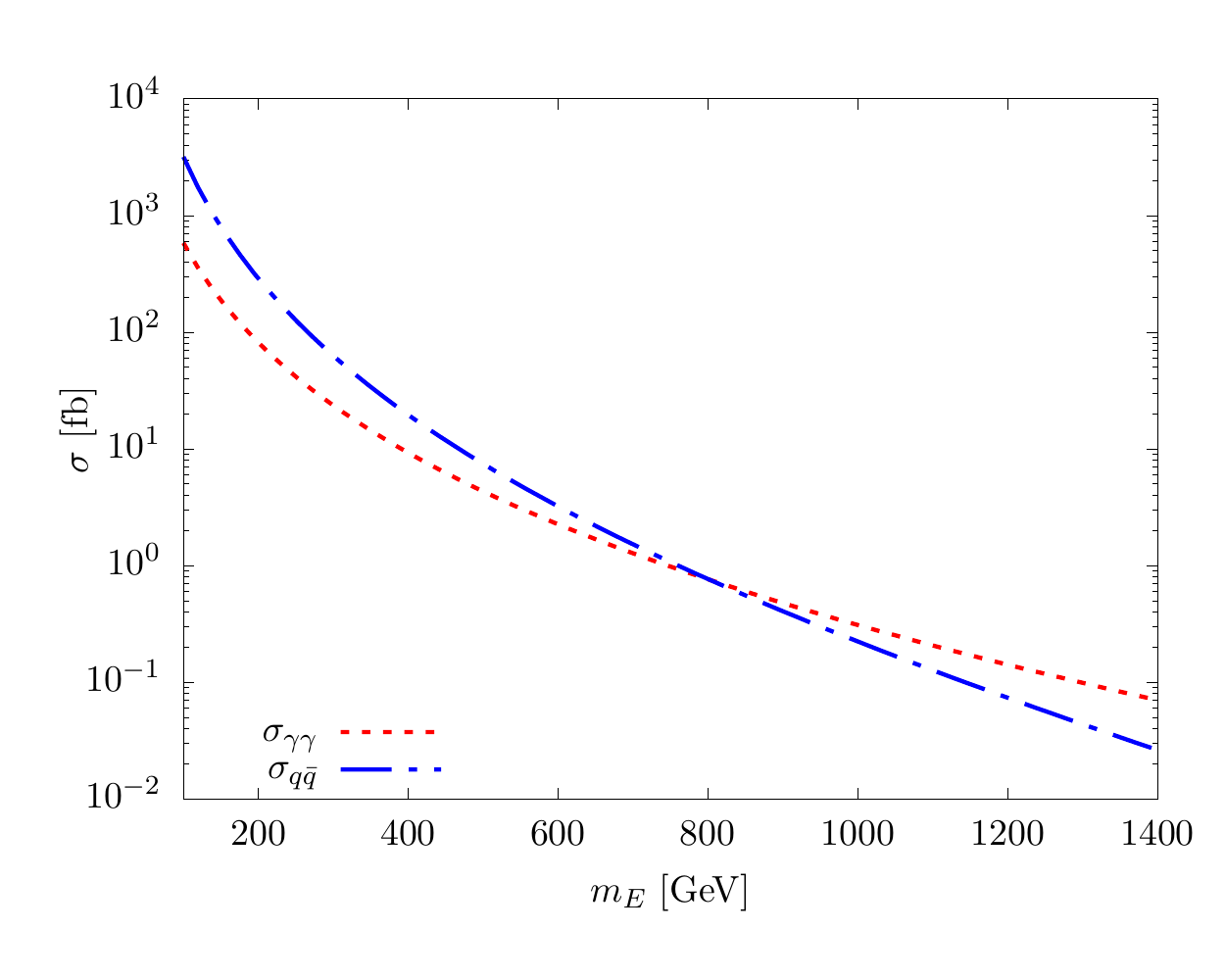}
  \includegraphics[width=0.48 \linewidth]{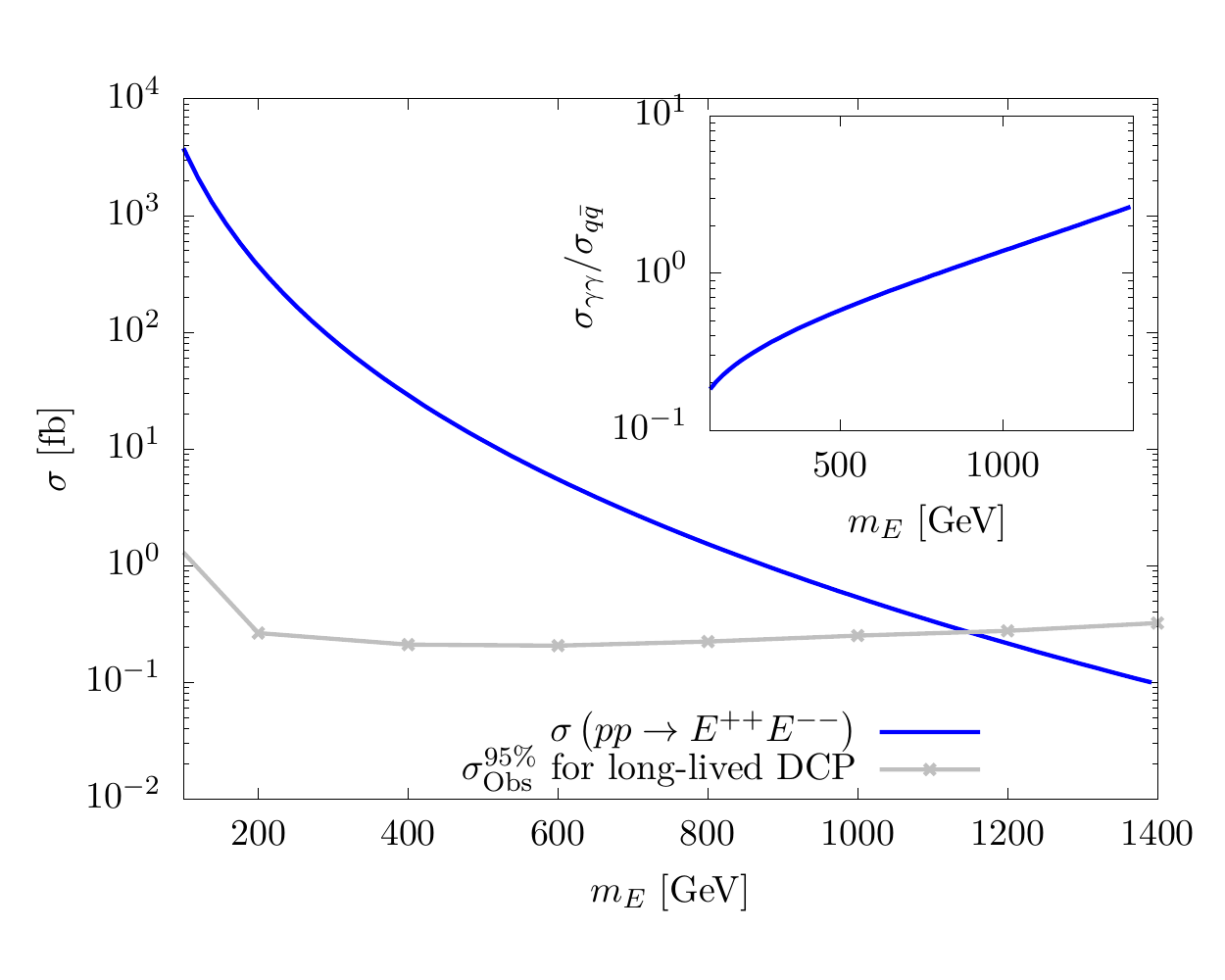}
  \mycaption{(Left panel) Drell-Yan ($\sigma_{q\bar q}$) and photon-fusion ($\sigma_{\gamma\gamma}$) contributions to the pair production of doubly charged fermions are presented as a function of $m_E$ at the LHC with 13 TeV center-of-mass energy. (Right panel) The model prediction for the total ($\sigma_{q\bar q}+\sigma_{\gamma\gamma}$) production cross-section of $E^{\pm\pm}$-pairs is presented. Inset shows the ratio of photon-fusion and Drell-Yan contribution. The gray solid line corresponds to the ATLAS observed 95\% CL upper limit on the pair production cross-section ($\sigma_{\rm Obs}^{95}$) of long-lived doubly-charged particles (DCPs) \cite{Aaboud:2018kbe}. }
\label{cross_E}
\end{figure}

In this section, we will discuss the collider signatures of the doubly charged fermion ($E^{\pm\pm}$). The Lagrangian (see Eq.~\ref{lag_kin},~\ref{lag_yuk} and \ref{lag_scalar}) of this model does not allow single production\footnote{The doubly charged fermion can be singly produced in association with a SM lepton and a heavy multi-charged scalar via the Yukawa interactions in Eq.~\ref{lag_yuk}. Such single production cross-sections are suppressed by the small Yukawa couplings  as well as the additional propagator and $2~\to ~3$ phase-space. However, we do not consider such process as the single production of $E^{\pm\pm}$ because of the presence of a heavy multi-charged scalar in the final state.} of $E^{\pm\pm}$. However, the doubly charged fermion can be pair produced via the gauge interactions with the SM photon and $Z$-boson (see Appendix~\ref{feyn}). The pair production of $E^{\pm\pm}$ at the hadron colliders takes place via  quark anti-quark initiated DY process  with a photon ($\gamma$) or $Z$-boson in the $s$-channel as shown in Fig.~\ref{prod} (left panel). Being electrically charged, $E^{\pm\pm}$ can also be pair produced via photo-fusion ($\gamma \gamma \to E^{++}E^{--}$) process as shown in Fig.~\ref{prod} (right panel). The photo-fusion (PF) of $E^{\pm\pm}$ pairs take place via the exchange of a $E^{\pm\pm}$ in the $t/u$-channel and hence, is relatively less suppressed by the parton center-of-mass energy as compared to the $s$-channel DY production. Moreover, $E^{\pm\pm}$ being a doubly charged particle, the PF cross-section is enhanced by a factor of $2^4$ at the Born level. However, it is also important to note that photons, being electromagnetically interacting, are associated with small parton density at the LHC. In fact, the parton density of the photon is so small that most of the older versions of parton distribution functions (PDF's) do not include photon as a parton. However, inclusion of the photon as a parton with an associated parton distribution function is necessary if one wants to include QED correction to the PDF. In the era of precision physics at the LHC when PDF's are determined upto NNLO in QCD, NLO QED corrections are equally important for the consistency of calculations. Moreover, for some processes, PF could become significant (or even dominant in some cases) at high energies.  In view of these facts, different groups (NNPDF, MRST, CTEQ {\em e.t.c.} \cite{Ball:2014uwa,Ball:2013hta,Martin:2004dh,Schmidt:2015zda}) have already included photon as a parton with an associated parton distribution function into their PDF sets. In this work, we have considered both DY and PF pair production of the doubly charged fermion. The total pair production cross-section at the LHC is given by,
\begin{eqnarray}
  \sigma(pp~\to X\bar X)~=&&\int dx_1 dx_2 \, f_{\gamma/p}\left(x_1,\mu_F^2\right)f_{\gamma/p}\left(x_2,\mu_F^2\right)\, \hat{\sigma}_{\gamma\gamma}\nonumber\\
  &+&\sum_{q,\bar q}\int dx_1 dx_2 \, f_{q/p}\left(x_1,\mu_F^2\right)f_{\bar q/p}\left(x_2,\mu_F^2\right)\, \hat{\sigma}_{q \bar q},
  \label{cross_section}
\end{eqnarray}
where, $f_{i/p}\left(x,\mu_F^2\right)$s are the parton distribution functions for the i$^{\rm th}$ parton, $\hat \sigma_{\gamma \gamma}$ and $\hat \sigma_{q \bar q}$ are the partonic  PF and DY pair production cross-sections, respectively, $\mu_F$ is the factorization scale. At the leading order, $\hat \sigma_{PF}\left(\gamma \gamma \to E^{++}E^{--}\right)$ and $\hat \sigma_{DY}\left(q \bar q \to E\bar E\right)$ are given by,
\begin{eqnarray}
  \frac{d\hat \sigma_{q\bar q}^{E\bar E}}{d\Omega}&=&\frac{\alpha_{EM}^2}{3 \hat s^3}\left[\left(Q_q-\frac{T_{3,q}-Q_q{\rm sin}^2\theta_W}{{\rm cos}^2\theta_W}\frac{\hat s}{\hat s - m_Z^2}\right)^2+Q_q^2\left(1+{\rm tan}^2\theta_W\frac{\hat s}{\hat s - m_Z^2}\right)^2\right]\nonumber\\
  &&\times \sqrt{1-\frac{4m_E^2}{\hat s}}\left[\left(m_E^2-\hat u\right)^2+\left(m_E^2-\hat t\right)^2+2m_E^2\hat s\right],\nonumber\\
   \frac{d\hat \sigma_{\gamma \gamma}^{E\bar E}}{d\Omega}&=&\frac{8\alpha_{EM}^2}{ \hat s}\sqrt{1-\frac{4m_E^2}{\hat s}}\left[\frac{\hat s{\left(\hat s+4m_E^2\right)-8m_E^4}}{\left(m_E^2-\hat t\right)\left(m_E^2-\hat u\right)}-\frac{4m_E^4}{\left(m_E^2-\hat t\right)^2}-\frac{4m_E^4}{\left(m_E^2-\hat u\right)^2}-2\right],
  \label{cross_section_fermion}
\end{eqnarray}
where, $\hat s$, $\hat t$ and $\hat u$ are the usual Mandelstam variables, $\alpha_{EM}$ and $\theta_W$ are the fine-structure constant and the Weinberg angle, respectively, whereas, $Q_q$ and $T_{3,q}$ refer to the charge and the 3$^{\rm rd}$ component of isospin of the corresponding quarks in the initial state.

We have developed a parton-level Monte-Carlo computer program for the numerical evaluation of the integration in Eq.~\ref{cross_section}. We have used  the {\bf NNPDF23LO} \cite{Ball:2014uwa} parton distribution functions with the factorization ($\mu_F$) and renormalization scales kept fixed at the subprocess center-of-mass energy $\sqrt {\hat s}$. In the left panel of Fig.~\ref{cross_E}, we have presented the DY ($\sigma_{q\bar q}$) and the PF ($\sigma_{\gamma \gamma}$) production cross-section of $E\bar E$-pairs as a function of doubly charged fermion mass ($m_E$) at the LHC with 13 TeV center-of-mass energy. The ensuing total ($\sigma_{q\bar q}+\sigma_{\gamma \gamma}$) pair production cross-section is presented in the right panel of the Fig.~\ref{cross_E}. In the inset of Fig.~\ref{cross_E} (right panel), we have presented the ratio of the PF and the DY production cross-sections as a function of $m_E$. Fig.~\ref{cross_E} (left panel and also the inset of the right panel) shows that DY production rate is larger than the PF production rate for the doubly charged fermion mass lighter than about 800 GeV. In this region, PF production constitutes a relatively large fraction (about 10\% to 50\% depending on the value of $m_E$) of the total cross-section and hence, can not be neglected. For large doubly charged fermion masses ($m_E > 800$ GeV), DY production, being a $s$-channel process, suffers larger suppression compared to the $t/u$-channel PF production and hence, $\sigma_{\gamma \gamma}$ dominates over $\sigma_{q\bar q}$.  The total pair production cross-section varies from a few pb to 0.1 fb as we vary $m_E$ over a range 100--1400 GeV. Once produced, the doubly charged fermion decays (directly or via cascade involving heavy multi-charged scalars) into the SM leptons and/or gauge bosons giving rise to multi-lepton final states at the LHC.

\subsection{Decay of $E^{\pm\pm}$}
\label{sec:decay_E}

The decays of the doubly charged fermion, which will be discussed in this section, play a crucial role in determining the signatures of $E^{\pm\pm}$ at the LHC. The Yukawa interactions in Eq.~\ref{lag_yuk} result into couplings involving a doubly charged fermion, a multi-charged scalar and a SM lepton (see Appendix~\ref{feyn}). Therefore, if kinematically allowed ($m_E~>~m_{\frac{5}{2}}$ and/or $m_{\frac{3}{2}}$), $E^{\pm\pm}$ undergoes 2-body decays into a multi-charged scalar in association with a SM lepton: $E^{\pm\pm}~\to~\phi^{\pm} l^{\pm},~\phi^{3\pm} l^{\mp}$ and $H_{a}^{\pm}\nu_l$, where $l$ includes all three generations of the SM leptons namely, electron ($e$), muon ($\mu$)  and tau ($\tau$), and $\nu_l$ is the $l$-neutrino. The partial decay widths for the 2-body decay modes of $E^{\pm\pm}$ are given by,
\begin{eqnarray}
    &&\Gamma\left({E^{\pm\pm}~\to~ \phi^{\pm(3\pm)}+l^{\pm(\mp)}}\right)~=~\frac{\left|f_{\frac{3}{2}\left(\frac{5}{2}\right)}\right|^2}{32 \pi}\,m_E\, \left(1-\frac{m^2_{\frac{3}{2}\left(\frac{5}{2}\right)}}{m^2_E}\right)^2,\nonumber\\
    &&\Gamma\left({E^{\pm\pm}~\to~ H_a^{\pm\pm}+\nu_l}\right)~=~\frac{\left|f_{\frac{5}{2}}O_{a1}\right|^2+\left|f_{\frac{3}{2}}O_{a2}\right|^2}{32 \pi}\, m_E \, \left(1-\frac{m^2_{H_a}}{m_E^2}\right)^2,
\end{eqnarray}
where, $O_{a1}$ and $O_{a2}$ are the elements of doubly charged scalar mixing matrix defined in Eq.~\ref{mixing}.

\begin{figure}
  \begin{center}

  \begin{tikzpicture}[line width=1.4 pt, scale=1.65,every node/.style={scale=0.9}]
    \draw[fermion,black] (-1.5,0.0) --(0.0,0.0);
    \draw[fermion,black] (0.0,0.0) --(0.8,0.6);
    \draw[scalar,black] (0.0,0.0) --(1.5,-1.0);
    \draw[fermion,black] (1.5,-1.0) --(2.5,-0.5);
    \draw[fermion,black] (1.5,-1.0) --(2.5,-1.5);
    \node at (-0.5,0.2) {$E^{\pm\pm}$};
    \node at (0.85,0.5) {$\nu_l$};
    \node at (0.5,-0.8) {${H_a}^{\pm\pm}$};
    \node at (2.0,-0.42) {$l^{\pm}$};
    \node at (2.0,-1.45) {$l^{\pm}$};
  \end{tikzpicture}

    \end{center}
  \caption{Feynman diagram showing the tree-level 3-body decay of the doubly charged fermion ($E^{\pm\pm}$).}
  \label{3-body}
\end{figure}
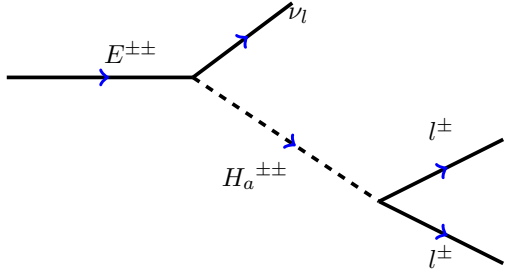

\begin{figure}[!ht]
  \centering
  \includegraphics[width=0.8 \linewidth]{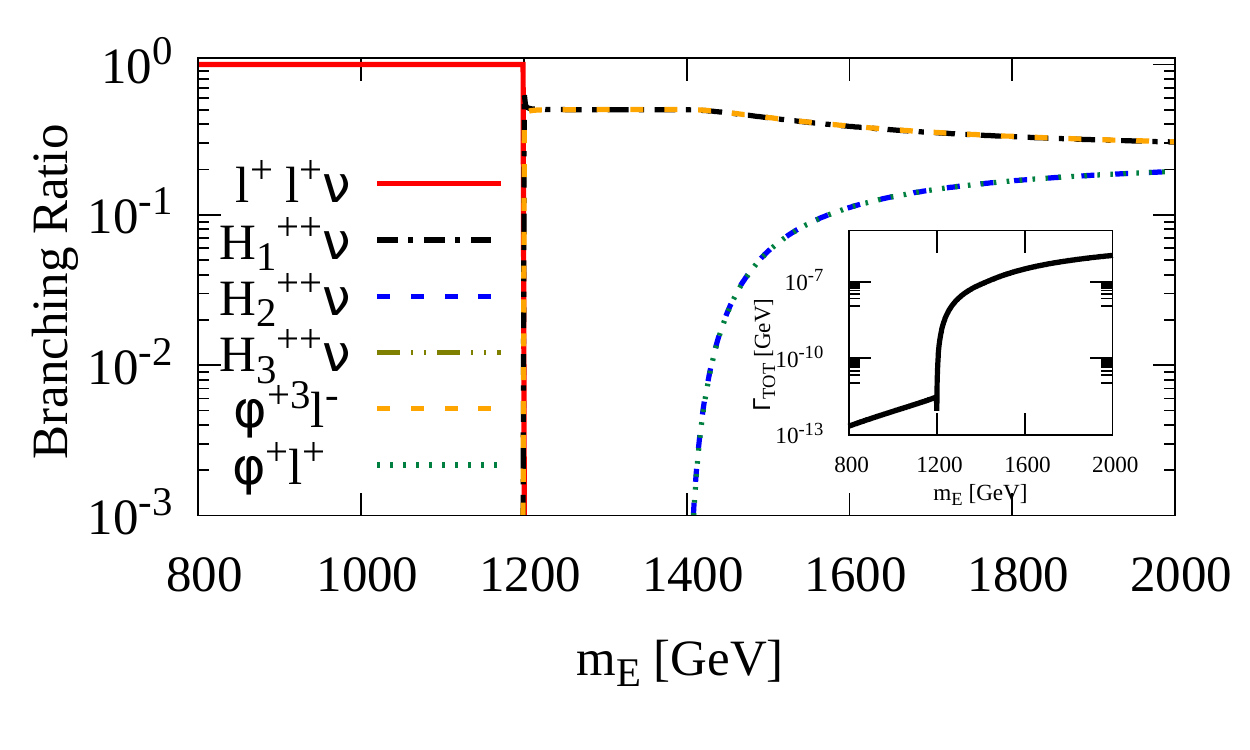}
\mycaption{Branching ratios of $E^{++}$ as a function of $m_E$ for $m_{\frac{5}{2}\left(\frac{3}{2}\right)}=1.2(1.4)~{\rm TeV},~m_{k}=1.5~{\rm TeV},~\mu=\mu^{\prime}=100~{\rm GeV},~f_{\frac{5}{2}\left(\frac{3}{2}\right)}=2\times 10^{-4},~\lambda=5\times 10^{-3}$ and $f_k=1.0$. The total decay width ($\Gamma_{TOT}$) of $E^{++}$ is presented in the inset.}
\label{br_E}
\end{figure}

\begin{figure}[!ht]
  \centering
  \includegraphics[width=0.9 \linewidth]{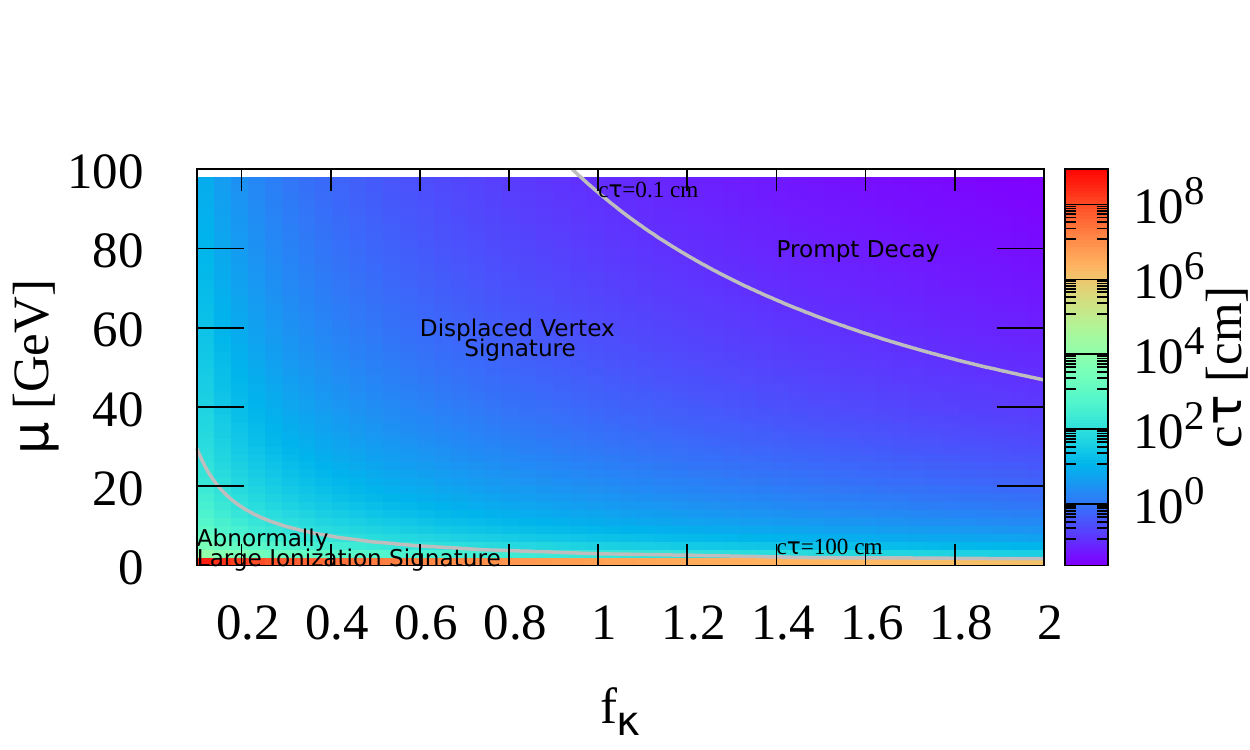}
\mycaption{The decay length (color gradient) of $E^{\pm\pm}$ is presented as a function of $f_k$ and $\mu$ for $m_E=800$ GeV. Other parameters are same as Fig.~\ref{br_E}.}
\label{length_E}
\end{figure}

If the 2-body decays of $E^{\pm\pm}$ are kinematically forbidden {\em i.e.,} $m_E~<~m_{\frac{3}{2}\left(\frac{5}{2}\right)}$, $E^{\pm\pm}$ undergoes tree-level 3-body decays into a neutrino in association with a pair of same-sign SM charged lepton. The 3-body decays proceed through an off-shell doubly charged scalar as depicted in Fig.~\ref{3-body}. The partial decay width of the 3-body decay is given by,
\begin{equation}
  \Gamma\left({E^{\pm\pm}~\to~ \nu_{l^\prime}l^{\pm}l^{\pm}}\right)~=~\frac{f_k^2}{512\pi^3}\, m_E\,\sum_{a=1}^3\,O_{a3}^2\left(\left|f_{\frac{5}{2}}O_{a1}\right|^2+\left|f_{\frac{3}{2}}O_{a2}\right|^2\right)\,I\left(\frac{m_{H_a}^2}{m_E^2}\right),
\label{eq:3-body}
\end{equation}
where,
\begin{equation}
I(x)~=~\int_0^1d\xi_1\int_0^1d\xi_2\,\frac{\xi_2\left(1-\xi_2\right)^2}{\left(\xi_2-x\right)^2}.\nonumber
\end{equation}
The branching ratios for the different decay modes of $E^{\pm\pm}$ are presented in Fig.~\ref{br_E} as a function of $m_E$.  We have assumed $f_{\frac{5}{2}}~=~f_{\frac{3}{2}}~=~2\times 10^{-4}$ and the other parameters are given by $m_{\frac{5}{2}\left(\frac{3}{2}\right)}=1.2(1.4)~{\rm TeV},~m_{k}=1.5~{\rm TeV},~\mu=\mu^{\prime}=100~{\rm GeV}~\lambda=5\times 10^{-3}$ and $f_k=1$\footnote{The Yukawa couplings involving a doubly charged singlet scalar and two SM right-handed leptons are not constrained from the upper bound on the absolute neutrino mass scale and hence, could be large. In fact, large $f_k$ is required to ensure prompt decay of $E^{\pm\pm}$ when 2-body decays are forbidden.}. Fig.~\ref{br_E} shows that $E^{\pm\pm}$ dominantly decays into a SM lepton plus a heavy charged scalar when kinematically allowed.  Same branching ratios for $H_1^{\pm\pm} \nu(H_2^{\pm\pm}\nu)$ and $\phi^{3\pm}l^\mp(\phi^{\pm}l^\pm)$ decay modes are a consequence of the fact that $H_1^{\pm\pm}(H_2^{\pm\pm})$ dominantly belongs to the scalar doublet which also includes $\phi^{3\pm}(\phi^{\pm})$ and hence, both decay widths are determined by the same Yukawa coupling (see Eq.~\ref{lag_yuk}).
$H_3^{\pm\pm}$ being dominantly the singlet doubly charged scalar ($k^{++}$), the decay $E^{\pm\pm}~\to~H_3^{\pm\pm}\nu$ is suppressed by the mixing ($O_{31}$ and $O_{32}$) in the doubly charged scalar sector and hence, is not visible in Fig.~\ref{br_E}. When the 2-body decays are kinematically forbidden, $E^{\pm\pm}$ undergoes 3-body decay into $l^\pm l^\pm\nu$ with 100\% branching fraction. In the inset of Fig.~\ref{br_E}, the total decay width ($\Gamma_{TOT}$) is presented as a function of $m_E$. When 2-body decays are kinematically allowed, the total decay width ${\cal O}\left({f_{\frac{5}{2}\left(\frac{3}{2}\right)}^2m_E}/{32\pi}\right)$ is large ($\Gamma_{TOT}>10^{-13}$ GeV) enough to ensure the prompt decay of $E\bar E$ pairs produced at the LHC. However, Eq.~\ref{eq:3-body} shows that the 3-body decays are suppressed by small Yukawa couplings ($f_{\frac{5}{2}\left(\frac{3}{2}\right)}$) as well as by one of the off-diagonal element of the doubly charged scalar mixing matrix, $O$. The inset of Fig.~\ref{br_E} shows that for $m_E~<~m_{\frac{5}{2}\left(\frac{3}{2}\right)}$ where only 3-body decays are kinematically allowed, the total decay width is suppressed but not suppressed enough to ensure displaced vertex or highly ionizing track signature at the LHC. However, this conclusion is highly dependent on the choice of  parameters\footnote{The large values for $\mu$, $\lambda$ and $f_{k}$ are assumed to ensure prompt decay of $E^{\pm\pm}$.} which are used to produce Fig.~\ref{br_E}. The 3-body decay widths depend on the Yukawa couplings and the doubly charged scalar mixings which are determined by $\mu,~\mu^{\prime}$  and $\lambda$. To identify the parts of parameter space ($f_{k}$--$\mu$ plane) which give rise to prompt decay, displaced vertex and abnormally large ionization signature \cite{Aaboud:2018kbe} at the LHC, in Fig.~\ref{length_E}, we have plotted the decay length (by color gradient) of $E^{\pm\pm}$ as a function of $f_{k}$ and $\mu$ for a fixed value of $m_E~=~800$ GeV. The values of other parameters are same as in Fig.~\ref{br_E}. The three regions of $f_{k}$--$\mu$ plane giving rise to prompt decay, displaced vertex and abnormally large ionization signature are clearly indicated in Fig.~\ref{length_E}.

\subsection{Collider signatures}

After discussing the production and decay of the doubly charged fermion, we are now equipped enough to study the signatures of $E^{\pm\pm}$ at the LHC with $\sqrt s~=~13$ TeV. The collider signatures of $E\bar E$-pairs at the LHC can be broadly categorized into two classes depending on the total decay width of $E^{\pm\pm}$. If the total decay width is large enough ($\Gamma_{TOT}~>~10^{-13}$ GeV), {\em i.e.,} the decay length is small enough ($<$1 mm), to ensure the decay of $E^{\pm\pm}$ inside the detector, the collider signatures are determined by the SM leptons/jets and missing energy resulting from the decay of $E\bar E$-pairs. However, if the doubly charged fermion is long-lived ({\em i.e.,} $\Gamma_{TOT}~<~10^{-16}$ GeV) and remains stable inside the LHC detectors {\em i.e.,} the decay length is larger than few meters,  production of $E\bar E$-pairs give rise to abnormally large ionization at the LHC detectors \cite{Aaboud:2018kbe}.

\subsubsection{Long-lived $E^{\pm\pm}$: Abnormally large ionization signature}
It has already been discussed in the previous section and shown in Fig.~\ref{length_E} that certain parts of the parameter space of this model give rise to a long-lived $E^{\pm\pm}$ which passes the entire LHC detectors without decaying. Being doubly charged, a long-lived $E^{\pm\pm}$ is highly ionizing, and thus leave a very characteristic signature of abnormally large ionization in the detector. This particular signature is quite interesting and clean because of the negligible background from the SM. The SM does not have any multi-charged particle and hence,  does not give rise to large ionization at the LHC. Such signatures have already been searched by the ATLAS collaboration \cite{Aaboud:2018kbe} with 36.1 fb$^{-1}$ integrated luminosity data and no such events were found. In absence of any observed events, 95\% confidence level (CL) upper limits on the pair-production cross-sections of long-lived multi-charged particles (MCPs) as a function of MCP masses and for different MCP charges are derived in Ref.~\cite{Aaboud:2018kbe}. In Fig.~\ref{cross_E} (right panel), we have also plotted the 95\% CL upper limits on the pair-production cross-sections of long-lived doubly-charged particles ($\sigma^{95\%}_{DCPs}$) along with the model prediction for the $E\bar E$-pair production cross-section. Fig.~\ref{cross_E} (right panel) shows that for a long-lived $E^{\pm\pm}$, doubly-charged fermion mass below about 1150 GeV is excluded from the ATLAS search for long-lived MCPs.

\subsubsection{Prompt decay of $E^{\pm\pm}$: Multi-leptons signature}
\label{E_coll}

The signatures of doubly charged fermion with prompt decay ({\em i.e.,} with large enough decay width to ensure the decay of $E^{\pm\pm}$ at the production vertex) depend on the allowed decay modes and branching ratios. As discussed in section~\ref{sec:decay_E}, if 2-body decays are kinematically possible ($m_E~>~m_{\frac{5}{2}\left(\frac{3}{2}\right)}$ and/or $m_{k}$), $E^{\pm\pm}$ dominantly decays into a heavy (multi-)charged scalar in association with a SM lepton. The (multi-)charged scalars decay further and the resulting collider signatures of $E\bar E$-pair production is determined by the subsequent decays of these (multi-)charged scalars which will be discussed in detail in the next section. If 2-body decays of the doubly charged fermion are kinematically forbidden {\rm i.e.,} $E^{\pm\pm}$ is lighter than the (multi-)charged scalars ($m_E~<~m_{\frac{5}{2}\left(\frac{3}{2}\right)}$ and $m_{k}$),  $E^{\pm\pm}$ dominantly decays into a pair of same-sign SM charged leptons in association with a neutrino. Therefore, for $m_E~<~m_{\frac{5}{2}\left(\frac{3}{2}\right)}$ and $m_{k}$, the production of $E\bar E$-pairs at the LHC gives rise to 4-leptons and two neutrinos in the final state. Neutrinos, being weakly interacting, remain elusive in the detector resulting in missing transverse energy signature: $pp~\to~E^{\pm\pm}E^{\mp\mp}~\to~4{\rm \text-leptons}~+~E_T\!\!\!\!\!\!/~$. In the context of the LHC with 13 TeV center-of-mass energy, we have studied 4-leptons plus missing energy final state as a signature of $E\bar E$-pair production. The 4-leptons signature has already been searched by the ATLAS collaboration \cite{Aaboud:2018zeb} as a signature of electroweakinos in the context of simplified R-parity conserving as well as R-parity violating supersymmetric scenarios. Ref.~\cite{Aaboud:2018zeb} uses 36.1 fb$^{-1}$ integrated luminosity data of the LHC running at 13 TeV center-of-mass energy. Data yields are found to be consistent with SM expectations. The consistency between data and the SM prediction results into a 95\% CL upper limit on the visible 4-leptons cross-section. We have used the ATLAS upper limit on the visible 4-leptons cross-section in the context of our model to constrain the mass of doubly charged fermion. We have closely followed the object (electrons, muons, jets, missing energy) reconstruction and event selection criteria used by the ATLAS collaboration in Ref.~\cite{Aaboud:2018zeb}.

\paragraph{Signal and Background}
\label{4l_back}
Several SM processes also result in the 4-leptons final state. The leading SM backgrounds for $4l$ arise from the hard-scattering processes (HSP) resulting in four or more leptons and $ZZ$ production followed by the leptonic decay of both the $Z$-bosons. Production of top anti-top ($t\bar t$) pairs in association with a pair of leptons ($pp~\to~t\bar t Z/\gamma^*$) contributes to the $4l$ background when the $t\bar t$-pairs decay leptonically. Production of tri-bosons ($ZZZ,~WZZ$, and $WWZ$) and Higgs boson also give rise to $4l$ final state. Backgrounds result from the production of $t\bar t t\bar t$, $t \bar t t W$, $t\bar t H$, $ZH$, $WH$ are highly suppressed due to small production cross-sections as well as by the leptonic branching ratios. The production of $t\bar t$, $Z$+jets, $t\bar t W$, $WZ$+jets, $WW$+jets, $WWW$+jets {\em e.t.c.} may also contribute to $4l$ background (reducible background) if one or more jets are misidentified as leptons. Since the probability of mistagging a jet as a lepton is small, the reducible $4l$ backgrounds are estimated \cite{Aaboud:2018zeb} to be negligible compared to the irreducible backgrounds. Therefore, in our analysis, we have not calculated these reducible backgrounds.

We have used a parton level\footnote{At the parton level, the production and subsequent decays of $E\bar E$-pairs give rise to purely leptonic final states without any additional quarks/gluons. Quarks/gluons might result from the initial state radiation (ISR). However, the signal selection (which will be discussed in the next section) relies only on leptons in the final state. Therefore, hadronization of quarks/gluons and subsequent decays of the hadrons will not have any significant effect on the calculation of signal and background cross-sections after the acceptance/selection cuts and hence, parton-level Monte-Carlo results can be trusted in this case.} Monte-Carlo program to simulate the production and subsequent decays of $E\bar E$-pairs. The phase space distributions of different kinematic variables for the signal are also computed in the framework of the same Monte-Carlo code. Different SM background processes are simulated at parton-level using MadGraph \cite{Alwall:2014hca}. We have used MadAnalysis \cite{Conte:2012fm} to study the MadGraph generated background events, compute different kinematic distributions, and impose cuts on different kinematic variables. Table~\ref{lob} shows a list of the SM processes which are simulated in MadGraph to estimate the SM contribution to $4l$-background. One should note that MadGraph@parton-level can not simulate ISR jets which might be important for some of the kinematic variables like, missing energy, effective mass {\em e.t.c.,}. To overcome this particular drawback of the MadGraph@parton-level, we have calculated the SM processes in association with some additional jets. For example, in the category of HSC/$ZZ$, we have calculated the SM production cross-sections of two positively charged and two negatively charged leptons in association with 0, 1, 2, and 3 additional quarks or gluons. The MadGraph@parton-level calculated background cross-sections are found to be consistent with the background numbers estimated by the ATLAS collaboration using sophisticated event simulations and detector level objects ($e,~\mu$, jets, missing energy) reconstructions.    
\begin{table}[h]
    \centering
    \begin{tabular}{c||c}
      \hline\hline
        Name &  Processes generated in MadGraph\\
      \hline\hline
      HSC/$ZZ$ & $2l^+2l^-$ + upto 3 jets \\
      $t\bar{t}ll$ & $t\bar{t}ll$ + upto 1 jet( leptonic decays)\\
      $VVZ$  &  $WWZ$ + upto 2 jets\\
      Higgs & $H$ + upto 2 jets, $WH$ + upto 2 jets, $ZH$ + upto 2 jets, $t\bar{t}H$ \\
      Others & $t\bar{t}t\bar{t}$, $t\bar{t}W^+W^-$\\
      \hline\hline
    \end{tabular}
    \caption{List to the SM processes which are calculated using MadGraph-MadAnalysis framework at the parton-level.}
    \label{lob}
\end{table}

\paragraph{Event Selection}
\label{4l_selection}

Since the SM contributes significantly to the final states similar to that we are interested in, one has to carefully examine and compare the phase space distributions of different kinematic variables for the signal as well as backgrounds and find some characteristics of our signal which are distinct from the SM processes. The characteristics of signal and background distributions will guide us to develop a systematic methodology of suppressing the SM backgrounds without drastically reducing the signal.

However, before going into the details of signal and background simulation and phase space distributions of different kinematic variables, it is important to list the  basic requirements for jets and isolated leptons to be visible as such. In this quest, it should be noted though that the LHC detectors have only a finite resolution. For any realistic detector, this is applicable to both transverse momentum ($p_T$) measurements as well as determination of the angle of motion. In our analysis, we have neglected the later\footnote{The angular resolution is, generically, far superior to the energy/momentum resolutions and too fine to be of any consequence at the level of sophistication of this analysis.} whereas, the former is simulated by smearing the energy with  Gaussian functions defined by an energy-dependent width\footnote{In general, width of the Gaussian smearing is also a function of the detector coordinates. However, we choose to simplify the task by assuming a flat resolution function equating it to the worst applicable for our range of interest.} as follows:
\beq
\frac{\sigma_E}{E} = \frac{a}{\sqrt{E}} \oplus b
\label{reso}
\eeq
where the errors are to be added in quadrature and
\beq
\barr{rclcrclcl}
a_\ell &= & 0.05 \ , &\quad& b_\ell &= &5.5 \times 10^{-3} \ & \qquad &
 {\rm for ~leptons,}
\\[1ex]
a_j &= &0.80 \ , & &  b_j &= &0.05 & & {\rm for~ partons}.
\earr
\eeq
In order to be visible at the LHC detectors, a jet or a lepton must have an adequately large transverse momentum and should fall well inside the rapidity coverage of the detector. To ensure the visibility of the jets and leptons at the LHC, we demand  
\beq
p_T({\rm jet}) > 20 \gev \ , \qquad p_T({\rm electron}) > 7 \gev \ , \qquad p_T({\rm muon}) > 5 \gev \ ,
\label{cut:pT}
\eeq
and
\beq
|\eta({\rm jet})| \leq 2.8 \ , \qquad |\eta({\rm electron}) | \leq 2.47 \\ , \qquad |\eta({\rm muon}) | \leq 2.7 \ .
\label{cut:eta}
\eeq
Furthermore, we demand the leptons and jets be well separated from each other by requiring
\beq
\Delta R_{ll} \geq 0.4 ~,~ \Delta R_{\ell j} \geq 0.4 ~{\rm and}~ \Delta R_{j \, j} \geq 0.4 \ .
\label{cut:jj-iso}
\eeq
where
$\Delta R \equiv \sqrt{(\Delta \eta)^2 + (\Delta \phi)^2} $. The detection and reconstruction efficiency of tau-leptons is significantly different from the electrons and muons. Therefore, we have only considered electrons and muons as leptons.  Unless specified otherwise, $l$ stands for electron and muon only ($l~\supset~ e,~\mu$) throughout the rest of this article. The requirements summarized by Eqns.~(\ref{cut:pT}--\ref{cut:jj-iso}) constitute our {\it acceptance cuts}. We tried to follow the acceptance and selection criteria used in the ATLAS search for $4l$~\cite{Aaboud:2018zeb} as closely as possible in framework of a parton-level Monte-carlo. With the set of {\em acceptance cuts} and {\em detector resolution} defined in Eqns.~(\ref{cut:pT}--\ref{cut:jj-iso}) and Eq.~\ref{reso}, respectively, we compute the $4l$\footnote{The signal and background require to have atleast 4-lepton (electron and/or muon only) in the final state. We do not impose any condition on the number of jets in the final state.} signal and background cross-sections at the LHC operating with $\sqrt s~=~13$ TeV and display them in Table~\ref{cut_flow}. Clearly, after the {\em acceptance cuts}, the SM backgrounds are order magnitude  large compared to the signal. Detailed analysis of different kinematic distributions is necessary to suppress the huge contributions from the SM background. Before moving to the discussion of different kinematic distributions and consequently, the event selection, it is important to define two phenomenologically important kinematic variables namely, the missing energy ($E_T\!\!\!\!\!\!/~$) and the effective mass ($M_{\rm eff}$) as follows,
\begin{eqnarray}
\not E_T &\equiv& \sqrt{ \bigg(\sum_{\rm vis.} p_x \bigg)^2
                 + \bigg(\sum_{\rm vis.} p_y \bigg)^2 }.\nonumber\\
M_{\rm eff} &\equiv& \not E_T + \sum_{i} p_T(l_i) + \sum_{i} p_T(j_i),
\label{meff}
\end{eqnarray}
where, the summation runs over all visible (consistent with the {\em acceptance cuts} listed in Eqns.~\ref{cut:pT}--\ref{cut:jj-iso}) leptons and jets.

\begin{figure}[!ht]
  \centering
  \includegraphics[width=0.48 \linewidth]{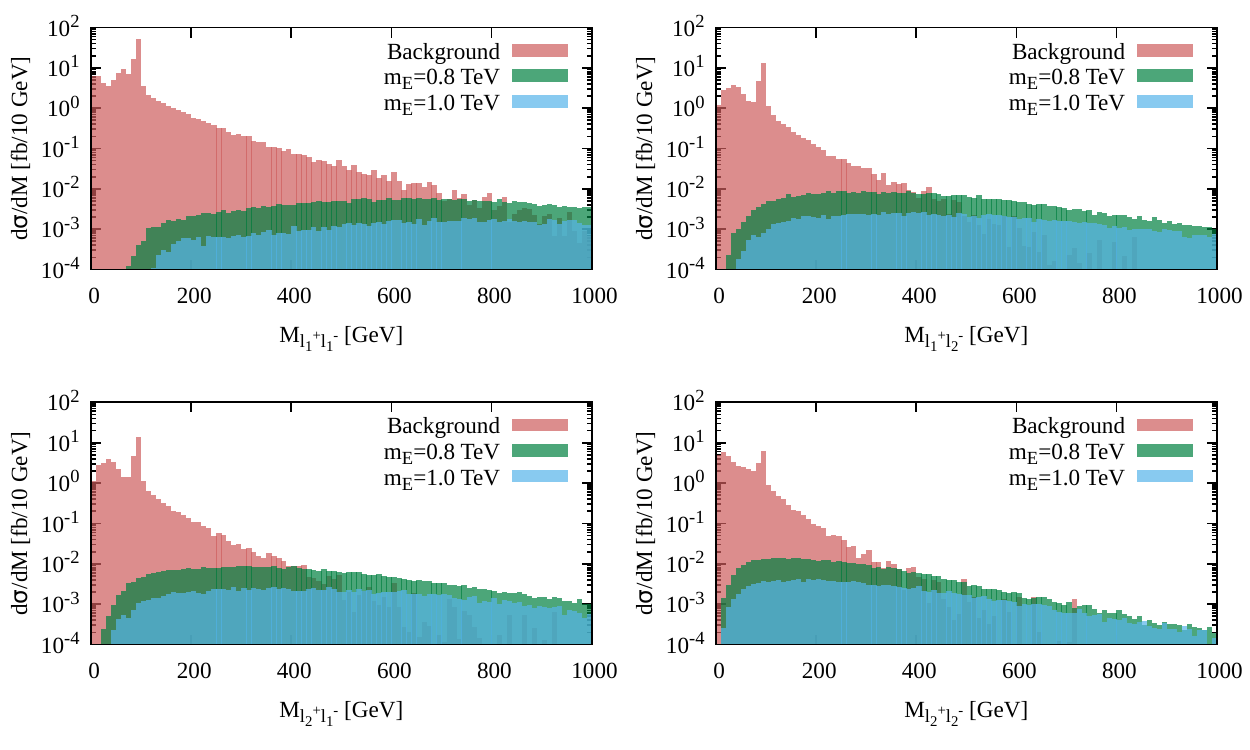}
  \includegraphics[width=0.48 \linewidth]{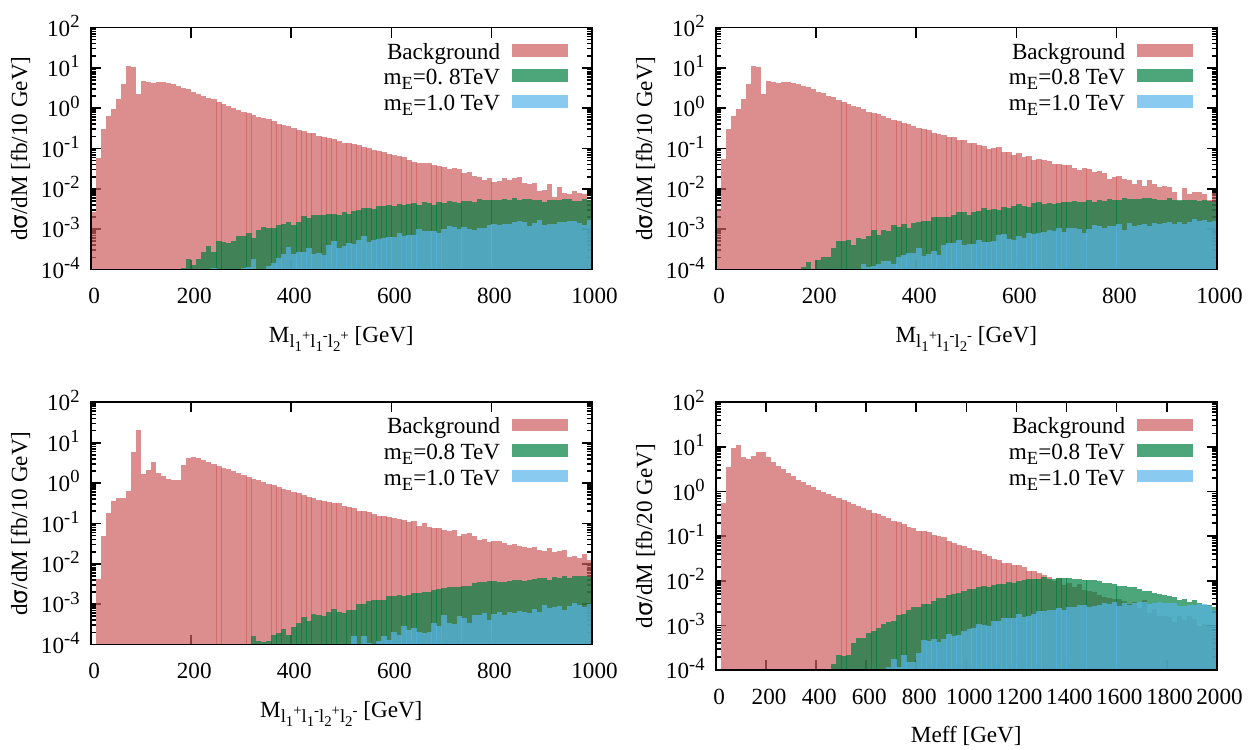}
\mycaption{(Left panel) Opposite sign dilepton invariant mass distributions (after ordering leptons according to their $p_T$ hardness, $p_T(l_{1}^{\pm})>p_T(l_2^\pm)$)  are presented after the acceptance cuts for both signal ($m_E=$ 0.8 and 1 TeV) and the SM background. Right panel top row shows Tri-lepton invariant mass ($M_{{\rm SFOS}+l}$) distributions. Four lepton invariant mass distribution ($M_{{\rm SFOS+SFOS}}$) and effective mass ($M_{eff}$) distribution are presented in the right panel bottom row.}
\label{dist_E_m}
\end{figure}

To design an event selection criteria for suppressing the background without significantly reducing the signal, one needs to understand the characteristics of the signal and background distributions. Table~\ref{cut_flow} (2$^{\rm nd}$ column) shows that after the {\em acceptance cuts}, dominant contribution to the background comes from HSP/$ZZ$ production. The leptonic decays of the $Z$-boson are characterized by a peak at $Z$-boson mass ($m_Z$) in the same-flavor, opposite-sign (SFOS) dilepton invariant mass distributions. Therefore, it is instructive to study each of the four possible SFOS dilepton invariant mass distributions ($M_{SFOS}$) namely, $M_{l_1^+ l_1^-}$, $M_{l_1^+ l_2^-}$, $M_{l_2^+ l_1^-}$ and $M_{l_2^+ l_2^-}$, constructed out of the momenta of the two leading\footnote{We have ordered the leptons according to their $p_T$ hardness. The positively charged and negatively charged lepton with higher (lower) $p_T$ are denoted by $l_1^+$ ($l_2^+$) and $l_1^-$ ($l_2^-$) respectively.}  positively and two leading negatively charged leptons.

  \begin{table}[h!]
    \centering
    \begin{tabular}{ c||c   }
      \hline\hline
      \multicolumn{2}{ c}{\em ATLAS cuts} \\
      \hline\hline
      Invariant mass cuts & Effective mass cuts \\\hline\hline
      $M_{\rm OS}~>~4$ GeV, $M_{\rm SFOS}~<~8.4$ or $M_{\rm SFOS}~>~10.4$, & \\
      $M_{\rm SFOS}~<~81.2$ or $M_{\rm SFOS}~>~101.2$ & \\
      $M_{\rm SFOS+l}~<~81.2$ or $M_{\rm SFOS+l}~>~101.2$, & $M_{\rm eff}~>~600$ GeV \\
      $M_{\rm SFOS+SPOS}~<~81.2$ or $M_{\rm SFOS+SPOS}~>~101.2$ & \\
      \hline\hline
    \end{tabular}
    \caption{Cuts implemented in the framework of our parton-level Monte-Carlo to adhere to the signal selection criteria proposed by ATLAS collaboration in Ref.~\cite{Aaboud:2018zeb}.}
    \label{ATLAS_cuts}
  \end{table}

In Fig.~\ref{dist_E_m} (four plots in the left panel), we show the SFOS dilepton invariant mass distributions for the SM background as well as for the signal with two different values of $m_E~=~0.8$ and 1.0 TeV. The $Z$-boson peak is clearly visible in the background SFOS invariant mass distributions. Therefore, one can easily suppress the background contributions $ZZ$ by imposing $Z$-veto {\em i.e.,}  excluding the parts of phase-space giving rise to $M_{SFOS}$ satisfying $|m_Z-M_{SFOS}|~<~10$ GeV. In our analysis, we have assumed $m_Z~=~90.1$ GeV and demanded $m_{SFOS}~<~81.2$ GeV or $m_{SFOS}~>~101.2$. To suppress the contributions from the leptonic decay of hadrons\footnote{Our calculation of signal and backgrounds are limited to parton-level and hence, the production of hadrons and their subsequent leptonic decays which might result in $4l$ final state were not considered. However, we have used ATLAS suggested cuts to suppress these contributions which are definitely present in a collider experiment.}, we further demand the invariant mass of opposite-sign dilepton $M_{OS}$ to be greater than 4 GeV and $M_{SFOS}$ to be outside the range 8.4--10.4 GeV. The set of cuts on the dilepton invariant mass discussed above are also used by the ATLAS collaboration in Ref.~\cite{Aaboud:2018zeb} and hence, fall in the category of {\em ATLAS cuts} defined in Table~\ref{ATLAS_cuts}. Fig.~\ref{dist_E_m} (left panel) shows that the signal SFOS dilepton distributions are shifted towards the larger invariant mass. This is a consequence of the fact that the signal leptons are coming from the 3-body decay of $E^{\pm\pm}$ with a mass of the order of TeV and hence, are usually associated with large transverse momentum. Therefore, one can introduce a lower bound on $M_{SFOS}$ to reduce the background significantly while minimally affecting the signal. In addition to the {\em ATLAS cuts} on $M_{SFOS}$, we demand $M_{SFOS}~>~150$ GeV. This falls into the category of our {\em proposed cuts} defined in Table~\ref{proposed_cuts}.

\begin{figure}[!t]
  \centering
  \includegraphics[width=0.8 \linewidth]{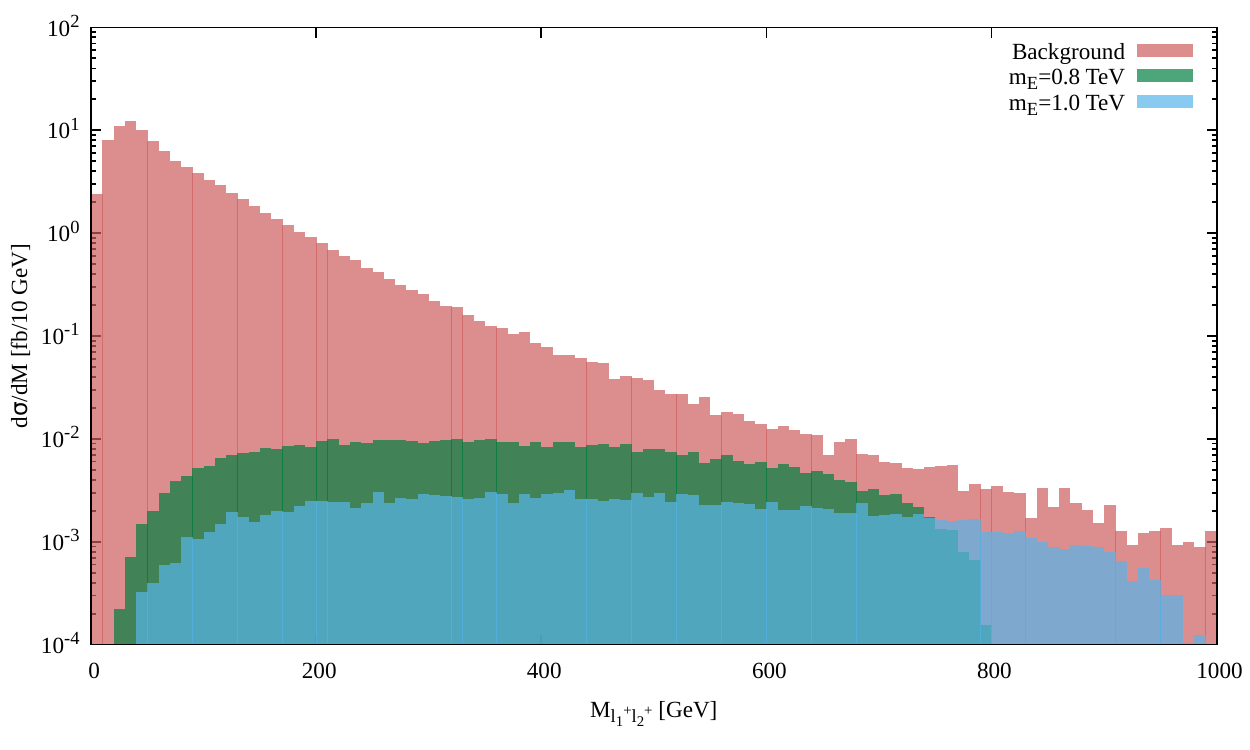}
\mycaption{Same sign dilepton invariant mass ($M_{l_1^+ l_2^+}$) distributions after the acceptance cuts for both signal ($m_E=$ 0.8 and 1 TeV) and the SM background.}
\label{SSD_E}
\end{figure}

  \begin{table}[h!]
    \centering
    \begin{tabular}{ c   }
      \hline\hline
          {\em Proposed Cuts} \\
      \hline\hline
      $M_{\rm OS}~>~150$ GeV,~~ $M_{\rm SFOS+l}~>~150$ GeV~~ and~~ $M_{\rm SFOS+SFOS}~>~200$ GeV \\
      \hline\hline
    \end{tabular}
    \caption{Additional cuts proposed to enhance the signal to background ratio in the context of this model.}
    \label{proposed_cuts}
  \end{table}

In Fig.~\ref{dist_E_m} (right panel), we display the (top row) tri-lepton ($M_{SFOS+l}$), (left plot of the bottom row) 4-lepton ($M_{SFOS+SFOS}$),  and (right plot of the bottom row) effective mass distributions. Interestingly, the SM background in the tri-lepton ($M_{l_1^+ l_1^{-} l_2^+}~{\rm and}~M_{l_1^+ l_1^{-} l_2^-}$) and 4-lepton ($M_{l_1^+ l_1^{-} l_2^+ l_2^-}$) invariant mass distributions are also associated with a peak at $m_Z$. These peaks arise due to the radiative $Z$-boson decays\footnote{The radiative decay of the $Z$-boson where a photon radiated from the $Z~\to~l^+l^-$ decay converts into a $l^+ l^-$ pair, is highly suppressed. However, the production cross-section of a single $Z$-boson at the LHC is huge. Therefore, despite being suppressed, the radiative $Z$-boson decays into 4-leptons contribute significantly to the $4l$ final state.} into 4-leptons. To suppress this background, we have used $Z$-veto on tri-lepton and 4-lepton invariant masses also. The $Z$-veto cuts on $M_{SFOS+l}$ and $M_{SFOS+SFOS}$ are summarized in Table.~\ref{ATLAS_cuts}. The Higgs peak as well as the kinematic threshold of $Z$-boson pair-production are also visible in the background $M_{SFOS+SFOS}$ distribution (see Fig.~\ref{dist_E_m}, left panel bottom right plot).  In view of the signal tri-lepton and 4-lepton invariant mass distributions, we propose additional lower bounds on $M_{SFOS+l}$ and $M_{SFOS+SFOS}$ which are listed in Table~\ref{proposed_cuts}. We present the effective mass ($M_{\rm eff}$), defined in Eq.~\ref{meff} as the scalar sum of the transverse momenta of all the visible particles, as well as the total missing transverse energy, distributions for the signal and background in Fig.~\ref{dist_E_m} (left panel bottom left plot). The signal leptons are arising from the decay of TeV scale particles and hence, the signal $M_{\rm eff}$ distribution is expected to peak at TeV as can be seen from the signal effective mass distributions in Fig.~\ref{dist_E_m}. Whereas, the background $M_{\rm eff}$ tends to have smaller values. As a result, effective mass is considered to be a powerful discriminator between the new physics signals and the SM background. We demand $M_{\rm eff}~>~600$ GeV which drastically reduces the background with minimal effect on the signal. Fig.~\ref{SSD_E} shows the invariant mass distribution of the same-sign (SS) dilepton pairs ($M_{SS}$) for the signal and background. Since any cut on $M_{SS}$ to suppress the background will also reduce the signal significantly, we have not imposed any cut on $M_{SS}$. Since the same-sign dilepton pairs arise from the decay $E^{\pm\pm}~\to~l^\pm l^\pm \nu$, the signal $M_{SS}$ is associated with a characteristic kinematic endpoint at $m_E$ which could be useful to determine $m_E$ after the discovery. In this work, we restrict ourselves only to the discovery potential of this model at the LHC and do not explore the possibilities of determining different parameters using kinematic variables. The ATLAS suggested signal selection criteria and our proposed cuts on top of the ATLAS cuts are presented in Table~\ref{ATLAS_cuts} and \ref{proposed_cuts}, respectively.

\begin{table}[t!]
    \centering
    \begin{tabular}{ c||c|c|c}
      \hline\hline
      \multicolumn{4}{ c}{The SM background and signal Cross-sections [fb] after different cuts} \\
      \hline\hline
      The SM background & \multicolumn{3}{ c}{Cuts} \\\cline{2-4}
      processes & Acceptance Cuts & ATLAS cuts & ATLAS + Proposed cuts\\\hline\hline
      HSP/$ZZ$ & 155.2 & $6.7 \times 10^{-2}$ & $1.9 \times 10^{-3}$\\
      $t\bar t l \bar l$ & 3.46 & 0.12 & $9.9 \times 10^{-3}$ \\
      Higgs & 1.35 & $9.1\times10^{-3}$& $2.1\times10^{-3}$\\
      VVZ & 0.64 & $7.6\times10^{-3}$ & --\\
      Others & $2.8\times10^{-2}$ & $1.3\times10^{-2}$ & $4.8 \times10^{-3}$\\\hline
      Total & 160.6 & $0.22~\left(\textcolor{red}{ 0.28^{+0.06}_{-0.06}}\right)$ & $1.9 \times10^{-2}$\\\hline\hline
      $m_E$ [TeV] & \multicolumn{3}{ c}{Signal Cross-Sections [fb]} \\\hline\hline
      0.8 & 0.5 & 0.46 & 0.31\\
      1.0 & 0.18 & 0.17 & 0.13\\\hline\hline
    \end{tabular}
    \caption{Various SM backgrounds and signal cross-sections are presented after {\em acceptance}, {\em ATLAS} and {\em ATLAS + proposed cuts}. Bracketed number in the {\em ATLAS cuts} column and the total background cross-section row is the total background cross-section after the {\em ATLAS cuts}  estimated by the ATLAS collaboration in Ref.~\cite{Aaboud:2018zeb}.}
    \label{cut_flow}
\end{table}

\paragraph{Results}

\begin{figure}[!ht]
  \centering
  \includegraphics[width=0.48 \linewidth]{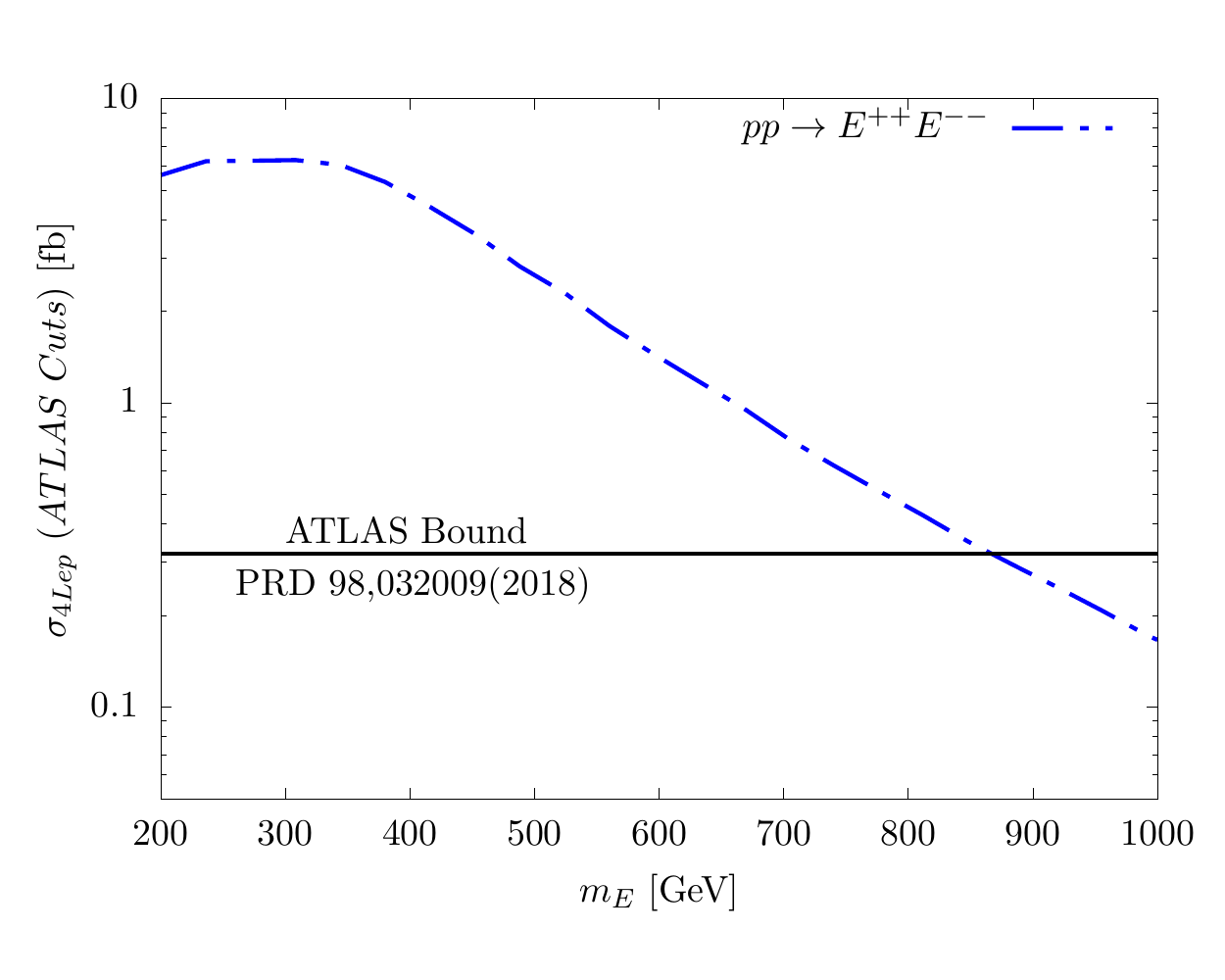}
  \includegraphics[width=0.48 \linewidth]{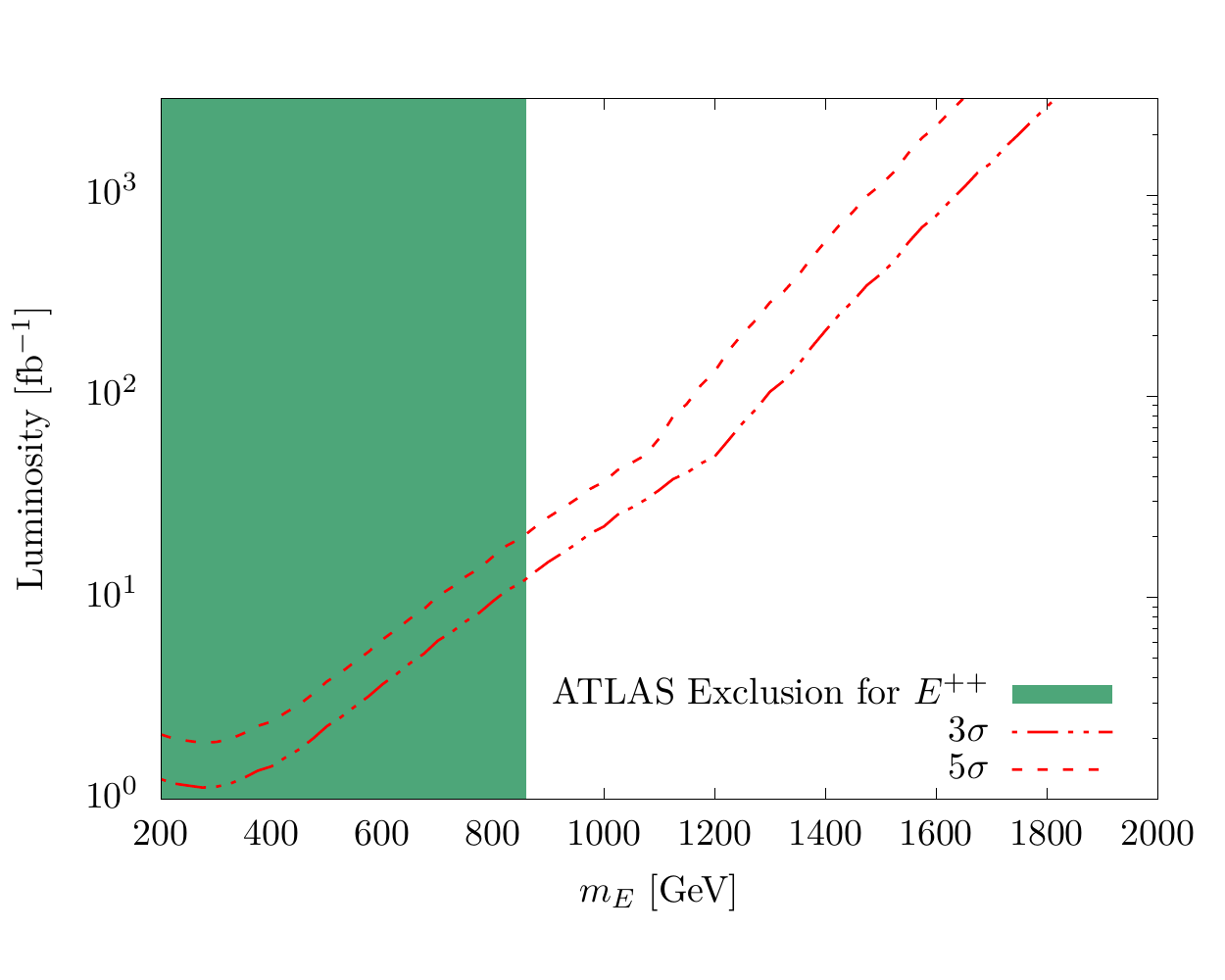}
\mycaption{(Left panel) Four lepton signal cross-section ($\sigma_{4Lep}$) as a function of $E^{\pm\pm}$ mass after the cuts (listed in Table~\ref{ATLAS_cuts}) used by the ATLAS collaboration in Ref.~\cite{Aaboud:2018zeb}. The black solid line corresponds to the ATLAS observed  95\% CL upper bound on the visible 4-lepton signal cross-section. (Right panel) Required luminosity for $3\sigma$ and $5\sigma$ discovery is plotted as a function of $m_E$ for the proposed event selection criteria (listed in Table~\ref{proposed_cuts}), }
\label{bound_E}
\end{figure}

The signal and the SM background cross-sections after the {\em ATLAS cuts} and {\em proposed cuts} listed in Table~\ref{ATLAS_cuts} and \ref{proposed_cuts}, respectively, are presented in table~\ref{cut_flow}. {\em ATLAS cuts} significantly reduce the background cross-section. Table~\ref{cut_flow} also shows that after the {\em ATLAS cuts}, the estimated background cross-section in the framework of our parton-level Monte-Carlo is consistent (within few percent) with the ATLAS estimation of the background. This consistency of the ATLAS and our analysis enables us to constrain the parameters (in this case $m_E$) of this model from the ATLAS model independent 95\% CL upper limit on the new physics contribution to $4l$ cross-section ($\sigma(4l)_{\rm vis}^{95}$) \cite{Aaboud:2018zeb} after the cuts in Table~\ref{ATLAS_cuts}. Fig.~\ref{bound_E} (left panel) shows the variation of signal $4l$ cross-section (after the {\em acceptance cuts} and {\em ATLAS cuts} in Table~\ref{ATLAS_cuts}) as a function of doubly charged fermion mass, $m_E$. The horizontal line in Fig.~\ref{bound_E} (left panel) corresponds to the ATLAS 95\% CL upper limit on the visible $4l$ cross-section. Fig.~\ref{bound_E} (left panel) clearly shows that for $m_E ~<~\sim 870$ GeV, the  contribution of $E \bar E$-pair production to visible $4l$ signal cross-section is larges than $\sigma(4l)_{\rm vis}^{95}$. Therefore, one can set a lower bound of about 870 GeV on the doubly charged fermion mass from the ATLAS search for the $4l$ final state.

In the last column of Table~\ref{cut_flow}, we have presented the background as well as the signal $4l$ cross-sections after applying the {\em proposed cuts} (listed in Table~\ref{proposed_cuts}) on top of the ATLAS cuts (listed in Table~\ref{ATLAS_cuts}). Total background cross-section is reduced by a factor of 10 as a result of applying the cuts in Table~\ref{proposed_cuts}. Whereas, the signal cross-sections are reduced by a factor of 1.5(1.3) only for $m_E~=~800(1000)$ GeV. In Fig.~\ref{bound_E} (right panel), the required integrated luminosities for the $3\sigma$ and $5\sigma$ discovery of the doubly charged fermion are presented as a function of $m_E$ at the LHC with $\sqrt s~=~13$ TeV. We define the signal to be observable with more than $S\sigma$ significance for a integrated luminosity ${\cal L}$ if,
\begin{equation}
  \frac{N_S}{\sqrt{N_S+N_B}} ~\ge~ S,
\end{equation}
where, $N_{S(B)}~=~\sigma_{S(B)}{\cal L}$ is the number of signal (background) events for an integrated luminosity ${\cal L}$. Fig.~\ref{bound_E} (right panel) shows that the LHC with 3000 fb$^{-1}$ integrated luminosity and 13 TeV center-of-mass energy will be able to probe $m_E$ upto about 1800 (1600) GeV at 3$\sigma$ (5$\sigma$) significance. The shaded region of Fig.~\ref{bound_E} (right panel) corresponds to the part of parameter-space which is already excluded from the ATLAS $4l$-search in Ref.~\cite{Aaboud:2018zeb}.

\subsection{Summary}
To summarize, we have discussed the production, decay, and the resulting collider signatures of the doubly charged fermion at the LHC with $\sqrt s~=~13$ TeV. In the scenario where $E^{\pm\pm}$ is lighter than the (multi-)charged scalars, $E^{\pm\pm}$ undergoes tree-level 3-body decays. Depending on the total decay width of the doubly charged fermion, the collider signatures of $E\bar E$-pair production at the LHC are broadly classified into two categories namely, the abnormally large ionization signature for long-lived $E^{\pm\pm}$ and multi-lepton (in particular, 4-lepton) signature for prompt $E^{\pm\pm}$ decay. Using the ATLAS results for long-lived MCP search, we obtain a lower bound of about  1150 GeV on the mass of long-lived $E^{\pm\pm}$. For prompt decay of $E^{\pm\pm}$, ATLAS $4l$ search with 36.1 fb$^{-1}$ data of the 13 TeV LHC  excludes $m_E$ below 870 GeV at 95\% CL. After investigating different characteristic kinematic distributions for the background as well as the signal, we proposed additional cuts to optimize the signal to the background ratio. With the proposed cuts, the discovery reach of the LHC with 3000 $fb^{-1}$ integrated luminosity data is estimated to be $m_E\sim$ 1800 (1600) GeV at 3$\sigma$ (5$\sigma$) significance.

\section{Phenomenology of scalars}
\label{scalar_phenomenology}

Being charged under both the  SM $SU(2)_L$ and $U(1)_Y$, the (multi-)charged scalars ($\phi^{3\pm},~\phi^\pm$ and $H_a^{\pm\pm}$) have gauge interactions (listed in Appendix~\ref{feyn}) with the SM gauge bosons, namely, the photon and $W/Z$-boson. Therefore, the (multi-)charged scalars can be pair produced or produced in association with another (multi-)charged scalar (associated production) at the LHC. We have computed the following\footnote{The production cross-sections of $H_a^{\pm\pm} H_b^{\mp\mp}$ with $a~\ne~b$ as well as $H_1^{\pm\pm}\phi^{\mp}$, $H_2^{\pm\pm}\phi^{3\mp}$, $H_3^{\pm\pm}\phi^{\mp}$ and $H_3^{\pm\pm}\phi^{3\mp}$ are suppressed by the mixings in the doubly-charged scalar sector and hence, not considered.} pair and associated productions of the (multi-)charged scalars at the LHC with $\sqrt s~=~13$ TeV.
  \begin{equation}
    pp~\to~\phi^{3\pm}\phi^{3\mp},~\phi^{\pm}\phi^{\mp},H_a^{\pm\pm}H_{a}^{\mp\mp},~H_{1}^{\pm\pm}\phi^{3\mp}~{\rm and~}~H_{2}^{\pm\pm}\phi^{1\mp}.
  \end{equation}
  At the LHC, the pair productions are quark anti-quark (photon-photon\footnote{In the context of $E\bar E$-production at the LHC, the importance of photoproduction was discussed in section~\ref{E_phenomenology} which holds true for the pair production of multi-charged scalars also.}) initiated processes, proceed through a $\gamma/Z$-boson (charged scalar) in the $s(t/u)$-channel. The photoproductions get an extra contribution from the quartic coupling involving two photons and two (multi-)charged scalars (see Appendix~\ref{feyn}). The associated productions get contribution  from the quark anti-quark initial state only and proceeds through a $W^\pm$-boson in the $s$-channel. Different production cross-sections at the LHC have been numerically computed by integrating the following parton-level differential cross-sections over the phase-space and parton-densities.
\begin{figure}[!t]
  \centering
  \includegraphics[width=0.48 \linewidth]{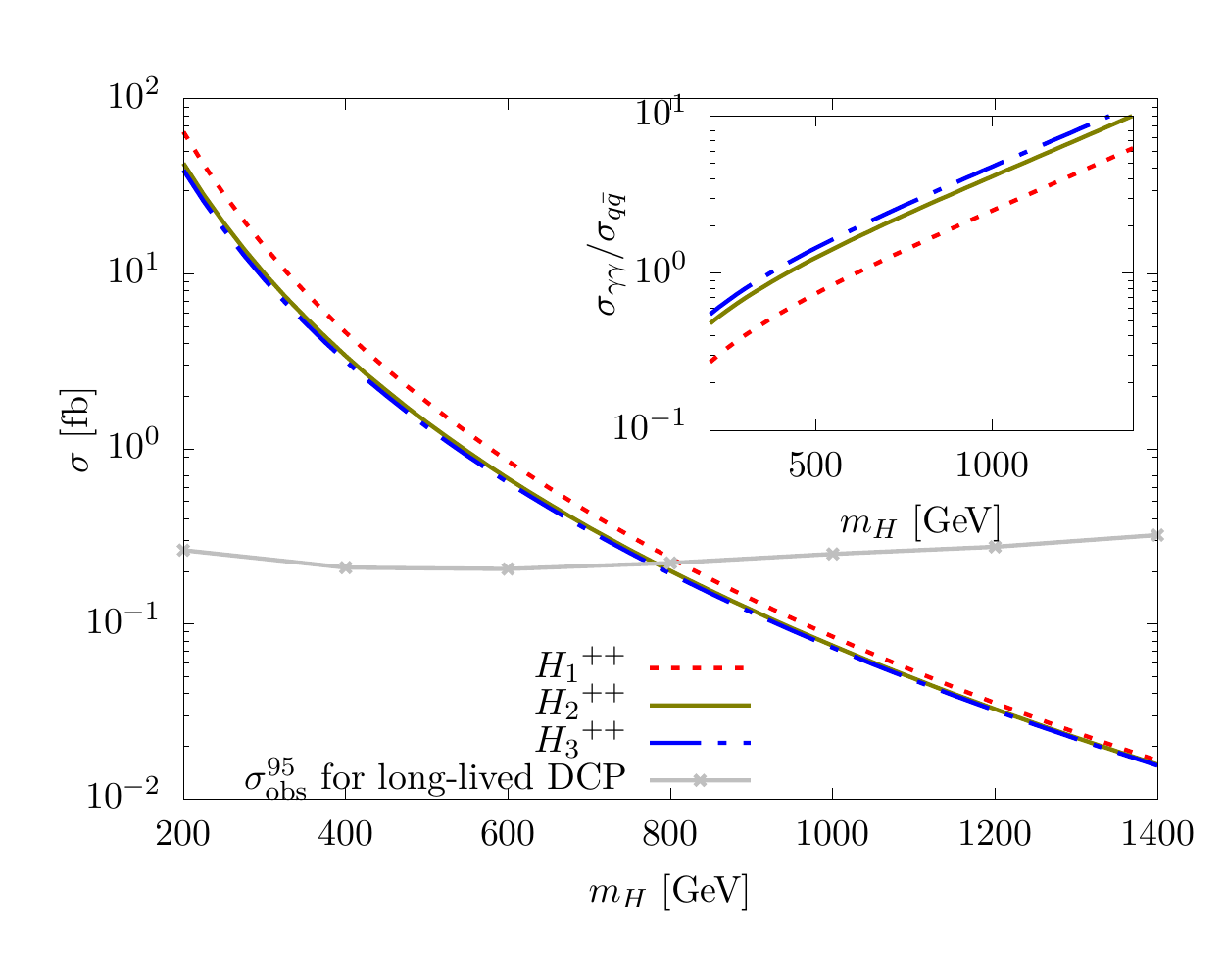}
  \includegraphics[width=0.48 \linewidth]{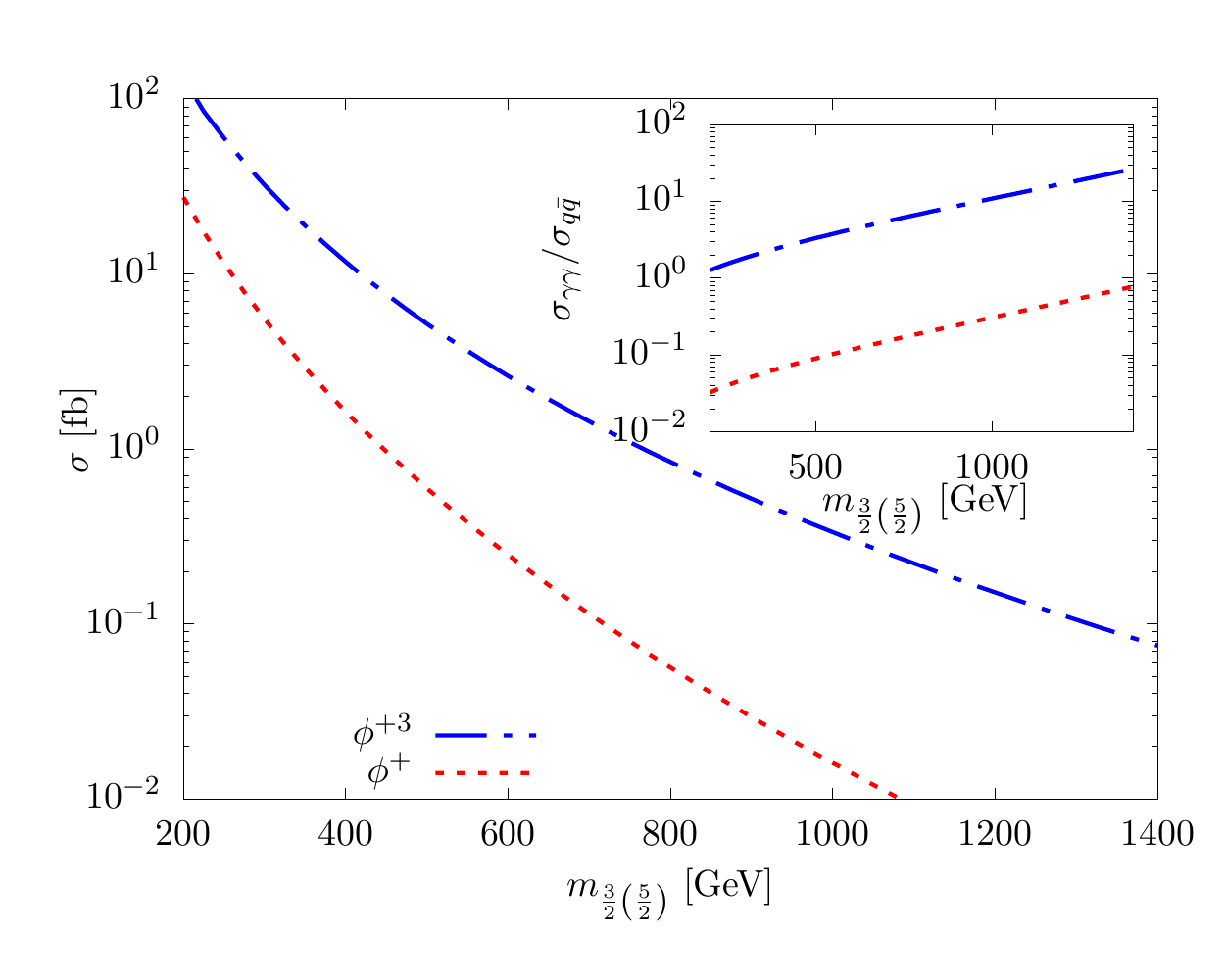}
  \mycaption{The total pair production cross-sections for the doubly-charged scalar pairs (left panel) as well as  triply-charged and singly-charged scalar pairs (right panel) are presented for the LHC with $\sqrt s~=~13$ TeV. The insets shows the ratio of PF and DY contributions. The grey solid lines correspond to the ATLAS observed 95\% CL cross-section upper limit ($\sigma_{\rm Obs}^{95}$) on long-lived (left panel) doubly-charged particles \cite{Aaboud:2018kbe}.}
\label{scalar_pair_prod}
\end{figure}
\begin{eqnarray}
 && \frac{d\hat \sigma_{q\bar q}^{SS^*}}{d\Omega}~=~\frac{\alpha_{EM}^2}{6 \hat s^3}\,\,\sqrt{1-\frac{4m_S^2}{\hat s}}\,\,\left(\hat u \hat t~-~m_{S}^4\right)\nonumber\\
  &&\times~ \left[
    \left\{
    Q_q Q_S + \frac{2\left(T_{3,q}-2Q_q{\rm sin}^2\theta_W \right)\left(T_{3,S}-Q_S{\rm sin}^2\theta_W \right)}{{\rm sin}^22\theta_W \left(1-\frac{m_Z^2}{\hat s}\right)}
    \right\}^2
    +
    4\left\{
    \frac{T_{3,q}\left(T_{3,S}-Q_S{\rm sin}^2\theta_W \right)}{{\rm sin}^22\theta_W \left(1-\frac{m_Z^2}{\hat s}\right)}
    \right\}^2
    \right]\,\,,\nonumber\\
  &&\frac{d \hat \sigma^{SS^*}_{\gamma \gamma}}{d\Omega}~=~\frac{Q_S^4\alpha^2_{EM}}{4\hat s}\,\,\sqrt{1-\frac{4m^2_{S}}{\hat s}}\,\,\left[\frac{(m^2_{S} + \hat u)^2}{(m^2_{S}-\hat u)^2}+ \frac{(m^2_{S}+\hat t)^2}{(m^2_{S} -\hat t)^2}+8\frac{m^4_{S}}{(m^2_{S} - \hat u)(m^2_{S}-\hat t)}\right]\,\,,\nonumber\\
  &&\frac{d \hat \sigma^{SS^{\prime}}_{q \bar{q^\prime}}}{d\Omega}~=~\frac{\alpha^2_{EM}}{48\hat s\,\, {\rm sin}^4\theta_W}\,\,\sqrt{1-\frac{4m^2_{S}}{\hat s}}\,\,\frac{\hat u \hat t-m_S^4}{\left(\hat s-m_W^2\right)^2}\,\,,
\end{eqnarray}
where, $\hat s$, $\hat t$ and $\hat u$ are the usual Mandelstam variables, $Q_q$ and $Q_S$ are the electric charges of the SM quark $q$ and charged scalar $S$, respectively, $m_S$ is the mass of $S$, and $T_{3,q}$ as well as $T_{3,S}$ are the weak isospin of quarks ($T_{3,q}~=~\frac{1}{2}\left(-\frac{1}{2}\right)$ for $q \supset u,~c~,t\left(d,~s,~b\right)$) and charged scalars ($T_{3,S}~=~\frac{1}{2}\left(-\frac{1}{2}\right)$ for $q \supset \phi^{3\pm},~H_2^{\pm\pm}\left(\phi^\pm,~H_1^{\pm\pm}\right)$ and $T_{3,S}~=~0$ for $H_3^{\pm\pm}$)\footnote{The  physical doubly charged scalars ($H_a^{\pm\pm}$) appear after the EWSB as mixtures of the $T_3~=~\frac{1}{2}~{\rm and}~-\frac{1}{2}$ components of the $Y~=~\frac{3}{2}~{\rm and}~\frac{5}{2}$ doublets, respectively,  and $Y~=~2$ singlet. Since the mixings in the doubly-charged scalar sector are constrained to be small, $H_{1(2)}^{\pm\pm}$ and $H_{3}^{\pm\pm}$ are dominantly the $T_3~=~-\frac{1}{2}\left(\frac{1}{2}\right)$ component of $Y~=~\frac{5}{2}\left(\frac{3}{2}\right)$ doublet and $Y~=~2$ singlet, respectively.}, respectively.

\begin{figure}[!t]
  \centering
  \includegraphics[width=0.8 \linewidth]{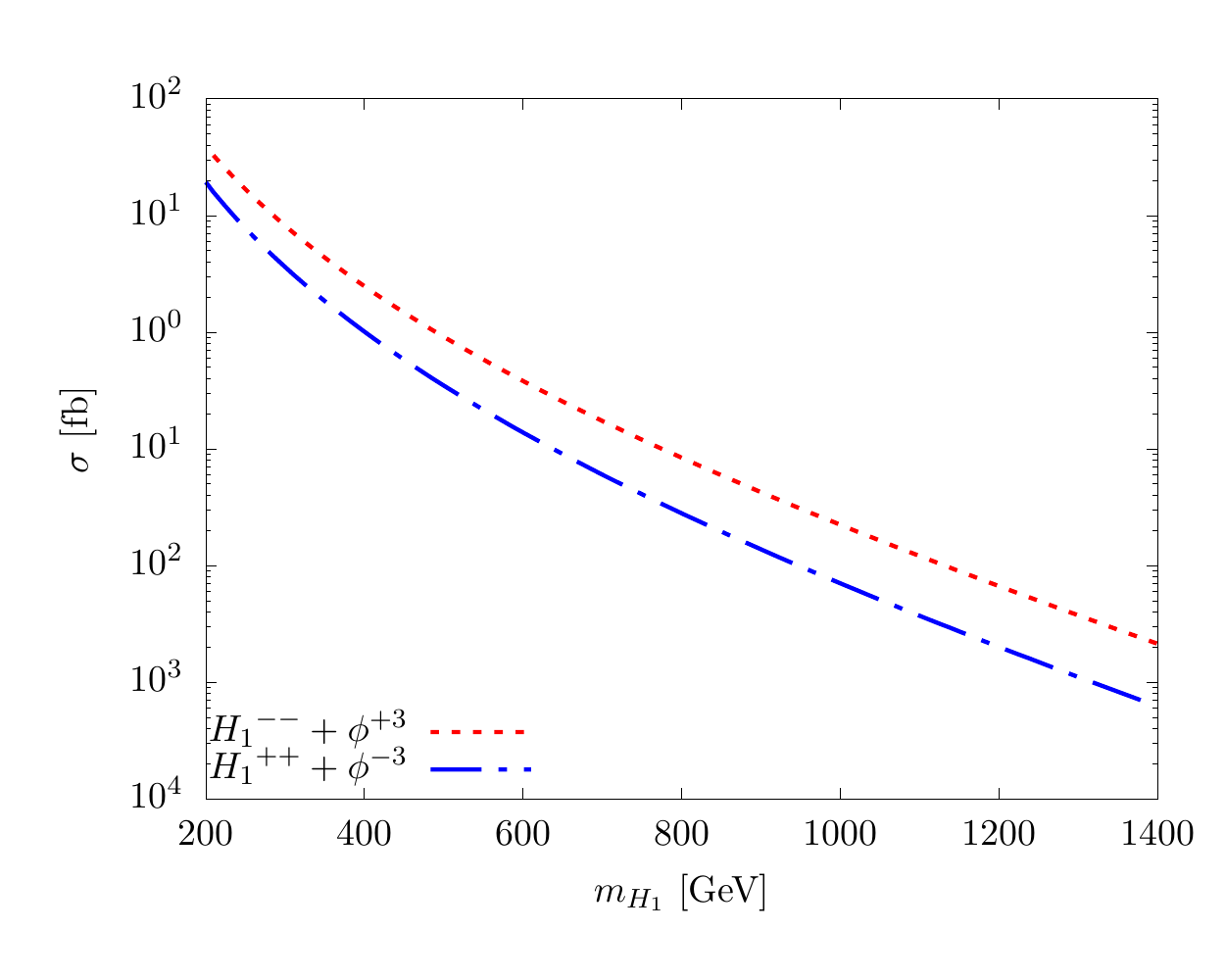}
\mycaption{Production cross-section of $H_1^{++}(H_1^{--})$ in association with a $\phi^{3-}(\phi^{3+})$ are plotted as a function of $m_{H_1}$ at the LHC with 13 TeV center-of-mass energy.}
\label{cross_asso}
\end{figure}

To evaluate the scalar pair and associated production cross-sections at the LHC with $\sqrt s~=~13$ TeV, we have numerically integrated Eq.~\ref{cross_section} over the {\bf NNPDF23LO} \cite{Ball:2014uwa} parton distribution functions. We fix the factorization ($\mu_F$) and renormalization scales at the subprocess center-of-mass energy $\sqrt {\hat s}$. The resulting scalar production cross-sections are presented in Fig.~\ref{scalar_pair_prod}(pair productions) and Fig.~\ref{cross_asso} (associated productions).  The insets in Fig.~\ref{scalar_pair_prod} show the ratio of the contributions from the PF and DY. Different doubly charged scalars, being the members of different scalar multiplets with different weak isospins, couple differently with the SM $Z$-boson. Therefore, the DY pair production cross-sections are different for different doubly-charged scalar pairs. However, production of different doubly-charged scalar pairs get same contribution from PF which is the dominant contribution in the large scalar mass region. Fig.~\ref{scalar_pair_prod} (right panel) shows that in the large mass region, $\sigma\left(\phi^{3\pm}\phi^{3\mp}\right)$ is more than an order of magnitude  bigger than $\sigma\left(\phi^{\pm}\phi^{\mp}\right)$. This can be attributed to the fact that the photo-production of $\phi^{3\pm}\phi^{3\mp}$ pairs are enhanced by a factor of $3^4$ compared to the photo-production of $\phi^{\pm}\phi^{\mp}$. In Fig.~\ref{cross_asso}, we have presented the production cross-sections of $H_1^{++}(H_1^{--})$ in association with a $\phi^{3-}(\phi^{3+})$  at the LHC with $\sqrt s~=~13$ TeV. The difference between $\sigma\left(H_1^{++}\phi^{3-}\right)$ and $\sigma\left(H_1^{--}\phi^{3+}\right)$ arises from the difference in the densities of the initial state partons. The associated productions of the  multi-charged scalars, being mediated by the $W$-boson in the $s$-channel, are completely determined by their $SU(2)_L$ charges. Therefore, the production cross-sections $\sigma\left(H_2^{++}\phi^{-}\right)$ and $\sigma\left(H_2^{--}\phi^{+}\right)$ are identical to $\sigma\left(H_1^{--}\phi^{3+}\right)$ and $\sigma\left(H_1^{++}\phi^{3-}\right)$, respectively, and hence, are not shown separately. After being produced at the LHC, the multi-charged scalars decays into the SM leptons and/or bosons giving rise to multi-lepton final states which will be discussed in the following.

\subsection{Decay of multi-charged scalars}

The collider signatures of the multi-charged scalars crucially depend on their decays which will be discussed in this section.

\subsubsection{Triply- and singly-charged scalar decay}
\begin{figure}
\begin{minipage}[t]{0.4\textwidth}
  \begin{tikzpicture}[line width=1.4 pt, scale=1.65,every node/.style={scale=0.9}]
    \draw[scalar,black] (-1.5,0.0) --(0.0,0.0);
    \draw[vector,black] (0.0,0.0) --(0.8,0.6);
    \draw[scalar,black] (0.0,0.0) --(1.5,-1.0);
    \draw[fermion,black] (1.5,-1.0) --(2.5,-0.5);
    \draw[fermion,black] (1.5,-1.0) --(2.5,-1.5);
    \node at (-0.5,0.3) {$\phi^{3\pm (\pm)}$};
    \node at (1.2,0.5) {$W^{\pm (\mp)}$};
    \node at (0.5,-0.8) {${H_a}^{\pm\pm}$};
    \node at (2.0,-0.44) {$l^{\pm}$};
    \node at (2.0,-1.47) {$l^{\pm}$};
  \end{tikzpicture}
\end{minipage}
\hspace{0.1\textwidth}
\begin{minipage}[t]{0.4\textwidth}
  \begin{tikzpicture}[line width=1.4 pt, scale=1.65,every node/.style={scale=0.9}]
    \draw[scalar,black] (-1.5,0.0) --(0.0,0.0);
    \draw[vector,black] (0.0,0.0) --(1.5,1.0);
    \draw[scalar,black] (0.0,0.0) --(1.5,-0.7);
    \draw[fermion,black] (1.5,1.0) --(2.5,1.5);
    \draw[fermion,black] (1.5,1.0) --(2.5,0.5);
    \node at (-0.5,0.3) {$\phi^{3\pm }$};
    \node at (0.5,0.8) {$W^{\pm}$};
    \node at (0.5,-0.8) {${H_1}^{\pm\pm}$};
    \node at (2.0,1.44) {$e^{\pm}$};
    \node at (2.0,0.6) {$\nu_e$};
  \end{tikzpicture}
\end{minipage}

  \caption{Feynman diagram showing the tree-level 3-body decays of the triply-(singly-)charged scalars ($\phi^{3\pm(\pm)}$).}
  \label{3-body-phi}
\end{figure}
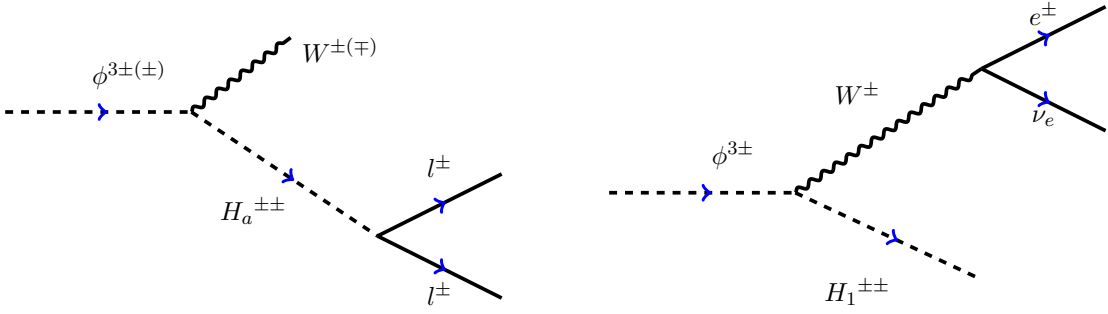

If kinematically allowed {\em i.e.,} $m_{\frac{5}{2}\left(\frac{3}{2}\right)}~>~m_E$, the triply-(singly-)charged scalar can decay into a doubly-charged fermion in association with a SM charged lepton: $\phi^{3\pm(\pm)}\to E^{\pm\pm} l^{\pm(\mp)}$. Other possible 2-body decays of the triply-(singly-)charged are the decays into one of the doubly-charged scalars and a $W$-boson: $\phi^{3\pm(\pm)}\to H_a^{\pm\pm} W^{\pm(\mp)}$. The partial decay widths for the 2-body $\phi^{3\pm}\left(\phi^{\pm}\right)$ decays are presented in the following:
\begin{eqnarray}
 \Gamma\left(\phi^{3\pm(\pm)}\to E^{\pm\pm} l^{\pm(\mp)}\right)&=&\frac{\left|f_{\frac{5}{2}\left(\frac{3}{2}\right)} \right|^2}{16\,\pi}\,m_{\frac{5}{2}\left(\frac{3}{2}\right)}\, \left(1-x_E\right),\nonumber\\
 \Gamma\left(\phi^{3\pm(\pm)}\to H_a^{\pm\pm} W^{\pm(\mp)}\right)&=&\frac{\alpha_{EM}\,\left|O_{a1(2)} \right|^2}{8\,{\rm sin}^2\theta_W}\,m_{\frac{5}{2}\left(\frac{3}{2}\right)}\,\,\lambda^{\frac{1}{2}}\left(1,x_{H_a},x_W\right)\,\,\eta\left(x_{H_a},x_W\right)\, ,
\end{eqnarray}
where, $\lambda(x,y,z)~=~x^2+y^2+z^2-2xy-2yz-2xz$, $\eta\left(x,y\right)~=~y-2-2x+\frac{\left(1-x\right)^2}{y}$ and $x_i~=~m_i^2/m_{\frac{5}{2}\left(\frac{3}{2}\right)}^2$. It is important to note that the smallness of neutrino masses implies close degeneracy between $\phi^{3\pm}\left(\phi^{\pm}\right)$ and $H_1^{\pm\pm}\left(H_2^{\pm\pm}\right)$ {\em i.e.,} $m_{\phi^{3\pm}\left(\phi^{\pm}\right)}~\approx~m_{H_1^{\pm\pm}\left(H_2^{\pm\pm}\right)}~\approx~m_{\frac{5}{2}\left(\frac{3}{2}\right)}$. Therefore, the 2-body decay  $\phi^{3\pm}\left(\phi^{\pm}\right)~\to~H_1^{\pm\pm}\left(H_2^{\pm\pm}\right)+W^{\pm}$ is kinematically forbidden. However, $\phi^{3\pm}$ undergoes 2-body decay into $H_{2(3)}^{\pm\pm}+W^{\pm}$ if $m_{\frac{5}{2}}~>~m_{\frac{3}{2}(k)}+m_W$. On the other hand, in the region of parameter space defined by $m_{\frac{3}{2}}~>~m_{\frac{5}{2}(k)}+m_W$, the decay $\phi^{\pm}~\to~H_{1(3)}^{\pm\pm}+W^{\mp}$ is allowed. If all the aforementioned 2-body decays are kinematically forbidden for $\phi^{3\pm}\left(\phi^{\pm}\right)$ {\em i.e.,} $m_{\frac{5}{2}\left(\frac{3}{2}\right)}~<~m_E,~m_{k}~{\rm and}~m_{\frac{3}{2}\left(\frac{5}{2}\right)}$,  the triply-(singly-)charged scalar undergoes tree-level 3-body decay into $l^{\pm}l^{\pm}W^{\pm(\mp)}$. The 3-body decay $\phi^{3\pm}\left(\phi^{\pm}\right)~\to l^{\pm}l^{\pm}W^{\pm(\mp)}$ proceeds through a off-shell doubly charged scalar as depicted in Fig.~\ref{3-body-phi}(left panel).
The partial width for the decay $\phi^{3\pm}\left(\phi^{\pm}\right)~\to l^{\pm}l^{\pm}W^{\pm(\mp)}$ is given by,
\begin{equation}
  \Gamma\left(\phi^{3\pm}\left(\phi^{\pm}\right)~\to l^{\pm}l^{\pm}W^{\pm(\mp)}\right)=\frac{\alpha_{EM}\left|f_k\right|^2}{128\,\pi^2\,{\rm sin}^2\theta_W}\,\,m_{\frac{5}{2}\left(\frac{3}{2}\right)}\,\,\sum_{a,b}\left(O_{a3}O_{b3}O_{a1(2)}O_{b1(2)}\right)\,\,I\left(x_{H_a},x_{H_b}\right),\nonumber
\end{equation}
\begin{equation}
  {\rm where},~x_i~=~\frac{m_{i}^2}{m^2_{\frac{5}{2}\left(\frac{3}{2}\right)}},~I\left(x,y\right)=\int_{x_W}^1 d\xi_1 \int_0^{\xi_2^{\rm max}} d\xi_2 \frac{\xi_2\left[\left(x_W-\xi_2\right)^2-2\left(x_W+\xi_2\right)+1\right]}{x_W\left(\xi_2-x_{H_a}\right)\left(\xi_2-x_{H_b}\right)},
  \label{3body_1}
\end{equation}
and $\xi_2^{\rm max}~=~(1-\xi_1)(\xi_1-x_W)/\xi_1$. The radiative corrections induce small (of the order of a GeV) mass splitting between the $\phi^{3\pm}$ and $H_{1}^{\pm\pm}$ with the triply-charged scalar being heavier than  $H_{1}^{\pm\pm}$.  Therefore, the triply-charged scalar can decay into a on-shell $H_{1}^{\pm\pm}$ in association with a $e^{\pm}\nu_e$ or $u\bar d$ pair via a off-shell $W$-boson as depicted in Fig.~\ref{3-body-phi}(right panel). The partial decay width for the decay $\phi^{3\pm}~\to e^{\pm} \nu_e H_1^{\pm\pm}$ is given by,
\begin{equation}
  \Gamma\left(\phi^{3\pm}~\to e^{\pm}\nu_e H_1^{\pm\pm}\right)~=~\frac{\alpha_{EM}^2O_{11}^2}{32\,\pi\,{\rm sin}^4\theta_W\,m_W^4}\,\,\left(m_{\phi^{3\pm}}-m_{H_1^{\pm\pm}}\right)^5.
  \label{3body_3}
\end{equation}
Note that the 3-body decay of $\phi^{3\pm}$ into $e^{\pm}\nu_e H_1^{\pm\pm}$ is suppressed by the splitting $\left(m_{\phi^{3\pm}}-m_{H_1^{\pm\pm}}\right)$ $\sim~1.8$ GeV and is estimated to be of the order of $10^{-12}$ GeV. Since $H_2^{\pm\pm}$ become slightly heavier than $\phi^{\pm}$ after the radiative corrections, the decay of singly-charged scalar into an on-shell $H_{2}^{\pm\pm}$ is kinematically forbidden.  
\begin{figure}[!t]
  \centering
  \includegraphics[width=1 \linewidth]{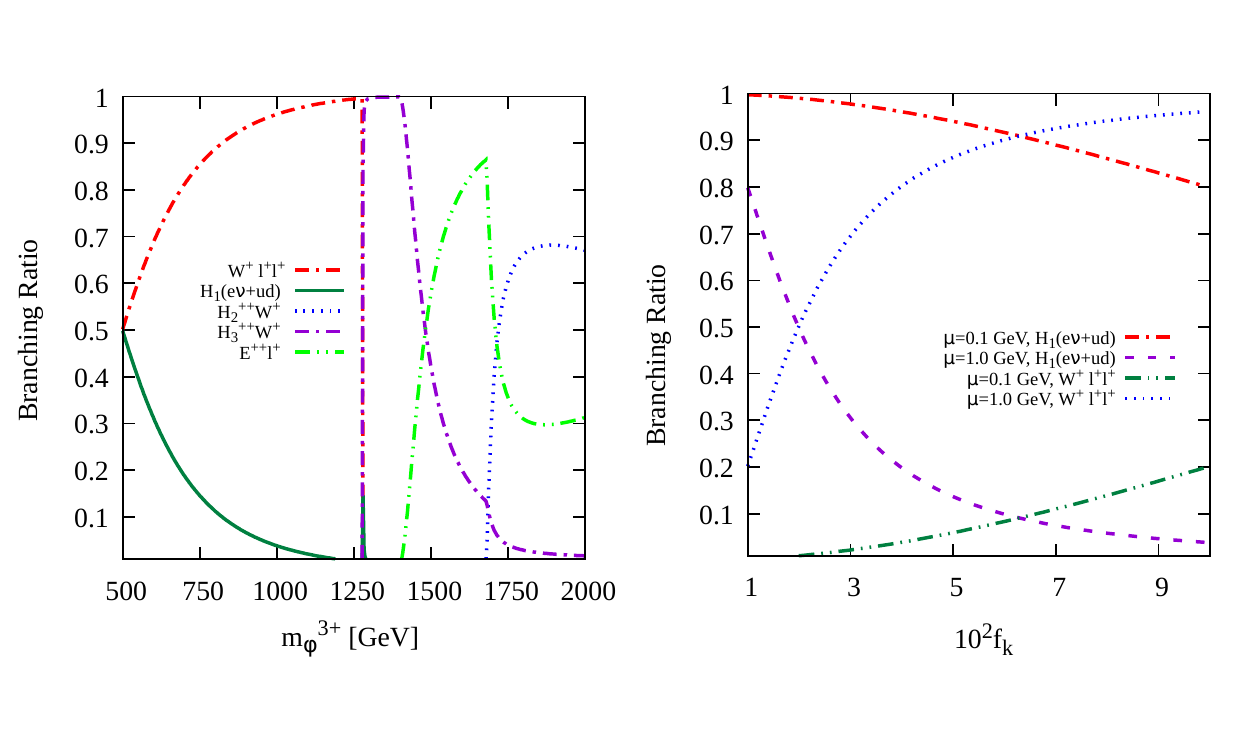}
\mycaption{(Left panel) The branching ratios of $\phi^{3+}$ is presented as a function of $m_{\phi^{3+}}$ for $m_{\frac{3}{2}}=1.6~{\rm TeV},~m_{k}=1.2~{\rm TeV},~m_{E}=1.4~{\rm TeV},~\mu=\mu^{\prime}=0.1~{\rm GeV},~f_{\frac{5}{2}\left(\frac{3}{2}\right)}=2\times 10^{-4},~\lambda=5\times 10^{-3}$ and $f_k=1.0$. (Right panel) Branching ratios of 3-body decays is shown as a function of $y_k$ for $m_{\frac{5}{2}}~=~1$ TeV and two different values of $\mu~=~0.1$ and 1 GeV.}
\label{br_phi3}
\end{figure}

Fig.~\ref{br_phi3}(left panel) shows the branching ratios of $\phi^{3\pm}$ into different 2- and 3-body decay modes. We have assumed $m_{\frac{3}{2}}=1.6~{\rm TeV},~m_{k}=1.2~{\rm TeV},~m_{E}=1.4~{\rm TeV},~\mu=\mu^{\prime}=0.1~{\rm GeV},~f_{\frac{5}{2}\left(\frac{3}{2}\right)}=2\times 10^{-4},~\lambda=5\times 10^{-3}$ and $f_k=1.0$. Obviously, the 2-body decays dominates over the 3-body decays as long as the 2-body decays are kinematically allowed {\em i.e.,} $m_{\phi^{3\pm}}~>~m_{E}$ or $m_{\frac{5}{2}(k)}$. Fig.~\ref{br_phi3}(left panel) also shows that for $m_{\phi^{3\pm}}~<~m_{E}$ and $m_{\frac{5}{2}(k)}$, the 3-body decay into a $W$-boson in association with a pair of same-sign lepton dominates over  the 3-body decay into a $H^{++}_1$ plus a $e^+\nu_e$ or $u\bar d$ pair. Eq.~\ref{3body_3} shows that the partial decay widths for $\phi^{3+}~\to~H_1^{++} e^+ \nu_e~{\rm and}~H_1^{++} u \bar d$ are completely\footnote{For small mixing in the doubly-charged scalar sector, $O_{11}~\approx~1$ and the radiative mass splitting $m_{\phi^{3\pm}}-m_{H_1^{\pm\pm}}~\approx~\frac{5}{2}\alpha_{EM}\,\,m_Z$.} determined by the SM parameters only. Whereas, $\Gamma\left(\phi^{3\pm}~\to l^{\pm}l^{\pm}W^{\pm}\right)$ depends on the Yukawa coupling $f_{k}$ as well as on the mixing in the doubly-charged scalar sector. In Fig.~\ref{br_phi3}(right panel), we have presented the 3-body decay branching ratios of $\phi^{3\pm}$ as function of $f_{k}$ for two different values of $\mu~=~0.1$ and 1 GeV. For smaller mixing ({\em i.e.,} smaller $\mu$) between $\phi^{\pm\pm}_{\frac{5}{2}}$ and $k^{\pm\pm}$, 3-body decay via an off-shell $W^*$-boson dominates. The partial decay width $\Gamma\left(\phi^{3\pm}~\to~l^{\pm}l^{\pm}W^{\pm}\right)$ become significant for larger $f_{k}$ and $\mu$. Decay branching ratios for the possible 2-body decays of $\phi^{\pm}$ are similar to that of $\phi^{3\pm}$ and hence, are not shown separately. However,  $\phi^{\pm}$ being lighter than $H_2^{\pm\pm}$, the 3-body decays $\phi^{\pm}$ into an on-shell $H_2^{\pm\pm}$ in association with a $e^{-}\bar \nu_e (e^{+}\nu_e)$ or $\bar u d(u\bar d)$ pair are kinematically forbidden. Therefore, for $m_{\frac{3}{2}}~<~m_{\frac{5}{2}},~m_{k}~{\rm or}~m_E$, $\phi^{\pm}$ decays into a $W$-boson in association with a same-sign dilepton pair with 100\% branching ratio.  

\begin{figure}[!ht]
  \centering
  \includegraphics[width= 1.0\linewidth]{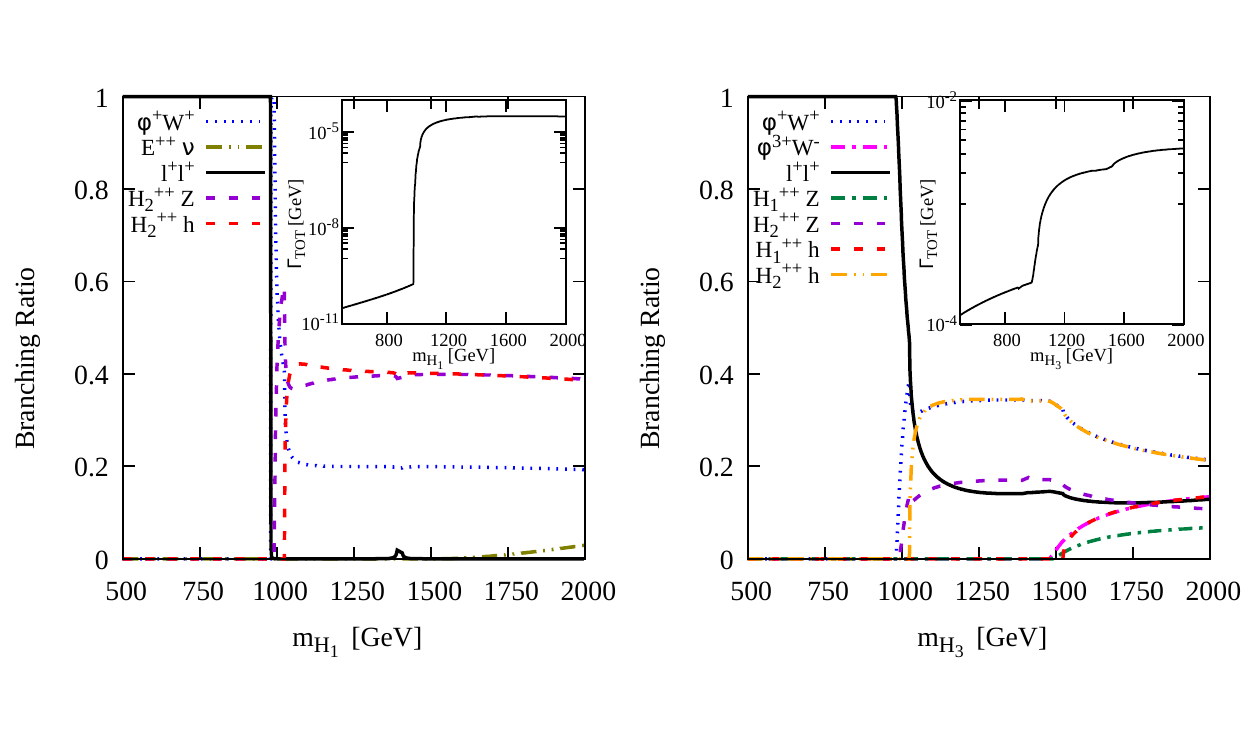}
  \mycaption{The branching ratios of $H_{1}^{\pm\pm}$ (left panel) and $H_3^{\pm\pm}$ (right panel) are presented as a function of the doubly-charged scalar mass. To calculate the branching ratios of $H_{1(3)}^{\pm\pm}$ in the left panel (right panel), we assume  $m_{\frac{3}{2}}=0.9~{\rm TeV},~m_{k\left(\frac{5}{2}\right)}=1.4~{\rm TeV},~m_{E}=1.5~{\rm TeV}$ and $f_{\frac{3}{2}}~=~f_{\frac{5}{2}}~=~2\times 10^{-4}$, $f_{k}~=~0.1\left(2\times 10^{-3}\right)$, $\lambda=~5\times 10^{-3}$ and $\mu~=~0.1(10)$ GeV. The insets show the total decay width ($\Gamma_{TOT}$).}
\label{h1_br}
\end{figure}

\subsubsection{Doubly-charged scalar decay}
\label{2_decay}
Doubly charged scalars, being charged under the $SU(2)_L$ and $U(1)_Y$, has gauge coupling with another (multi-)charged scalar and a $W/Z$-boson. If kinematically allowed, the doubly-charged scalars undergo 2-body decay into a lighter (multi-)charged scalar and a $W/Z$-boson: $H_a^{\pm\pm}~\to~\phi^{3\pm(\pm)}W^{\mp(\pm)}$ and $H_a^{\pm\pm}~\to~H_b^{\pm\pm}Z$. After the EWSB, the scalar potential in Eq.~\ref{lag_scalar} gives rise to interactions involving two doubly-charged scalar and a SM Higgs boson (see Appendix~\ref{feyn} for details). Therefore, doubly-charged scalar can decay into a SM Higgs in association with another doubly-charged scalar: $H_a^{\pm\pm}~\to~H_b^{\pm\pm} h$. The doubly-charged scalars can also decay into $E^{\pm\pm}\nu_l$ or $l^\pm l^\pm$-pairs via the Yukawa interactions in Eq.~\ref{lag_yuk}: $H_a^{\pm\pm}~\to~E^{\pm\pm}\nu_l~{\rm or}~l^\pm l^\pm$. The partial decay widths for the abovementioned 2-body decays are given by,
\begin{eqnarray}
 && \Gamma\left(H_a^{\pm\pm}\to\phi^{3\pm(\pm)}W^{\mp(\pm)}\right)~=~\frac{\alpha_{EM}\, O_{a1(2)}^2}{8\,{\rm sin^2}\theta_W}\,m_{H_a}\,\lambda^{\frac{1}{2}}\left(1,x_{\frac{5}{2}\left(\frac{3}{2}\right)},x_W\right)\,\eta\left(x_{\frac{5}{2}\left(\frac{3}{2}\right)},x_W\right)\, ,\nonumber\\
 && \Gamma\left(H_a^{\pm\pm}\to H_b^{\pm\pm}Z\right)~=~\frac{\alpha_{EM}\,\left(O_{a1}O_{b1}-O_{a2}O_{b2}\right)^2}{4\,{\rm sin^2}2\theta_W}\,m_{H_a}\,\lambda^{\frac{1}{2}}\left(1,x_{H_b},x_Z\right)\,\eta\left(x_{H_b},x_Z\right),\nonumber\\
  &&  \Gamma\left(H_a^{\pm\pm}\to H_b^{\pm\pm}h\right)~=~\frac{{\cal C}_{ab}^2}{16\,\pi\,m_{H_a}}\,\lambda^{\frac{1}{2}}\left(1,x_{H_b},x_h\right),\nonumber\\
  && \Gamma\left(H_a^{\pm\pm}\to E^{\pm\pm}\nu\right)~=~\frac{\left|f_{\frac{5}{2}}\right|^2\,O_{a1}^2\,+\,\left|f_{\frac{3}{2}}\right|^2\,O_{a2}^2}{16\,\pi}\,m_{H_a}\,\left(1-x_E\right)^2\, ,\nonumber\\
  && \Gamma\left(H_a^{\pm\pm}\to l^{\pm}l^\pm\right)~=~\frac{\left|f_{k}\right|^2\,O_{a3}^2}{16\,\pi}\,m_{H_a}\, ,
  \label{h_w}
\end{eqnarray}
where, $\lambda\left(x,y,z\right)$, $\eta\left(x,y\right)$ are already defined in the previous section and
\begin{equation}
  {\cal C}_{ab}~=~\mu\,\left\{\frac{1}{\sqrt 2}\left(O_{a2}-O_{a1}\right)O_{b3}+O_{a3}\left(O_{b2}-O_{b1}\right)\right\}+\lambda v\left(O_{a1}O_{b2}+O_{a2}O_{b1}\right).
\end{equation}

\begin{figure}[t]
  \centering
  \includegraphics[width= 1.0\linewidth]{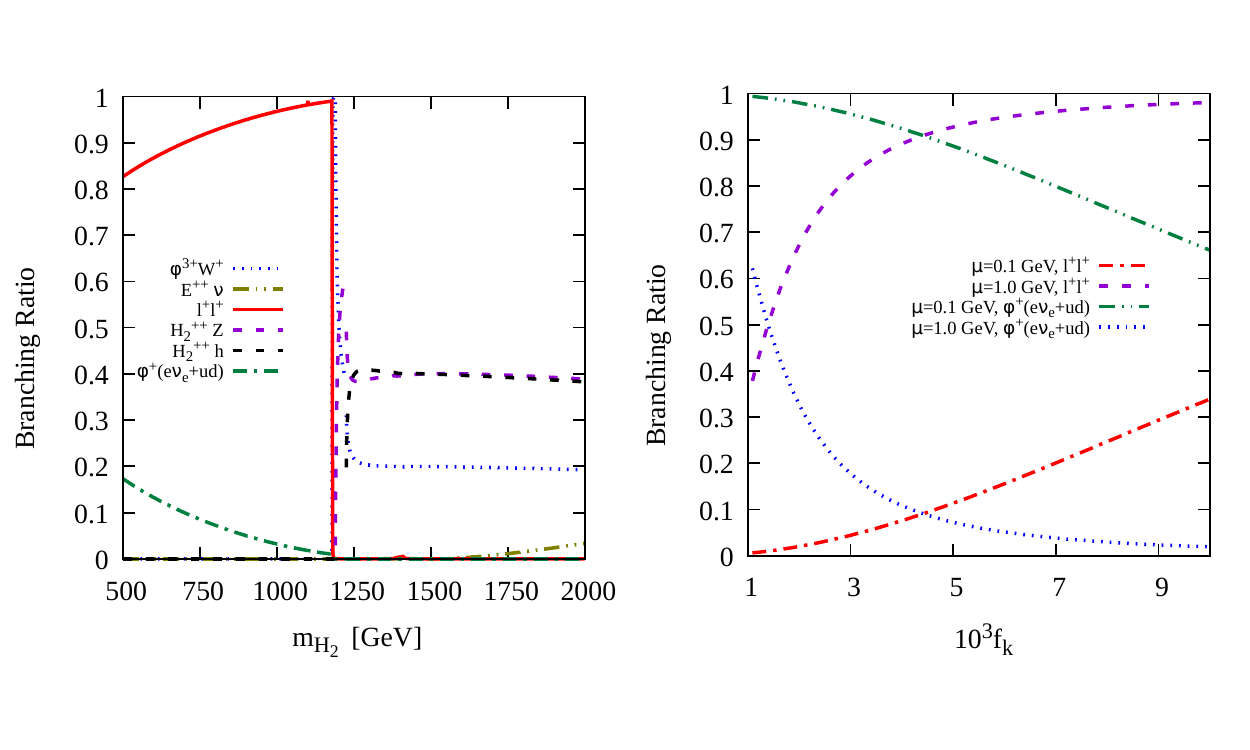}
  \mycaption{(Left panel) The branching ratios of $H_2^{\pm\pm}$ are presented as a function of $m_{H_2}$ for $m_{\frac{5}{2}}=1.1~{\rm TeV},~m_{k}=1.4~{\rm TeV},~m_{E}=1.5~{\rm TeV},~\mu=\mu^{\prime}=0.1~{\rm GeV},~f_{\frac{5}{2}\left(\frac{3}{2}\right)}=2\times 10^{-4},~\lambda=5\times 10^{-3}$ and $f_k=5\times 10^{-2}$. (Right panel) Branching ratios of two possible decay modes of $H_2^{\pm\pm}$ for $m_{\frac{3}{2}}~<~m_{\frac{5}{2}},~m_{k}$ and $m_E$ are presented as a function of $f_k$ for $m_{\frac{3}{2}}~=~0.8$ TeV and two different values of $\mu~=~0.1$ and 1 GeV.}
\label{h2_br}
\end{figure}

The left panel (right panel) of Fig.~\ref{h1_br} shows the branching ratios of $H_1^{\pm\pm}\left(H_3^{\pm\pm}\right)$ as a function of the doubly-charged scalar mass. $H_{1}^{\pm\pm}$ being lighter than $\phi^{3\pm}$, can not decay into $\phi^{3\pm}$. However, depending on the choice of mass parameters {\em i.e.,} $m_{\frac{5}{2}\left(\frac{3}{2}\right)},~m_k$ and $m_E$, the decays of $H_1^{\pm\pm}\left(H_3^{\pm\pm}\right)$ into other (multi-)charged scalars and fermion are allowed. If the decays of $H_{1}^{\pm\pm}\left(H_{3}^{\pm\pm}\right)$ into lighter (multi-)charged scalars or fermions are kinematically forbidden {\em i.e.,} $m_{\frac{5}{2}\left(k\right)}~<~m_{\frac{3}{2}},~m_{k\left(\frac{5}{2}\right)}$ and $m_E$, $H_{1}^{\pm\pm}\left(H_{3}^{\pm\pm}\right)$ dominantly decays into a same-sign dilepton. In the insets of Fig.~\ref{h1_br}, we have presented the total decay width. It is important to note that $\Gamma\left(H_{1}^{\pm\pm}~\to~l^\pm l^\pm\right)$ (see Eq.~\ref{h_w}) is suppressed by the Yukawa coupling $f_k$ as well as by the small mixing in the doubly charged scalar sector, whereas, $\Gamma\left(H_{3}^{\pm\pm}~\to~l^\pm l^\pm\right)$ is suppressed only by the Yukawa coupling $f_k$. The insets of Fig.~\ref{h1_br} shows that the chosen values of $f_k~=~0.1(2\times 10^{-3}),~\mu~=~0.1(10)$ GeV and $\lambda~=~5\times 10^{-3}$ ensure prompt decay of $H_{1(3)}^{\pm\pm}$ at the LHC. However, it can be easily estimated from Eq.~\ref{h_w} that with the same set of values for $\mu$ and $\lambda$, $f_{k}~<~10^{-3}\left(10^{-9}\right)$ gives rise to a long-lived $H_{1(3)}^{\pm\pm}$ which remains stable inside the detector.

The mass of the doubly-charged scalar $H_2^{\pm\pm}$ is slightly larger than the mass of $\phi^{\pm}$. The radiative mass splitting between $H_2^{\pm\pm}$ and $\phi^{\pm}$ is given by $m_{H_2^{\pm\pm}}-m_{\phi^{\pm}}~\approx~\frac{3}{2}\alpha_{EM}\,\,m_Z~\approx~1.1$ GeV. Therefore, $H_2^{++}$ can decay into an on-shell $\phi^{+}$ in association with a $e^+\nu_e$ or $u \bar d$ pair. The tree-level 3-body decay $H_2^{++}~\to~\phi^{+}e^+\nu_e$ proceeds through an off-shell $W$-boson and the partial decay width is given by,
\begin{equation}
  \Gamma\left(H_2^{\pm\pm}~\to~e^\pm\nu_e\phi^{\pm}\right)~=~\frac{\alpha_{EM}^2O_{22}^2}{32\,\pi\,{\rm sin}^4\theta_W\,m_W^4}\,\,\left(m_{H_2^{\pm\pm}}-m_{\phi^{\pm}}\right)^5.
  \label{3body_h2}
\end{equation}
The decay $H_2^{++}~\to~\phi^{+}e^+\nu_e$ is suppressed by the small mass splitting between $H_2^{\pm\pm}$ and $\phi^{\pm}$. However, if the above-mentioned 2-body decays of $H_2^{\pm\pm}$ are kinematically forbidden or suppressed (by the Yukawa parameters and/or mixing), the 3-body decay could be important and will have important consequences at the LHC which will be discussed in section~\ref{coll_scalar}. In Fig.~\ref{h2_br}(left panel), we have presented the branching ratios of $H_2^{\pm\pm}$ as a function of $m_{H_2}$ for $m_{\frac{5}{2}}=1.1~{\rm TeV},~m_{k}=1.4~{\rm TeV},~m_{E}=1.5~{\rm TeV},~\mu=\mu^{\prime}=0.1~{\rm GeV},~f_{\frac{5}{2}\left(\frac{3}{2}\right)}=2\times 10^{-4},~\lambda=5\times 10^{-3}$ and $f_k=5\times 10^{-2}$. Fig.~\ref{h2_br}(left panel) shows that for $m_{\frac{3}{2}\left(k\right)}~<~m_{\frac{5}{2}},~m_{k}$ and $m_E$, the possible decay modes of $H_2^{\pm\pm}$ are $H_2^{++}~\to~l^+ l^+$ and $H_2^{++}~\to~\phi^{+}+e^{+}\nu_e/u \bar d$ with the same-sign dileptonic decay being the dominant one. However, the partial decay widths of the  same-sign dileptonic decays of $H_2^{\pm\pm}$ (see Eq.~\ref{h_w}) are proportional to the $f_k^2$ as well as $O_{23}^2$. Therefore, suppressed mixing between $\phi^{\pm\pm}_{\frac{3}{2}}$ and $k^{\pm\pm}$ ({\em i.e.,} smaller $\mu$) and/or smaller $f_k$  result into suppressed  $\Gamma\left(H_2^{\pm\pm}\to l^{\pm}l^\pm\right)$ and hence, enhanced branching ratio for $H_2^{++}~\to~\phi^{+}+e^{+}\nu_e/u \bar d$. In Fig.~\ref{h2_br}(right panel), we have presented the  branching ratios of $H_2^{++}~\to~l^+ l^+$ and $H_2^{++}~\to~\phi^{+}+e^{+}\nu_e/u \bar d$ as a function of $f_k$ for $m_{\frac{3}{2}}~=~0.8$ TeV and two different values of $\mu~=~0.1$ and 1 GeV. Fig.~\ref{h2_br}(right panel) shows that for smaller values of $f_k$ and/or $\mu$, the kinematically suppressed 3-body decays, $H_2^{++}~\to~\phi^{+}+e^{+}\nu_e/u \bar d$, dominate over the Yukawa and mixing suppressed 2-body decays, $H_2^{++}~\to~l^+ l^+$.  After the detailed discussion about the possible decay modes and branching ratios of the (multi-)charged scalars, we are now prepared enough to study their collider signatures which will be discussed in the following.

\subsection{Collider signatures}
\label{coll_scalar}
In this section, we will discuss the collider signatures of (multi-)charged scalars at the LHC with $\sqrt s~=~13$ TeV. We will also study the impact of existing LHC searches on the parameter space of our model. In the context of the LHC phenomenology, we are particularly interested on the lightest scalar multiplet because it will be the lightest one that will be copiously produced and hence, more easily discovered. Therefore, the signatures of this model at the LHC can be classified according to the  hierarchy between the masses of different scalar multiplets and the doubly-charged fermion. In our analysis, we consider four possible scenarios defined in the following:
\begin{itemize}
\item {\em Scenario I} assumes $m_{\frac{3}{2}}~\ll~m_{\frac{5}{2}(k)},~m_E$ and hence, $\phi^{\pm}$ and $H_2^{\pm\pm}$, being the lightest among the (multi-)charged scalars and fermion, are the most important in the context of the LHC experiment. The collider phenomenology of {\em Scenario I} is determined by $m_{\frac{3}{2}}$, $\mu$ and $f_k$. The pair and associated production of $\phi^{\pm}/H_2^{\pm\pm}$ at the LHC are determined by $m_{\frac{3}{2}}$. Whereas, $\mu$ determines the $\phi^{\pm\pm}_{\frac{3}{2}}$--$k^{\pm\pm}$ mixing and hence, controls the branching ratios of $H_2^{\pm\pm}$. The branching ratios of $H_2^{\pm\pm}$ also depends on $f_{k}$ (see Fig.~\ref{h2_br}).  For our numerical calculations, we have assumed $m_{\frac{5}{2}(k)},~m_E~\sim~2.5$ TeV and varied $m_{\frac{3}{2}}$ over a range between 300--1500 GeV. Although, the collider phenomenology of {\em Scenario I} is almost insensitive on the values of $f_{\frac{3}{2}\left(\frac{5}{2}\right)}~{\rm and}~\lambda$,  we have assumed $f_{\frac{3}{2}\left(\frac{5}{2}\right)}~=~2\times 10^{-4}$ and $\lambda~=~5\times 10^{-3}$.
\item {\em Scenario II} is defined by the mass hierarchy $m_{\frac{5}{2}}~\ll~m_{\frac{3}{2}(k)},~m_E~\sim~2.5$ TeV. In this case, the signatures at the LHC are governed by the production and decay of $\phi^{3\pm}$ and $H_1^{\pm\pm}$ which are the lightest among the (multi-)charged scalars and fermion. The collider phenomenology of {\em Scenario II} is mainly determined by $m_{\frac{5}{2}}$, $\mu$ and $f_k$.    The values of other collider insensitive parameters are chosen same as in the case of {\em Scenario I}.
\item {\em Scenario III} assumes $m_{k}~\ll~m_{\frac{3}{2}\left(\frac{5}{2}\right)},~m_E~\sim~2.5$ TeV and hence, $H_3^{\pm\pm}$ being the lightest one, determines the collider signatures. If $H_3^{\pm\pm}$ is the lightest, it  decays into a same-sign dilepton pair with 100\% branching ratio. Therefore, the LHC phenomenology of {\em Scenario III} is completely determined by the mass of $H_3^{\pm\pm}$ and hence, $m_k$.
\item {\em Scenario IV} corresponds to $m_E~\ll~m_{\frac{5}{2}\left(\frac{3}{2}\right)},~m_{k}~\sim~2.5$ TeV and hence, results into a doubly-charged fermion as the lightest one. The phenomenology of {\em Scenario IV} has already been discussed in section~\ref{E_phenomenology}.
\end{itemize}

\begin{figure}[t]
  \centering
  \includegraphics[width= 0.8\linewidth]{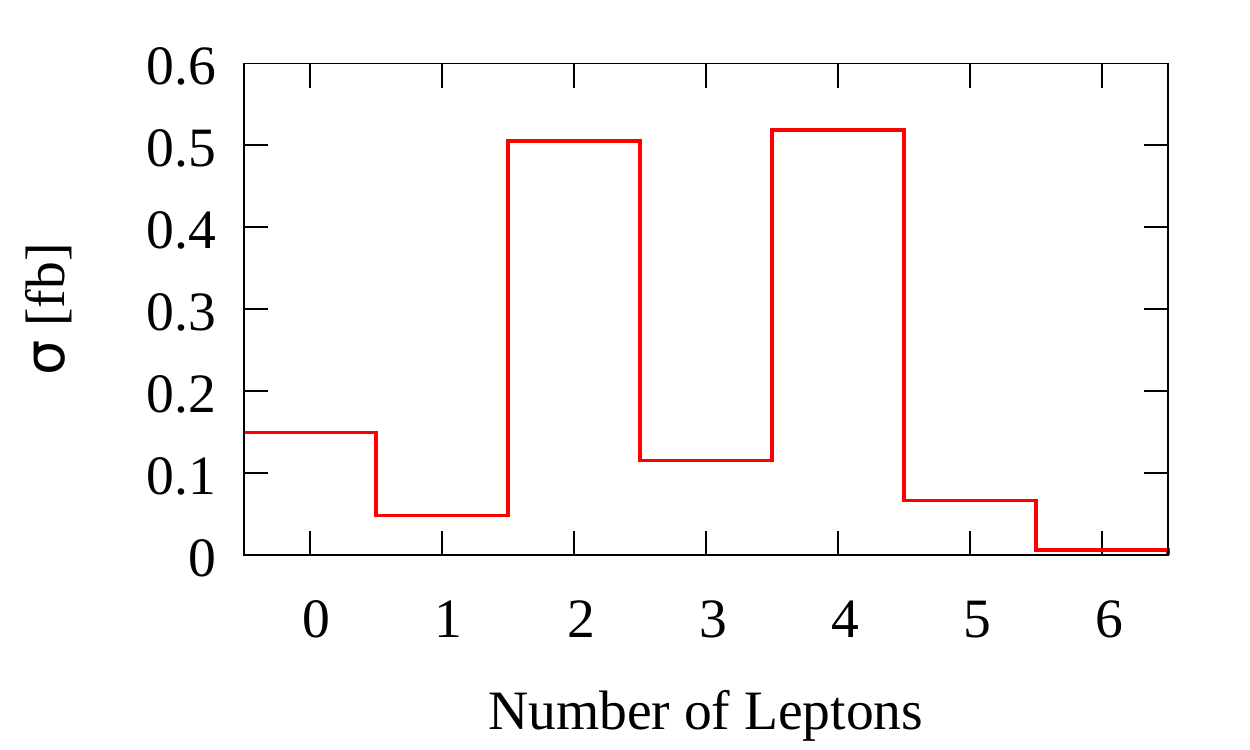}
\mycaption{Distribution of no. of leptons (electron and muon only) after imposing the {\em acceptance cuts} listed in Eqns.~\ref{cut:pT}--\ref{cut:jj-iso} is presented for {\em Scenario I} with $m_{\frac{3}{2}}~=~0.8$ TeV, $\mu~=~0.1$ GeV and $f_{k}~=~10^{-3}$.}
\label{nlep}
\end{figure}

In this section, we will study the LHC phenomenology of {\em Scenario I, II} and {\em III}. Before going into the detailed discussion about the prompt decay signatures of the above-mentioned scenarios, we will discuss the possibility of these (multi-)charged scalars being long-lived and hence, abnormally large ionization signatures at the LHC. In presence of the tree-level 3-body decays via an off-shell $W$-boson, $H_2^{\pm\pm}$ and $\phi^{3\pm}$ always undergo prompt decays at the LHC. However, $H_{1(3)}^{\pm\pm}$ can be long-lived in certain parts of parameter space (see section~\ref{2_decay}). In Fig.~\ref{scalar_pair_prod} (left panel), we have plotted the ATLAS observed 95\% CL upper limits on the pair-production cross-sections of long-lived doubly-charged particles ($\sigma^{95\%}_{DCPs}$) along with the model predictions for the doubly-charged scalar pair production cross-sections. Fig.~\ref{scalar_pair_prod} (left panel) shows that for a long-lived $H_{1(3)}^{\pm\pm}$, $m_{H_{1(3)}}$ below about 800 GeV is excluded from the ATLAS search for long-lived MCPs.

In {\em Scenario I} and {\em II}\footnote{The phenomenology of {\em Scenario III} is straight forward because of the presence of a characteristic same-sign dilepton invariant mass peak and hence, will be discussed separately.}, the prompt decays of (multi-)charged scalars result in one or more leptons, $W$-boson in the final state. Therefore, the pair and associated productions of $\phi^{\pm}~{\rm and}~H_2^{\pm\pm}$ (in the case of {\em Scenario I}) as well as $\phi^{3\pm}~{\rm and}~H_1^{\pm\pm}$  (in the case of {\em Scenario II}) give rise to multi-lepton, jets and $E_T\!\!\!\!\!\!/~$ signatures at the LHC. For example,  in the context of {\em Scenario I}, $\phi^{\pm}$ dominantly decays into a pair of same-sign leptons in association with a $W$-boson. Whereas, depending on the choice of $f_k$ and $\mu$, both same-sign dileptonic decays and 3-body decays into $\phi^{+}e^+\nu_e/u\bar d$ are possible for $H_2^{++}$. Therefore, depending on the subsequent decays of the $W$-bosons, the pair and associated production of $\phi^{\pm}~{\rm and}~H_2^{\pm\pm}$ may result into up to 6 leptons in the final state. For {\em Scenario I} with $m_{\frac{3}{2}}~=~0.8$ TeV, $\mu~=~0.1$ GeV and $f_{k}~=~10^{-3}$, Fig.~\ref{nlep} shows the no. of lepton (electron and muon only) distribution\footnote{In order to impose the cuts and generate the distributions, we have used a parton-level Monte-Carlo computer code. The technical details of our simulation were already discussed in section~\ref{E_coll}.} after the {\em acceptance cuts} listed in Eqns.~\ref{cut:pT}--\ref{cut:jj-iso}. Six lepton final state arises only from the pair production of  $\phi^{\pm}$ followed by the leptonic decay of both $W$-bosons and hence, is highly suppressed. Final states with odd no. of leptons are also suppressed because of the fact that only the production of $\phi^{\pm}\phi^{\mp}$ and $\phi^{\pm}H_2^{\mp\mp}$ contribute to the odd lepton final state when one $W$-boson decay leptonically. Whereas, dilepton and 4-lepton final states get contributions from all possible combinations of $\phi^{\pm}~{\rm and}~H_2^{\pm\pm}$ pair and associated productions and hence, are enhanced. The dilepton signature suffers from huge SM background and thus, we choose to study 4-lepton final states as a signature of (multi-)charged scalars in our model. In section~\ref{E_coll}, 4-lepton searches (existing ATLAS search as well as our proposed search) at the LHC have already been discussed in detail in the context of the doubly-charged fermion. In the next section, we will study the implications of those 4-lepton searches (discussed in section~\ref{E_coll}) in the context of the (multi-)charged scalars.  

\subsubsection{Four-lepton signature}

Without going into the details of 4-lepton search strategies (which have already been discussed in details in section~\ref{E_coll}), in Table~\ref{select_scalar}, $4l$ signal cross-sections after different cuts are presented for {\em Scenario I} and {\em II} for different values of $m_{\frac{3}{2}}$ and $m_{\frac{5}{2}}$, respectively.  The definitions of the {\em acceptance cuts}, the {\em ATLAS cuts} and the {\em proposed cuts} can be found in Eqns.~\ref{cut:pT}--\ref{cut:jj-iso}, Table~\ref{ATLAS_cuts} and Table~\ref{proposed_cuts}, respectively. Whereas, the potential sources of $4l$ contributions from the SM processes (the SM backgrounds) have been discussed in section~\ref{4l_back} and the numerical values of the SM background cross-sections after different cuts are presented in Table~\ref{cut_flow}.

\begin{table}[h!]
    \centering
    \begin{tabular}{ c||c|c}
      \hline\hline
      \multicolumn{3}{ c}{$4l$ signal Cross-sections [fb] after different cuts} \\
      \hline\hline
      Mass [GeV]   & ATLAS cuts & ATLAS + Proposed cuts\\\hline\hline
      $m_{\frac{3}{2}}$ & \multicolumn{2}{ c}{\em Scenario I} \\\hline
      650 & 0.32 & 0.27\\
      750 & 0.16 & 0.15\\\hline\hline
      $m_{\frac{5}{2}}$ & \multicolumn{2}{ c}{\em Scenario II} \\\hline
      750 & 0.34& 0.27\\
      850 & 0.2 & 0.17\\\hline\hline
    \end{tabular}
    \caption{$4l$ signal cross-sections for {\em Scenario I} and {\em II} after the ATLAS (tabulated in Table~\ref{ATLAS_cuts}) and ATLAS + proposed cuts (tabulated in Table~\ref{proposed_cuts}) are presented for different values of $m_{\frac{3}{2}}$ and $m_{\frac{5}{2}}$, respectively.}
    \label{select_scalar}
\end{table}

\begin{figure}[!ht]
  \centering
  \includegraphics[width=0.48 \linewidth]{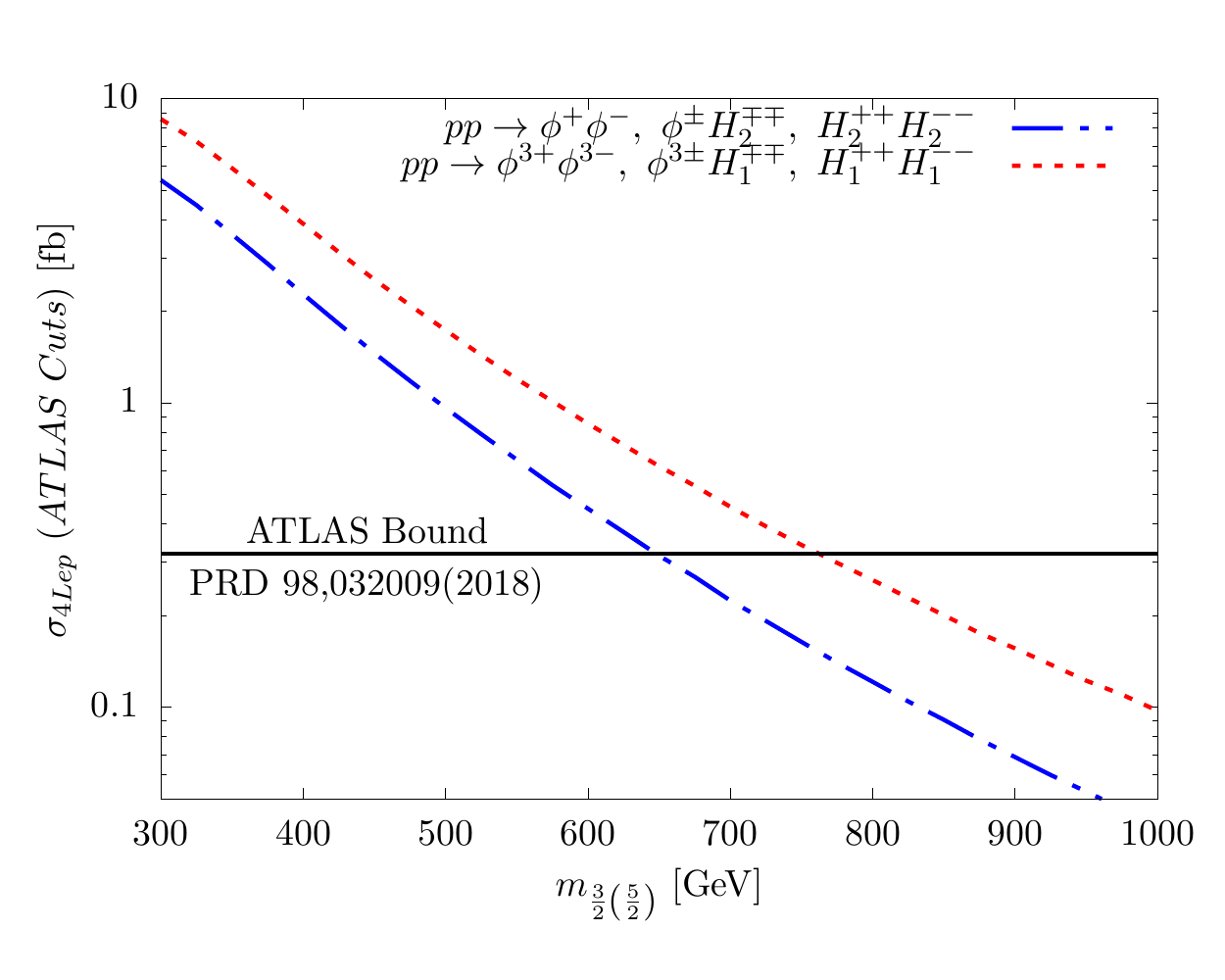}
  \includegraphics[width=0.48 \linewidth]{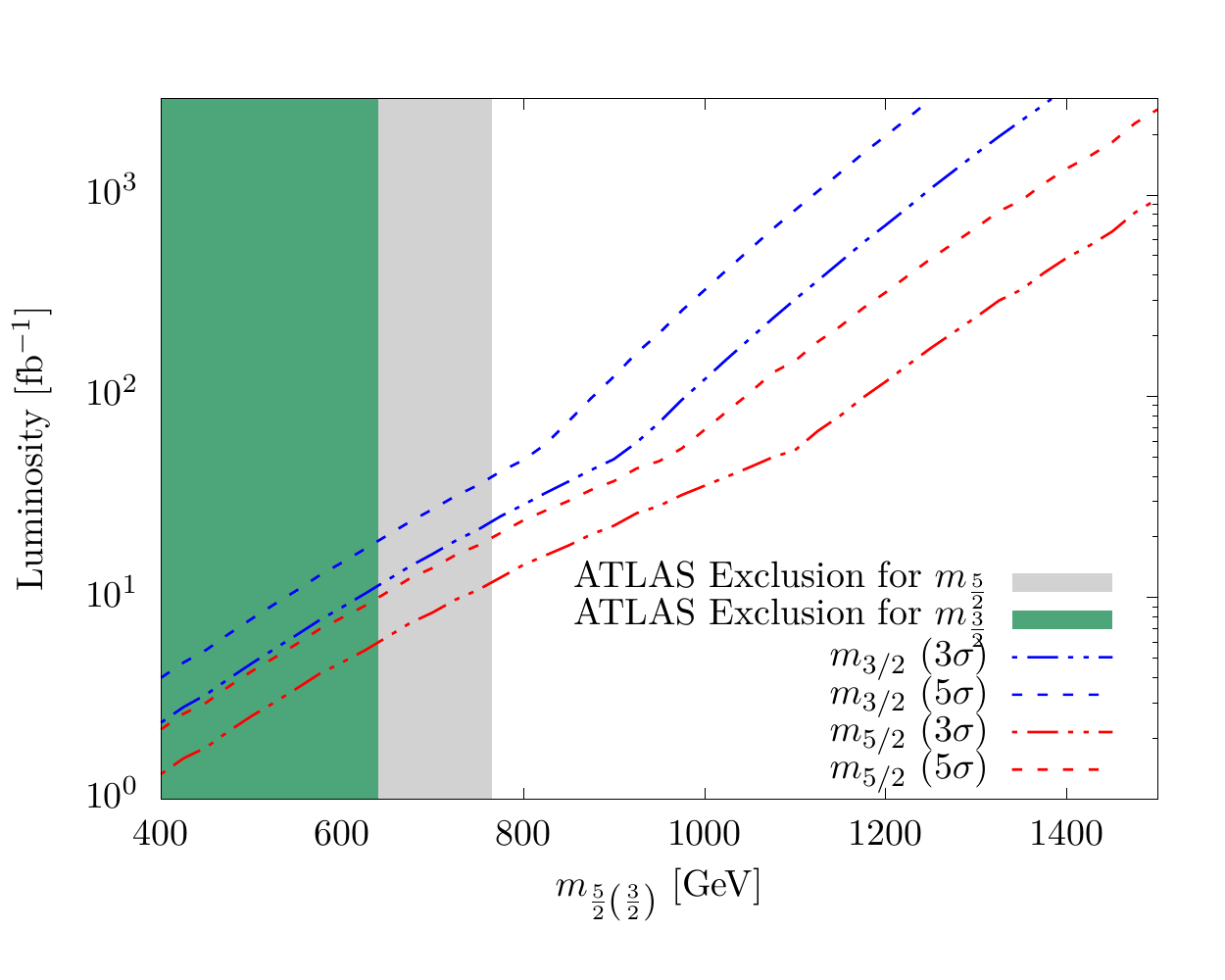}
\mycaption{(Left panel) Four lepton signal cross-section ($\sigma_{4Lep}$) as a function of $m_{\frac{3}{2}\left(\frac{5}{2}\right)}$ after the selection cuts (listed in Table~\ref{ATLAS_cuts}) used by the ATLAS collaboration in Ref.~\cite{Aaboud:2018zeb}. The black solid line corresponds to the observed  95\% CL upper bound on the 4-lepton signal cross-section. (Right panel) With the proposed event selection criteria (see Table~\ref{proposed_cuts}), required luminosities for $3\sigma$ and $5\sigma$ discovery of {\em Scenario I}({\em II}) are plotted as a function of $m_{\frac{3}{2}\left(\frac{5}{2}\right)}$.}
\label{bound_scalar}
\end{figure}

In Fig.~\ref{bound_scalar} (left panel), we have presented the $4l$ signal cross-sections (after the {\em acceptance cuts} and {\em ATLAS cuts}) for {\em Scenario I} and {\em II} as a function of $m_{\frac{3}{2}}$ and $m_{\frac{5}{2}}$, respectively. The horizontal line in Fig.~\ref{bound_scalar} (left panel) corresponds to the ATLAS 95\% CL upper limit on the visible $4l$ cross-section ($\sigma(4l)_{\rm vis}^{95}$) \cite{Aaboud:2018zeb}. Fig.~\ref{bound_scalar} (left panel) clearly shows that for $m_{\frac{3}{2}\left(\frac{5}{2}\right)} ~<~\sim 650(760)$ GeV, contribution from {\em Scenario I}({\em II}) to the visible $4l$ signal cross-section is larger than $\sigma(4l)_{\rm vis}^{95}$. Therefore, one can set a lower bound of about 650(760) GeV on $m_{\frac{3}{2}\left(\frac{5}{2}\right)}$ from the ATLAS search for the $4l$ final state in Ref.~\cite{Aaboud:2018zeb}.

In Fig.~\ref{bound_scalar} (right panel), the required integrated luminosities for the $3\sigma$ and $5\sigma$ discovery of {\em Scenario I}({\em II}) are presented as a function of $m_{\frac{3}{2}\left(\frac{5}{2}\right)}$ at the LHC with $\sqrt s~=~13$ TeV.  We found that the LHC with 3000 fb$^{-1}$ integrated luminosity and 13 TeV center-of-mass energy will be able to probe $m_{\frac{3}{2}\left(\frac{5}{2}\right)}$ upto about 1250 (1530) GeV at 3$\sigma$ significance. The shaded region of Fig.~\ref{bound_scalar} (right panel) corresponds to the part of parameter-space which is already excluded from the ATLAS $4l$-search in Ref.~\cite{Aaboud:2018zeb}.

\subsubsection{Search for the doubly-charged scalars}

\begin{figure}[t]
  \centering
  \includegraphics[width= 0.8\linewidth]{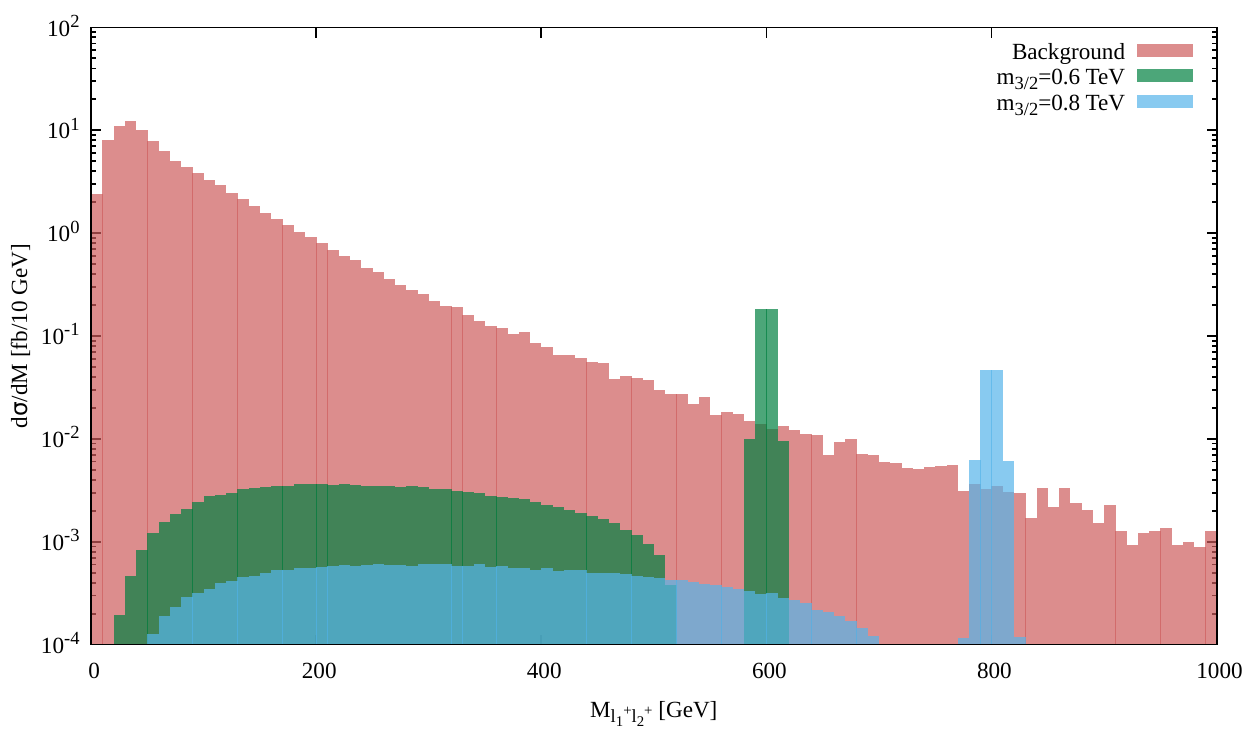}
  \mycaption{Same sign dilepton invariant mass ($M_{l_1^+ l_2^+}$) distributions for both signal ($m_{\frac{3}{2}}=$ 0.6 and 0.8 TeV) and the SM background are presented for {\em Scenario I} after imposing the {\em acceptance cuts} (listed in Eqns.~\ref{cut:pT}--\ref{cut:jj-iso}) on the $4l$ final state (see section~\ref{4l_selection} for details). We have assumed $\mu~=~1$ GeV and $f_k~=~10^{-2}$.}
\label{SSD_phi}
\end{figure}

One of the possible decay modes of the doubly-charged scalars is $H_a^{\pm\pm}~\to~l^\pm l^\pm$.  Therefore, some of the 4-lepton signal events, discussed in the previous section, might contain a same-sign pair of leptons resulting from the decay of a doubly-charged scalar. Though the branching ratios of the same-sign dileptonic decays of the doubly-charged scalars depend on the choice of $f_k$ and $\mu$, it is instructive to search for invariant mass peaks in the same-sign dilepton invariant mass distributions\footnote{Final states with a pair of positively charged and a pair of negatively charged lepton falls into the category of 4-lepton final state defined in section~\ref{4l_selection}. Therefore, two same-sign dilepton invariant mass distributions can be constructed.}. For example, in {\em Scenario I}, the doubly-charged scalar, $H_2^{\pm\pm}$, dominantly decays into $l^\pm l^\pm$ for larger $\mu$ and $f_k$ and hence,  one expects to observe characteristic same-sign dilepton invariant mass peaks as a signature of $H_2^{\pm\pm}$-pair production.  The pair productions $H_1^{\pm \pm}H_1^{\mp \mp}$ and  $H_3^{\pm \pm}H_3^{\mp \mp}$ in {\em Scenario II} and {\em III}, respectively, always result into spectacular same-sign dilepton invariant mass peaks (see Fig.~\ref{h1_br} for the branching ratios of $H_{1(3)}^{\pm\pm}$). In {\em Scenario II}, the production of $\phi^{3\pm}\phi^{3\mp}$ and $\phi^{3\pm}H_1^{\pm\pm}$ followed by the decay of $\phi^{3\pm}$ into an on-shell $H_1^{\pm\pm}$ also contribute to the same-sign dilepton invariant mass peaks. Fig.~\ref{SSD_phi} shows the same-flavour, same-sign (positive) dilepton invariant mass ($M_{l_1^+ l_2^+}$) distributions for both the signal ({\em Scenario I} with $\mu~=~1$ GeV and $f_k~=~10^{-2}$, $m_{\frac{3}{2}}=$ 0.6 and 0.8 TeV) and the SM background after imposing the {\em acceptance cuts}. The signal $M_{l_1^+ l_2^+}$-distributions are characterized by a spectacular peak at $m_{\frac{3}{2}}$ (resulting from the production and same-sign dileptonic decay of $H_{2}^{\pm\pm}$) followed by a kinematic edge (resulting from the production and decay of $\phi^{\pm}$) at $m_{\frac{3}{2}}-m_W$.

\begin{figure}[!ht]
  \centering
  \includegraphics[width= 0.8\linewidth]{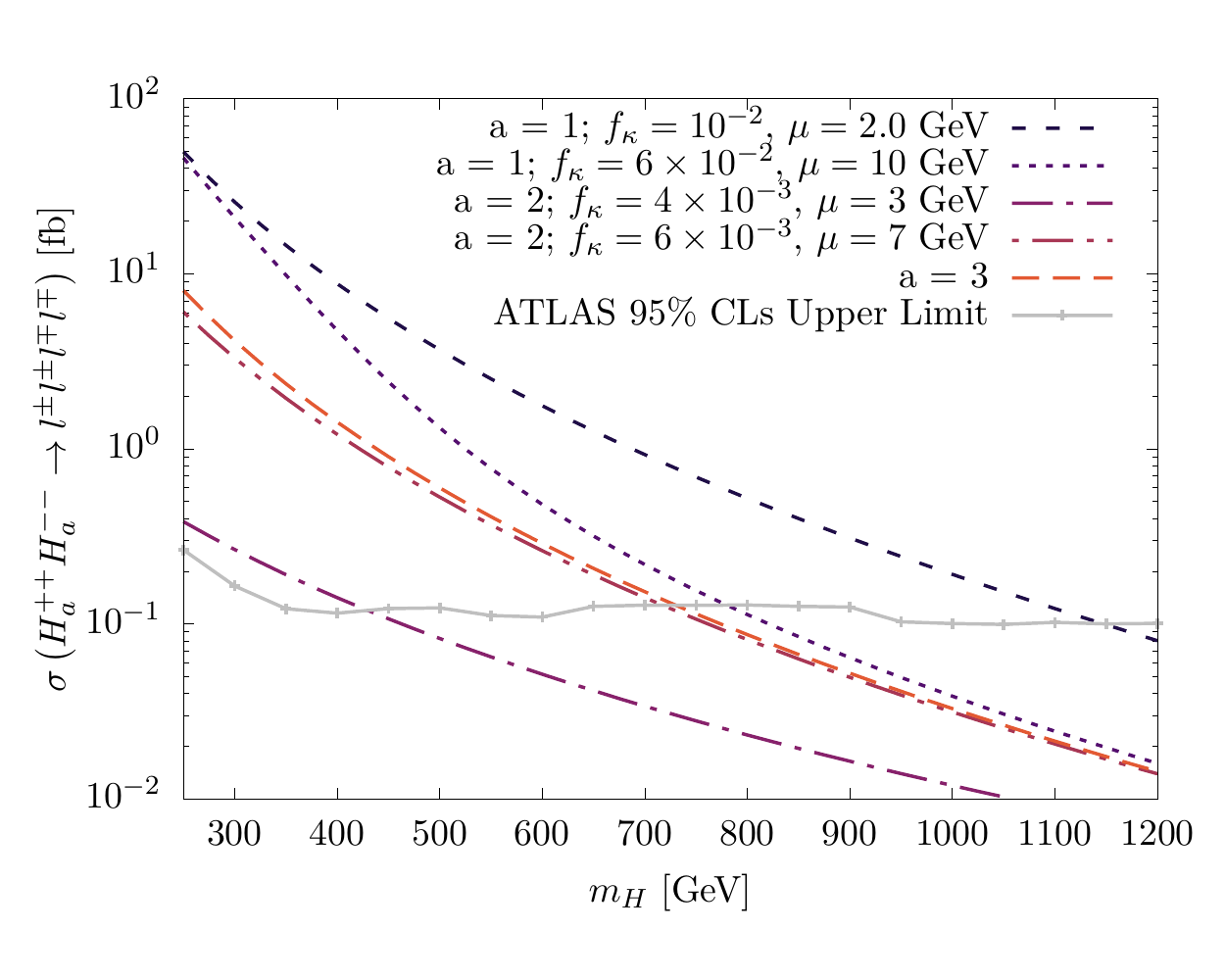}
\mycaption{The pair-production cross-sections of doubly-charged scalars ($H_a^{\pm\pm}$ where $a\ni 1,2,3$) decaying into $l^\pm l^\pm$ ($l\ni e,~\mu$) are presented as a function of doubly-charged scalar mass ($m_{H_a}$) for different values of $\mu$ and $f_k$. The grey solid line corresponds to the ATLAS observed \cite{Aaboud:2017qph} model independent 95\% CL upper limit on the doubly-charged scalar pair production cross-section.}
\label{bound_h_cross}
\end{figure}

Leptonic final states with a pair of prompt, isolated, highly energetic leptons with the same electric charge are very rare in the framework of the SM. However, such final states might have a significant rate in the framework of different BSM scenarios and thus, are extensively studied by the ATLAS~\cite{Aaboud:2017qph,ATLAS:2017iqw,Ucchielli:2017qad,ATLAS:2014kca,Nuti:2014eaa,ATLAS:2012hi} and the CMS~\cite{Chatrchyan:2012ya,CMS:2017pet,CMS:2016cpz} collaborations of the LHC experiments. To constrain the parameter space of our model, we are interested in the existing LHC searches for a doubly-charged scalar decaying into a pair of same-sign leptons. In this work, we consider the ATLAS search \cite{Aaboud:2017qph} for an invariant mass peak in the observed invariant mass of same-charge lepton pairs with 36.1 fb$^{-1}$ integrated luminosity data of the LHC running at $\sqrt s~=~$13 TeV. The ATLAS analysis in Ref.~\cite{Aaboud:2017qph} is aimed to search for a doubly-charged scalar that is not only present in our model but also arises in a large variety of BSM scenarios~\cite{Zee:1985id,Babu:1988ki,Nebot:2007bc,Gunion:1989in,Pati:1974yy,Mohapatra:1974hk,Senjanovic:1975rk,Gunion:1989ci,ArkaniHamed:2002qx,Muhlleitner:2003me,Perez:2008zc,Georgi:1985nv}. However, the observed data is found to be consistent with the SM predictions in all the two, three and four lepton signal regions\footnote{To constrain the parameter space of our model, we have used the ATLAS 95\% CL upper bound (model-independent) on the pair production cross-section of doubly-charged scalars decaying into a pair of same-charge leptons. The definitions of different signal regions and event selection criteria, which lead to the above-mentioned bound, are not discussed here. We refer the interested readers to Ref.~\cite{Aaboud:2017qph} for details.} considered in Ref~.\cite{Aaboud:2017qph}. In the absence of evidence for a signal, the pair production cross-section of doubly-charged scalar, which decays into a pair of same-sign electrons/muons with a 100\% branching ratio, is excluded down to about 0.1 fb at 95\% CL. In the context of our model, the ATLAS model-independent bounds on the pair production cross-sections of doubly-charged scalars are utilized to constrain the masses of $H_1^{\pm\pm}$, $H_2^{\pm\pm}$ and $H_3^{\pm\pm}$ in {\em Scenario II, I} and {\em III}, respectively.

\begin{figure}[!ht]
  \centering
  \includegraphics[width= 1.0\linewidth]{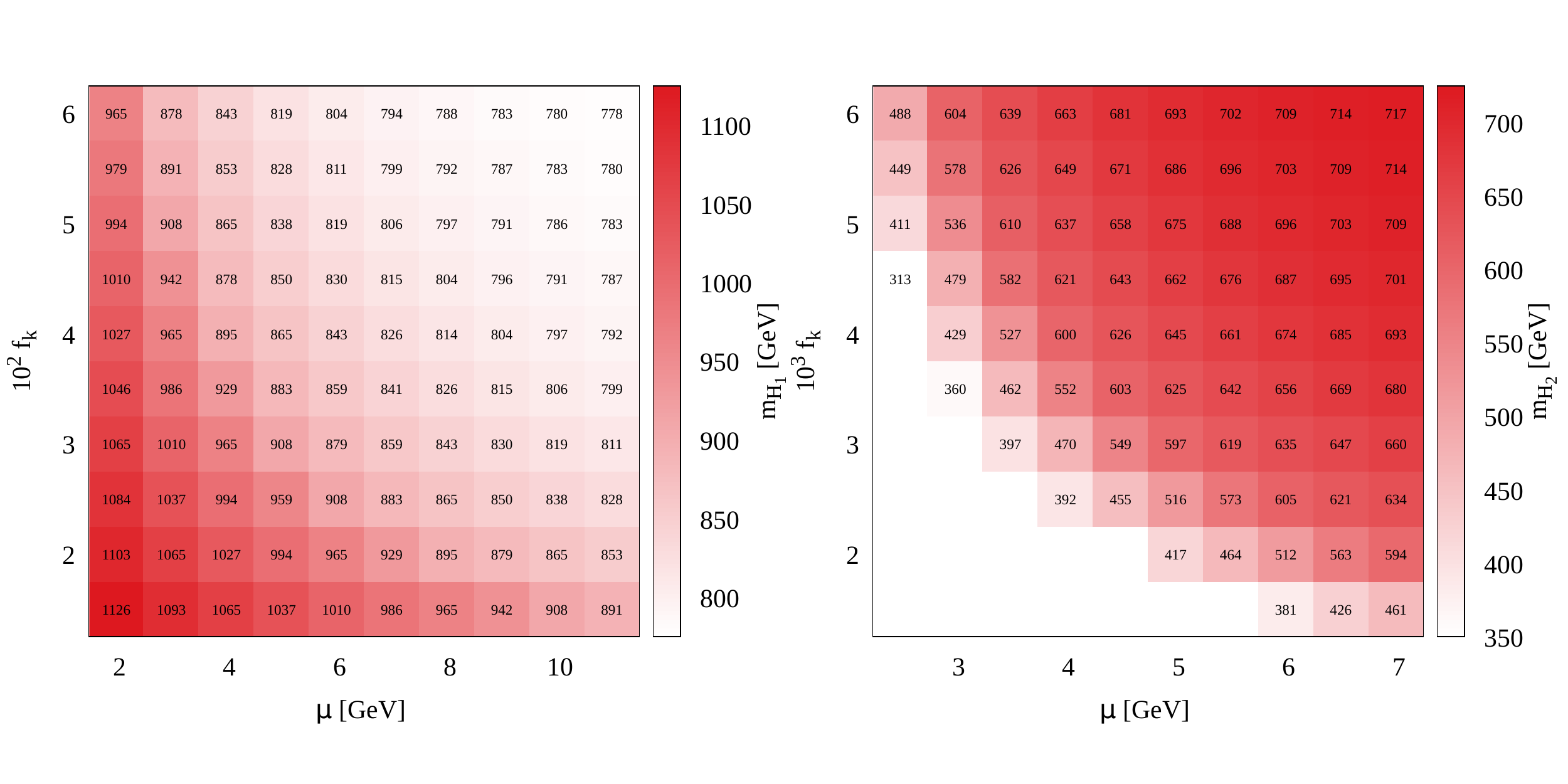}
\mycaption{The observed 95\% CL lower limits (color gradient) on the masses of the doubly-charged scalars in {\em Scenario I}(right panel) and {\em Scenario II}(left panel) are presented as a function of $\mu$(x--axis) and $f_k$(y--axis)}
\label{bound_h}
\end{figure}

Fig.~\ref{bound_h_cross} shows the model predictions for the doubly-charged scalar pair-production cross-sections for {\em Scenario I}($a=2$), {\em II}($a=1$) and {\em III}($a=3$) as a function of $m_{H_a}$ for different values of $\mu$ and $f_k$. The ATLAS observed 95\% CL upper bound on the pair-production cross-section of the doubly-charged scalars decaying into pair of same-sign leptons is also depicted in Fig.~\ref{bound_h_cross}. The model predictions for the pair productions of doubly-charged scalars followed by the same-sign dileptonic decays significantly depend on the values of $\mu$ and $f_k$ for {\em Scenario I} and {\em II}.  This can be attributed to the fact that in presence of a second decay mode (namely, $H_2^{\pm\pm}~\to~\phi^{\pm}W^{*\pm}$), the same-sign dileptonic decay branching ratios of $H_{2}^{\pm\pm}$ in {\em Scenario I} depend on $\mu$ and $f_k$. In particular, larger $\mu$ and $f_k$ correspond to larger branching ratios for $H_2^{\pm\pm}~\to ~ l^\pm l^\pm$ and hence, stronger bound on $m_{H_2}$ from the ATLAS search for same-sign dilepton invariant mass peak. Fig.~\ref{bound_h_cross} shows that $m_{H_2}$ below $\sim$ 720(430) GeV is excluded for $\mu~=~7$ GeV and $f_k~=~6\times 10^{-3}$($\mu~=~3$ GeV and $f_k~=~3.5\times 10^{-3}$). In {\em Scenario II}, the same-sign dileptonic decays are the only allowed decay modes for $H_{1}^{\pm\pm}$ and hence,  $H_1^{\pm\pm}~\to~l^\pm l^\pm$ has  100\% branching ratio. However, the pair-production of $H_1^{\pm\pm}$ gets extra contributions from the pair and associated productions of $\phi^{3\pm}$ followed by the decay of $\phi^{3\pm}$ into an on-shell $H_1^{\pm\pm}$. The $\mu$ and $f_k$ dependence in the theoretical predictions for the $H_1^{\pm\pm}$ ({\em Scenario II}) pair-productions in Fig.~\ref{bound_h_cross} arises from the  $\mu$ and $f_k$ of $\phi^{3\pm}~\to ~ H_1^{\pm\pm}W^{*\pm}$ branching ratio (see left panel of Fig.~\ref{br_phi3}). Fig.~\ref{bound_h_cross} shows that the ATLAS bound on the doubly-charged scalar pair production cross-section in Ref.~\cite{Aaboud:2017qph} sets a lower bound of about 780(1130) GeV on $m_{H_1}$ for  $\mu~=~10$ GeV and $f_k~=~6\times 10^{-2}$($\mu~=~2$ GeV and $f_k~=~10^{-2}$). It is clear from the previous discussion that for {\em Scenario I} and {\em Scenario II}, the lower bounds on the masses of doubly-charged scalars crucially depends on the values of $\mu$ and $f_k$. Therefore, in Fig.~\ref{bound_h}, we have tabulated the lower bounds on $m_{H_1}$ (left panel) and $m_{H_2}$ (right panel) as a function of $\mu$ and $f_k$. Fig.~\ref{bound_h} shows that for {\em Scenario I}({\em Scenario II}), the lower bound on $m_{H_2}(m_{H_1})$ can be as low(high) as 313(1150) GeV for smaller values of $\mu$ and $f_k$.

\subsection{Summary}
The collider phenomenology of (multi-)charged scalars is discussed in this section. At the LHC with $\sqrt s~=~13$ TeV, the pair and associated production rate of (multi-)scalars (with masses of the order of few hundred GeVs to TeV) are large enough to study their signatures. After being produced at the LHC,  $\phi^{3\pm}$ and $H_2^{\pm\pm}$ undergo prompt decay. Whereas, for smaller values of $\mu,~\lambda$ and $f_k$, the decay length of $\phi^{\pm}$, $H_1^{\pm\pm}$ and $H_3^{\pm\pm}$ could be large enough to ensure the decay of these scalars out side the LHC detector. ATLAS search for abnormally large ionization signature to probe long-lived multi-charged particles excludes $m_{H_{1(3)}}$ below about 800 GeV for long-lived $H_{1(3)}^{\pm\pm}$. The prompt decays of the (multi-)charged scalars at the LHC give rise to interesting multi-lepton final states with  characteristics kinematic distributions. We have studied 4-lepton final states and the important results are summarized in the following:
\begin{itemize}
\item In {\em Scenario I}({\em II}), $m_{\frac{3}{2}\left(\frac{5}{2}\right)}$ is excluded below 650(760) GeV at 95\% CL from the ATLAS search for 4-lepton signatures in Ref.~\cite{Aaboud:2018zeb}. With our proposed 4-lepton signal selection criteria optimized for this model, the expected reach of the LHC with 3000 fb$^{-1}$ is estimated to be 1250(1530) GeV for $m_{\frac{3}{2}\left(\frac{5}{2}\right)}$ in the context of {\em Scenario I}({\em II}).
\item The ATLAS search \cite{Aaboud:2017qph} for doubly-charged scalars using the observed invariant mass of same-sign lepton pairs at the 13 TeV LHC with 36.1 fb$^{-1}$ integrated luminosity data puts an upper limit of about 730 GeV on $m_{H_3}$ in {\em Scenario III}. Whereas, the observed upper limits on  $m_{H_{2(1)}}$ in {\em Scenario I}({\em II}) varies in range 313--720(1150--780) GeV as we vary the values of $\mu$ and $f_k$ from smaller to larger.
\item In the context of {\em Scenario I}, It is important to note the complementarity between the ATLAS searches in Ref.~\cite{Aaboud:2018zeb} and Ref.~\cite{Aaboud:2017qph}. For larger values of $\mu$ and $f_k$, stronger bound on $m_{H_2}$ (and hence, $m_{\frac{3}{2}}$) results from the ATLAS search for same-sign dilepton invariant mass peak in Ref.~\cite{Aaboud:2017qph}. Whereas, the 4-lepton search in Ref.~\cite{Aaboud:2018zeb} yields the most stringent bound on $m_{\frac{3}{2}}$ for the smaller values $\mu$ and $f_k$.
\item For {\em Scenario II}, however, the strongest bound on $m_{\frac{5}{2}}$ always arises from same-sign dilepton invariant mass peak search in Ref.~\cite{Aaboud:2017qph} regardless of the values of $\mu$ and $f_k$.  
\end{itemize}

\section{Conclusion and Outlook}
\label{conclusion}

We studied the phenomenology of a model that generates neutrino masses at a 1-loop level. To realize the Weinberg operator at 1-loop level, the model includes additional scalar doublets and singlet as well as fermion singlets in the framework of the SM gauge symmetry. Usually, loop induced neutrino mass models require some additional symmetry to forbid tree-level seesaw contributions to the Weinberg operator. However, in our model, the additional fields and their gauge quantum numbers are chosen in such a way that the couplings, which give rise to the Weinberg operator at tree-level, are absent and hence, the tree-level contributions to the neutrino masses are forbidden without any additional symmetry. Apart from effortlessly explaining neutrino oscillation data with Yukawa couplings of the order of the SM charged lepton Yukawa couplings, the model can explain the discrepancy between the experimental measurement and the SM prediction of muon magnetic moment $(g-2)_\mu$ and gives rise to interesting signatures at the collider experiments.

In this work, we have studied the neutrino and collider phenomenology of this model. After fitting the neutrino masses and mixings, we studied the constraints resulting from the upper bound on the absolute neutrino mass scale. In this model, the absolute values of the neutrino masses depend on the Yukawa couplings, masses of heavy fermions/scalars and the mixings between the doubly-charged scalars {\em i.e.,} on the cubic ($\mu,~\mu^\prime$) and quartic ($\lambda$) terms in the scalar potential. We found that the lower region of the $\mu$--$\lambda$ plane is consistent with the upper bound on the absolute neutrino mass scale for Yukawa couplings of the order of $10^{-3}$--$10^{-4}$ and TeV scale masses of the newly introduced scalars and fermions. We studied the production, decay, and the resulting collider signatures of these TeV scale fermion/scalars in the context of the LHC experiment.

Being (multi-)charged, the heavy-fermion and the scalars are pair-produced via Drell-Yan or photon-fusion processes at the hadron colliders. We have shown that at the LHC with $\sqrt s~=~13$ TeV, the photoproduction, being enhanced by a factor of $Q^4$ where $Q$ is the electric charge of the final state fermion/scalars, contributes significantly (even dominantly in some cases) to the total pair-production cross-sections and hence, cannot be neglected. The associated productions of the TeV scale scalars have also been considered. The signatures of these fermion and scalars at the LHC crucially depend on their subsequent decays. Depending on the total decay widths of these TeV scale particles, we have classified the collider signatures into two categories.\\
{\bf \em Prompt Decay Signatures:} If the particle decay width is large enough to ensure prompt decay at the LHC, we studied multi-lepton (in particular, 4-lepton) signatures. Bounds on the masses of the new scalars and fermion are derived from ATLAS search for $4l$~\cite{Aaboud:2018zeb} at the LHC with $\sqrt s~=~13$ TeV and 36.1 fb$^{-1}$ integrated luminosity. We found that the doubly-charged fermion mass below 870 GeV is excluded. Whereas, the bounds on the masses of the multi-charged scalars vary between 650--760 GeV depending on the hypercharges. The ATLAS $4l$ search strategy in Ref.~\cite{Aaboud:2018zeb} is designed to search for electroweakinos in the context of supersymmetric scenarios. We have proposed a new set of kinematic cuts to maximize the $4l$ signal to background ratio in the context of our model and shown that the reach of the LHC could be significantly improved with the proposed event selection criteria. The production and subsequent same-sign dileptonic decays of the doubly-charged scalars give rise to characteristic same-sign dilepton invariant mass peak signature which has already been studied by the ATLAS collaboration in Ref.~\cite{Aaboud:2017qph}. We recast the ATLAS results in Ref.~\cite{Aaboud:2017qph} in the context of our model and obtain bounds on masses of the doubly-charged scalars. In the context of our model, we found complementarity between ATLAS searches in Ref.~\cite{Aaboud:2018zeb} and \cite{Aaboud:2017qph}.\\
{\bf \em Highly Ionizing Charge Tracks:} The 2-body gauge decays for the lightest of the newly introduced TeV scale scalars/fermion are kinematically forbidden. The remaining Yukawa induced 2-body decays or tree-level 3-body decays are suppressed by several factors like, the Yukawa couplings, mixings in the doubly-charged scalar sector, phase-space {\em e.t.c.} Note that the upper bound on the absolute neutrino mass scale gives rise to stringent constraints on the Yukawa couplings and mixings in the doubly-charged scalar ({\em i.e.,} the values of $\mu,~\mu^{\prime}$ and $\lambda$) sector. As a consequence, the doubly-charged fermion or scalars become long-lived in a significant part of the allowed parameter space. A long-lived multi-charged scalar/fermion is highly ionizing, and thus leave a very characteristic signature of abnormally large ionization at the LHC. Using the results from the ATLAS search \cite{Aaboud:2018kbe} for abnormally large ionization signatures, we obtain lower bounds of about 1150 GeV and 800 GeV on the masses of the long-lived doubly-charged fermion and scalars, respectively.

We mention in closing that we obtained bounds on the parameter-space a collider testable radiative seesaw model. Our collider analysis is limited to parton-level Monte-Carlo and hence, cannot simulate ISR, FSR, and subsequent hadronization of the final state quarks and gluons. As a result, we have only considered purely leptonic final states. In this work, we have discussed the collider signatures of the lightest non-SM scalars/fermion and derived the collider bounds on their masses from the LHC with $\sqrt s~=~13$ TeV and 36.1 fb$^{-1}$ integrated luminosity data. It is important to note that the next-to-lightest non-SM scalars/fermion dominantly decay into the lightest non-SM scalars/fermion in association with an SM gauge boson ($W/Z$-boson) or a Higgs boson. Therefore, the production and subsequent decays of the next-to-lightest non-SM scalars/fermion in the framework of this model give rise to interesting di-boson (Higgs, $W/Z$-boson) in association with multiple leptons signatures at the LHC. This is beyond the scope of this article to consider all those possible final states. However, this article will surely pave the way for such phenomenological studies.       

\subsubsection*{Acknowledgments}
We thank Dr. Aruna Kumar Nayak and Saiyad Ashanujjaman for their technical help. The simulations were performed on SAMKHYA, the high performance computing facility at IOP, Bhubaneswar. K.G. acknowledges the support from the DST/INSPIRE Research Grant [DST/INSPIRE/04/2014/002158] and SERB Core Research Grant [CRG/2019/006831].

\section*{Appendix}
\appendix
\section{List of Feynman Rules}
\label{feyn}
\begin{minipage}[t]{0.24\textwidth}
  \begin{center}
    \begin{tikzpicture}[line width=1.4 pt, scale=1.45 ,every node/.style={scale=0.9}]    
      \draw[vector,black] (-1.0, 0.0) --(0.0, 0.0);
      \draw[scalar,black] (0,0.0) --(1.0,-1.0);
      \draw[scalar,black] (0,0.0) --(1.0,1.0);
      \node at (-0.70,0.3) {$A_{\mu}$};
      \node at (0.5,0.8) {$H^{++}_a$};
      \node at (0.5,-0.8) {$H^{--}_b$};
      \node at (0.5,0.3) {$p_1$};
      \node at (0.5,-0.3) {$p_2$};
      \node at (0.0,-1.30) {$:  i 2e \delta_{ab} \left(p_1-p_2 \right)_{\mu}$};
    \end{tikzpicture}\\
    \vspace{7mm}
    \begin{tikzpicture}[line width=1.4 pt, scale=1.45 ,every node/.style={scale=0.9}] 
      \draw[vector,black] (-1.0, 0.0) --(0.0, 0.0);
      \draw[scalar,black] (0,0.0) --(1.0,1.0);
      \draw[scalar,black] (1.0,-1.0)  --(0,0.0);
      \node at (-0.70,0.3) {$W^{+(-)}_{\mu}$};
      \node at (0.5,-0.8) {$H^{++}_a$};
      \node at (0.5,0.8) {$\phi^{+3(1)}$};
      \node at (0.5,0.3) {$p_1$};
      \node at (0.5,-0.3) {$p_2$};
      \node at (0.0,-1.30) {$: i \frac{e O_{a1(2)}}{\sqrt{2} \rm{sin} {\theta_w}} \left(p_1-p_2 \right)_{\mu}$};
    \end{tikzpicture}\\
    \vspace{14mm}
    \begin{tikzpicture}[line width=1.4 pt, scale=1.45 ,every node/.style={scale=0.9}] 
      \draw[fermion,black] (0.0, 0.0) --(1.0, 1.0);
      \draw[fermionbar,black] (0.0, 0.0) --(1.0, -1.0);
      \draw[scalar,black] (-1.0,0.0) --(0.0,0.0);
      \node at (-0.70,0.3) {$H^{++}_a$};
      \node at (0.5,0.8) {$E^{++}_{\alpha}$};
      \node at (0.5,-0.8) {$\nu_{\beta}$};
      \node at (0.0,-1.3) {$:- i( y^{\alpha \beta}_{\frac{5}{2}} O_{a1}P_L + y^{\alpha \beta}_{\frac{3}{2} } O_{a2}P_R)$};
    \end{tikzpicture}\\
    \vspace{5mm}
    \begin{tikzpicture}[line width=1.4 pt, scale=1.45,every node/.style={scale=0.9}] 
      \draw[vector,black] (-1.0,1.0) --(0.0,0.0);
      \draw[vector,black] (-1.0,-1.0)  --(0.0,0.0);
      \draw[scalar,black] (0,0.0) --(1.0,1.0);
      \draw[scalar,black] (0,0.0) --(1.0,-1.0);
      \node at (-0.5,-0.8) {$A_{\nu}$};
      \node at (-0.5,0.8) {$A_{\mu}$};
      \node at (0.5,-0.8) {$\phi^{Q-}$};
      \node at (0.5,0.8) {$\phi^{Q+}$};
      \node at (0.0,-1.30) {$: i2e^2Q^2 g_{\mu \nu}$};
    \end{tikzpicture}
  \end{center}
\end{minipage}
\hspace{0.12\textwidth}
\begin{minipage}[t]{0.24\textwidth}
  \begin{center}
    \begin{tikzpicture}[line width=1.4 pt, scale=1.45 ,every node/.style={scale=0.9}] 
      \draw[vector,black] (-1.0, 0.0) --(0.0, 0.0);
      \draw[scalar,black] (0,0.0) --(1.0,1.0);
      \draw[scalar,black] (0,0.0) --(1.0,-1.0);
      \node at (-0.70,0.3) {$A_{\mu}$};
      \node at (0.5,-0.8) {$\phi^{Q-}$};
      \node at (0.5,0.8) {$\phi^{Q+}$};
      \node at (0.5,0.3) {$p_1$};
      \node at (0.5,-0.3) {$p_2$};
      \node at (0.0,-1.30) {$:ieQ \left(p_1-p_2 \right)_{\mu}$};
    \end{tikzpicture}\\
    \vspace{7mm}
    \begin{tikzpicture}[line width=1.2 pt, scale=1.45,every node/.style={scale=0.9}]
      \draw[vector,black] (-1.0, 0.0)  --(0.0, 0.0);
      \draw[scalar,black] (0,0.0)  --(1.0,-1.0);
      \draw[scalar,black] (0,0.0) --(1.0,1.0);
      \node at (-0.70,0.3) {$Z_{\mu}$};
      \node at (0.5,0.8) {$H^{++}_a$};
      \node at (0.5,-0.8) {$H^{--}_b$};
      \node at (0.5,0.3) {$p_1$};
      \node at (0.5,-0.3) {$p_2$};
      \node at (0.0,-1.30) {$: i\frac{e}{\rm{sin}{2 \theta_w}}[4\rm{cos}^2{ \theta_w}\delta_{ab}- O_{a1} O_{b1}$};
      \node at (0.0,-1.60) {$ +O_{a2} O_{b2} ] \left( p_1-p_2 \right)_{\mu}$};
    \end{tikzpicture}\\
     \vspace{7mm}

    \begin{tikzpicture}[line width=1.4 pt, scale=1.45,every node/.style={scale=0.9}]
      \draw[fermion,black] (0.0, 0.0) --(1.0, 1.0);
      \draw[fermionbar,black] (0.0, 0.0)--(1.0, -1.0);
      \draw[scalar,black] (-1.0,0.0) --(0.0,0.0);
      \node at (-0.70,0.3) {$H^{++}_a$};
      \node at (0.5,0.8) {$l^{+}_{\alpha}$};
      \node at (0.5,-0.8) {$l^{-}_{\beta}$};
      \node at (0.0,-1.3) {$:- i y^{\alpha \beta}_\kappa O_{a3} P_R$};
    \end{tikzpicture}\\
    \vspace{7mm}
    \begin{tikzpicture}[line width=1.4 pt, scale=1.45 ,every node/.style={scale=0.9}] 
      \draw[vector,black] (0.0,0.0) --(-1.0,1.0);
      \draw[vector,black] (0.0,0.0) --(-1.0,-1.0);
      \draw[scalar,black] (0,0.0) --(1.0,1.0);
      \draw[scalar,black] (0,0.0) --(1.0,-1.0);
      \node at (-0.5,-0.8) {$A_{\nu}$};
      \node at (-0.5,0.8) {$A_{\mu}$};
      \node at (0.5,+0.8) {$H^{++}_a$};
      \node at (0.5,-0.8) {$H^{--}_b$};
      \node at (0.0,-1.30) {$: i8e^2 \delta_{ab} g_{\mu \nu}$};
    \end{tikzpicture}
  \end{center}
\end{minipage}
\hspace{0.12\textwidth}
\begin{minipage}[t]{0.24\textwidth}
  \centering
  \begin{tikzpicture}[line width=1.4 pt, scale=1.45 ,every node/.style={scale=0.9}] 
    \draw[vector,black] (-1.0, 0.0) --(0.0, 0.0);
    \draw[scalar,black] (0,0.0) --(1.0,1.0);
    \draw[scalar,black] (0,0.0) --(1.0,-1.0);
    \node at (-0.70,0.3) {$Z_{\mu}$};
    \node at (0.5,-0.8) {$\phi^{Q-}$};
    \node at (0.5,0.8) {$\phi^{Q+}$};
    \node at (0.5,0.3) {$p_1$};
    \node at (0.5,-0.3) {$p_2$};
    \node at (0.0,-1.30) {$: ie\frac{\left(Q \rm{cos}{2 \theta_w}-2 \right)}{\rm{sin}{2 \theta_w}} \left(p_1-p_2 \right)_{\mu}$};
  \end{tikzpicture}\\
  \vspace{7mm}
  \begin{tikzpicture}[line width=1.4 pt, scale=1.45,every node/.style={scale=0.9}]
    \draw[fermion,black] (0.0, 0.0) --(1.0, 1.0);
    \draw[fermionbar,black] (0.0, 0.0) --(1.0, -1.0);
    \draw[scalar,black] (-1.0,0.0) --(0.0,0.0);
    \node at (-0.70,0.3) {$\phi^{+3(1)}$};
    \node at (0.5,0.8) {$E^{++}_{\alpha}$};
    \node at (0.5,-0.8) {$l^{\mp}_{\beta}$};
    \node at (0.0,-1.30) {$:  i(-i) y^{\alpha \beta}_{\frac{5}{2}(\frac{3}{2})} P_R( P_L)$};
  \end{tikzpicture}\\
  \vspace{7mm}
  \begin{tikzpicture}[line width=1.4 pt, scale=1.45,every node/.style={scale=0.9}]
    \draw[scalar,black] (-1.0, 0.0) --(0.0, 0.0);
    \draw[scalar,black] (0.0, 0.0) --(1.0, -1.0);
    \draw[scalar,black] (0,0.0) --(1.0,1.0);
    \node at (-0.70,0.3) {$H^{++}_{a}$};
    \node at (0.5,0.8) {$H$};
    \node at (0.5,-0.8) {$H^{++}_{b}$};
    \node at (0,-1.30) {$: i [O_{a1} (-\frac{1}{\sqrt{2}}O_{b3}\mu^\prime+O_{b2} \lambda v)$};
      \node at (0,-1.6) {$+O_{a2} (\frac{1}{\sqrt{2}}O_{b3}\mu+O_{b1} \lambda v )$};
      \node at (0,-1.9) {$+O_{a3} (O_{b2}\mu-O_{b1}\mu^\prime ) ]$};
  \end{tikzpicture}\\
  \vspace{5mm}
  \begin{tikzpicture}[line width=1.4 pt, scale=1.45 ,every node/.style={scale=0.9}] 
    \draw[vector,black] (0.0,0.0) --(-1.0,1.0);
    \draw[vector,black] (0.0,0.0) --(-1.0,-1.0);
    \draw[scalar,black] (0,0.0) --(1.0,1.0);
    \draw[scalar,black] (0,0.0) --(1.0,-1.0);
    \node at (-0.5,-0.8) {$Z_{\nu}$};
    \node at (-0.5,0.8) {$Z_{\mu}$};
    \node at (0.5,0.8) {$\phi^{Q+}$};
    \node at (0.5,-0.8) {$\phi^{Q-}$};
    \node at (0.0,-1.30) {$: i2e^2 \frac{\left( Q \rm{cos}{2 \theta_w}-2 \right)^2}{ \rm{sin}^2{2 \theta_w}}g_{\mu\nu}$};
  \end{tikzpicture}
\end{minipage}

\begin{minipage}[t]{0.22\textwidth}
   \begin{center}
    \begin{tikzpicture}[line width=1.4 pt, scale=1.45 ,every node/.style={scale=0.9}] 
      \draw[vector,black] (0.0,0.0) --(-1.0,1.0);
      \draw[vector,black] (0.0,0.0) --(-1.0,-1.0);
      \draw[scalar,black] (0,0.0) --(1.0,1.0);
      \draw[scalar,black] (0,0.0) --(1.0,-1.0);
      \node at (-0.5,-0.9) {$A_{\nu}$};
      \node at (-0.5,0.9) {$Z_{\mu}$};
      \node at (0.5,0.9) {$\phi^{Q+}$};
      \node at (0.5,-0.9) {$\phi^{Q-}$};
      \node at (0.0,-1.35) {$: i2Qe^2\frac{\left(Q\rm{cos}{2 \theta_w} -2 \right)}{ \rm{sin}{2 \theta_w}} g_{\mu\nu}$};
    \end{tikzpicture}\\
    \vspace{13mm}
    \begin{tikzpicture}[line width=1.4 pt, scale=1.45 ,every node/.style={scale=0.9}] 
      \draw[scalar,black] (0.0,0.0) --(1.0,1.0);
      \draw[scalar,black] (0,0.0) --(1.0,-1.0);
      \draw[vector,black] (0.0,0.0) --(-1.0,1.0);
      \draw[vector,black] (0.0,0.0)  --(-1.0,-1.0);
      \node at (0.50,0.8) {$H^{--}_a$};
      \node at (0.5,-0.8) {$\phi^{+3(1)}$};
      \node at (-0.5,0.8) {$W^{\mp}_\nu$};
      \node at (-0.50,-0.8) {$A_\mu$};
      \node at (0.0,-1.30) {$: i \frac{5(3)e^2 O_{a1(2)}}{\sqrt{2} \rm{sin}\theta_w}g_{\mu \nu}$};
    \end{tikzpicture}\\
    \vspace{7mm}
    \begin{tikzpicture}[line width=1.4 pt, scale=1.45,every node/.style={scale=0.9}]
      \draw[scalar,black] (0.0, 0.0) --(-1.0, 1.0) ;
      \draw[scalar,black] (-1.0, -1.0) --(0.0, 0.0);
      \draw[scalar,black] (0.0, 0.0) --(1.0, -1.0);
      \draw[scalar,black] (0,0.0)  --(1.0,1.0);
      \node at (-0.5,0.8) {$H$};
      \node at (0.5,0.8) {$H$};
      \node at (0.5,-0.8) {$ H^{++}_{a}$};
      \node at (-0.5,-0.8) {$ H^{++}_{b}$};
      \node at (0.0,-1.30) {$: i \left[O_{a1} O_{b2}+O_{a2} O_{b1} \right]\lambda$};
    \end{tikzpicture}\\
    \vspace{7mm}
    \begin{tikzpicture}[line width=1.4 pt, scale=1.45 ,every node/.style={scale=0.9}] 
      \draw[fermion,black] (0.0, 0.0) --(1.0, 1.0);
      \draw[fermion,black] (0.0, 0.0) --(1.0, -1.0);
      \draw[vector,black] (-1.0,0.0) --(0.0,0.0);
      \node at (-0.70,0.3) {$A_{\mu}$};
      \node at (0.5,0.8) {$E^{++}_{\alpha}$};
      \node at (0.5,-0.8) {$E^{--}_{\beta}$};
      \node at (0.0,-1.3) {$:i2e \delta_{\alpha \beta} \gamma_{\mu}$};
    \end{tikzpicture}\\
   \end{center}
\end{minipage}
\hspace{0.1\textwidth}
\begin{minipage}[t]{0.22\textwidth}
  \begin{center}
    \begin{tikzpicture}[line width=1.4 pt, scale=1.45 ,every node/.style={scale=0.9}] 
      \draw[vector,black] (0.0,0.0) --(-1.0,1.0);
      \draw[vector,black] (0.0,0.0) --(-1.0,-1.0);
      \draw[scalar,black] (0,0.0) --(1.0,1.0);
      \draw[scalar,black] (0,0.0) --(1.0,-1.0);
      \node at (-0.5,-0.8) {$Z_{\nu}$};
      \node at (-0.5,0.8) {$Z_{\mu}$};
      \node at (0.5,0.8) {$H^{++}_a$};
      \node at (0.5,-0.8) {$H^{--}_b$};
      \node at (0.0,-1.30) {$:  i \frac{2e^2}{ \rm{sin}^22\theta_w}[16{\rm sin}^4\theta_W\delta_{ab}$};
        \node at (0.0,-1.60) {$+8{\rm sin}^2\theta_W\left(O_{a1}O_{b1}-O_{a2}O_{b2}\right)$};
        \node at (0.0,-1.90) {$+\left(O_{a1}O_{b1}+O_{a2}O_{b2}\right)]$};
    \end{tikzpicture}\\
    \vspace{7mm}
    \begin{tikzpicture}[line width=1.4 pt, scale=1.45 ,every node/.style={scale=0.9}] 
      \draw[scalar,black] (0.0,0.0) --(1.0,1.0);
      \draw[scalar,black] (0,0.0) --(1.0,-1.0);
      \draw[vector,black] (0.0,0.0) --(-1.0,1.0);
      \draw[vector,black] (0.0,0.0)  --(-1.0,-1.0);
      \node at (0.5,0.8) {$H^{--}_a$};
      \node at (0.5,-0.8) {$\phi^{+3(1)}$};
      \node at (-0.5,0.8) {$W^{\mp}_\nu$};
      \node at (-0.50,-0.8) {$Z_\mu$};
      \node at (0.0,-1.30) {$: -i\frac{5(3)e^2 O_{a1(2)}}{\sqrt{2} \rm{cos}\theta_w}g_{\mu \nu}$};
    \end{tikzpicture}\\
    \vspace{7mm}
    \begin{tikzpicture}[line width=1.4 pt, scale=1.45 ,every node/.style={scale=0.9}] 
      \draw[vector,black] (0,0) --(-1.0,1.0);
      \draw[vector,black] (0,0)  --(-1.0,-1.0);
      \draw[scalar,black] (0,0.0) --(1.0,1.0);
      \draw[scalar,black] (0,0.0) --(1.0,-1.0);
      \node at (-0.5,0.8) {$W^{+}_{\nu}$};
      \node at (-0.5,-0.8) {$W^{-}_{\mu}$};
      \node at (0.5,0.8) {$H^{++}_a$};
      \node at (0.5,-0.8) {$H^{--}_b$};
      \node at (0.0,-1.30) {$: i\frac{e^2}{\left(1- \rm{cos}2\theta_w \right)}\left(\delta_{ab}-O_{a3}O_{b3} \right)g_{\mu \nu}$};
    \end{tikzpicture}\\
    \vspace{7mm}
    \begin{tikzpicture}[line width=1.4 pt, scale=1.45 ,every node/.style={scale=0.9}] 
      \draw[fermion,black] (0.0, 0.0) --(1.0, 1.0);
      \draw[fermion,black] (0.0, 0.0) --(1.0, -1.0);
      \draw[vector,black] (-1.0,0.0) --(0.0,0.0);
      \node at (-0.70,0.3) {$Z_{\mu}$};
      \node at (0.5,0.8) {$E^{++}_{\alpha}$};
      \node at (0.5,-0.8) {$E^{--}_{\beta}$};
      \node at (0.0,-1.3) {$:-i2e \delta_{\alpha \beta} \rm{tan}\theta_{w}  \gamma_{\mu}$};
    \end{tikzpicture}
  \end{center}
\end{minipage}
\hspace{0.12\textwidth}
\begin{minipage}[t]{0.22\textwidth}
    \begin{center}
    \begin{tikzpicture}[line width=1.4 pt, scale=1.45 ,every node/.style={scale=0.9}] 
      \draw[vector,black] (0.0,0.0) --(-1.0,1.0);
      \draw[vector,black] (0.0,0.0) --(-1.0,-1.0);
      \draw[scalar,black] (0,0.0) --(1.0,1.0);
      \draw[scalar,black] (0,0.0) --(1.0,-1.0);
      \node at (-0.5,-0.8) {$A_{\nu}$};
      \node at (-0.5,0.8) {$Z_{\mu}$};
      \node at (0.5,0.8) {$H^{++}_a$};
      \node at (0.5,-0.8) {$H^{--}_a$};
      \node at (0.0,-1.30) {$: i \frac{4e^2}{\rm{sin}{2 \theta_w}} [4\rm{cos}^2{\theta_w}\delta_{ab}  $};
      \node at (0.0,-1.60) {$-O_{a1}O_{b1}+O_{a2}O_{b2} ]g_{\mu\nu}$};
    \end{tikzpicture}\\
    \vspace{7mm}
    \begin{tikzpicture}[line width=1.4 pt, scale=1.45 ,every node/.style={scale=0.9}] 
      \draw[vector,black] (0.0,0.0) --(-1.0,1.0);
      \draw[vector,black] (0.0,0.0) --(-1.0,-1.0);
      \draw[scalar,black] (0,0.0) --(1.0,1.0);
      \draw[scalar,black] (0,0.0) --(1.0,-1.0);
      \node at (-0.5,0.8) {$W^{+}_{\nu}$};
      \node at (-0.5,-0.8) {$W^{-}_{\mu}$};
      \node at (0.5,0.8) {$\phi^{Q+}$};
      \node at (0.5,-0.8) {$\phi^{Q-}$};
      \node at (0.0,-1.30) {$: i\frac{e^2}{ \left(1- \rm{cos}2\theta_w \right) } g_{\mu \nu}$};
    \end{tikzpicture}
    \end{center}
\end{minipage}

\bibliographystyle{JHEP}
\bibliography{v0_spell.bbl}

\providecommand{\href}[2]{#2}\begingroup\raggedright\begin{thebibliography}{100}

\bibitem{Weinberg:1979sa}
S.~Weinberg, \emph{{Baryon and Lepton Nonconserving Processes}},
  \href{https://doi.org/10.1103/PhysRevLett.43.1566}{\emph{Phys. Rev. Lett.}
  {\bfseries 43} (1979) 1566}.

\bibitem{Klein:2019iws}
C.~Klein, M.~Lindner and S.~Ohmer, \emph{{Minimal Radiative Neutrino Masses}},
  \href{https://doi.org/10.1007/JHEP03(2019)018}{\emph{JHEP} {\bfseries 03}
  (2019) 018} [\href{https://arxiv.org/abs/1901.03225}{{\ttfamily
  1901.03225}}].

\bibitem{Minkowski:1977sc}
P.~Minkowski, \emph{{$\mu \to e\gamma$ at a Rate of One Out of $10^{9}$ Muon
  Decays?}}, \href{https://doi.org/10.1016/0370-2693(77)90435-X}{\emph{Phys.
  Lett.} {\bfseries 67B} (1977) 421}.

\bibitem{Yanagida:1979as}
T.~Yanagida, \emph{{Horizontal gauge symmetry and masses of neutrinos}},
  {\emph{Conf. Proc.} {\bfseries C7902131} (1979) 95}.

\bibitem{GellMann:1980vs}
M.~Gell-Mann, P.~Ramond and R.~Slansky, \emph{{Complex Spinors and Unified
  Theories}}, {\emph{Conf. Proc.} {\bfseries C790927} (1979) 315}
  [\href{https://arxiv.org/abs/1306.4669}{{\ttfamily 1306.4669}}].

\bibitem{Mohapatra:1979ia}
R.~N. Mohapatra and G.~Senjanovic, \emph{{Neutrino Mass and Spontaneous Parity
  Nonconservation}},
  \href{https://doi.org/10.1103/PhysRevLett.44.912}{\emph{Phys. Rev. Lett.}
  {\bfseries 44} (1980) 912}.

\bibitem{Magg:1980ut}
M.~Magg and C.~Wetterich, \emph{{Neutrino Mass Problem and Gauge Hierarchy}},
  \href{https://doi.org/10.1016/0370-2693(80)90825-4}{\emph{Phys. Lett.}
  {\bfseries 94B} (1980) 61}.

\bibitem{Schechter:1980gr}
J.~Schechter and J.~W.~F. Valle, \emph{{Neutrino Masses in SU(2) x U(1)
  Theories}}, \href{https://doi.org/10.1103/PhysRevD.22.2227}{\emph{Phys. Rev.}
  {\bfseries D22} (1980) 2227}.

\bibitem{Wetterich:1981bx}
C.~Wetterich, \emph{{Neutrino Masses and the Scale of B-L Violation}},
  \href{https://doi.org/10.1016/0550-3213(81)90279-0}{\emph{Nucl. Phys.}
  {\bfseries B187} (1981) 343}.

\bibitem{Lazarides:1980nt}
G.~Lazarides, Q.~Shafi and C.~Wetterich, \emph{{Proton Lifetime and Fermion
  Masses in an SO(10) Model}},
  \href{https://doi.org/10.1016/0550-3213(81)90354-0}{\emph{Nucl. Phys.}
  {\bfseries B181} (1981) 287}.

\bibitem{Mohapatra:1980yp}
R.~N. Mohapatra and G.~Senjanovic, \emph{{Neutrino Masses and Mixings in Gauge
  Models with Spontaneous Parity Violation}},
  \href{https://doi.org/10.1103/PhysRevD.23.165}{\emph{Phys. Rev.} {\bfseries
  D23} (1981) 165}.

\bibitem{Cheng:1980qt}
T.~P. Cheng and L.-F. Li, \emph{{Neutrino Masses, Mixings and Oscillations in
  SU(2) x U(1) Models of Electroweak Interactions}},
  \href{https://doi.org/10.1103/PhysRevD.22.2860}{\emph{Phys. Rev.} {\bfseries
  D22} (1980) 2860}.

\bibitem{Perez:2008ha}
P.~Fileviez~Perez, T.~Han, G.-y. Huang, T.~Li and K.~Wang, \emph{{Neutrino
  Masses and the CERN LHC: Testing Type II Seesaw}},
  \href{https://doi.org/10.1103/PhysRevD.78.015018}{\emph{Phys. Rev. D}
  {\bfseries 78} (2008) 015018}
  [\href{https://arxiv.org/abs/0805.3536}{{\ttfamily 0805.3536}}].

\bibitem{Foot:1988aq}
R.~Foot, H.~Lew, X.~G. He and G.~C. Joshi, \emph{{Seesaw Neutrino Masses
  Induced by a Triplet of Leptons}},
  \href{https://doi.org/10.1007/BF01415558}{\emph{Z. Phys.} {\bfseries C44}
  (1989) 441}.

\bibitem{Babu:2013pma}
K.~Babu and J.~Julio, \emph{{Predictive Model of Radiative Neutrino Masses}},
  \href{https://doi.org/10.1103/PhysRevD.89.053004}{\emph{Phys. Rev. D}
  {\bfseries 89} (2014) 053004}
  [\href{https://arxiv.org/abs/1310.0303}{{\ttfamily 1310.0303}}].

\bibitem{Sierra:2014rxa}
D.~Aristizabal~Sierra, A.~Degee, L.~Dorame and M.~Hirsch, \emph{{Systematic
  classification of two-loop realizations of the Weinberg operator}},
  \href{https://doi.org/10.1007/JHEP03(2015)040}{\emph{JHEP} {\bfseries 03}
  (2015) 040} [\href{https://arxiv.org/abs/1411.7038}{{\ttfamily 1411.7038}}].

\bibitem{Okada:2015vwh}
H.~Okada and Y.~Orikasa, \emph{{Radiative neutrino model with an inert triplet
  scalar}}, \href{https://doi.org/10.1103/PhysRevD.94.055002}{\emph{Phys. Rev.
  D} {\bfseries 94} (2016) 055002}
  [\href{https://arxiv.org/abs/1512.06687}{{\ttfamily 1512.06687}}].

\bibitem{Nomura:2016run}
T.~Nomura, H.~Okada and Y.~Orikasa, \emph{{Radiative neutrino mass in
  alternative left--right model}},
  \href{https://doi.org/10.1140/epjc/s10052-017-4657-4}{\emph{Eur. Phys. J. C}
  {\bfseries 77} (2017) 103}
  [\href{https://arxiv.org/abs/1602.08302}{{\ttfamily 1602.08302}}].

\bibitem{Nomura:2016ask}
T.~Nomura and H.~Okada, \emph{{An Extended Colored Zee-Babu Model}},
  \href{https://doi.org/10.1103/PhysRevD.94.075021}{\emph{Phys. Rev. D}
  {\bfseries 94} (2016) 075021}
  [\href{https://arxiv.org/abs/1607.04952}{{\ttfamily 1607.04952}}].

\bibitem{Nomura:2016dnf}
T.~Nomura, H.~Okada and Y.~Orikasa, \emph{{Radiative neutrino model with
  $SU(2)_L$ triplet fields}},
  \href{https://doi.org/10.1103/PhysRevD.94.115018}{\emph{Phys. Rev. D}
  {\bfseries 94} (2016) 115018}
  [\href{https://arxiv.org/abs/1610.04729}{{\ttfamily 1610.04729}}].

\bibitem{Nomura:2017vzp}
T.~Nomura and H.~Okada, \emph{{Radiative neutrino mass in an alternative
  $U(1)_{B-L}$ gauge symmetry}},
  \href{https://doi.org/10.1016/j.nuclphysb.2019.02.025}{\emph{Nucl. Phys. B}
  {\bfseries 941} (2019) 586}
  [\href{https://arxiv.org/abs/1705.08309}{{\ttfamily 1705.08309}}].

\bibitem{Cepedello:2017lyo}
R.~Cepedello, M.~Hirsch and J.~Helo, \emph{{Lepton number violating
  phenomenology of d = 7 neutrino mass models}},
  \href{https://doi.org/10.1007/JHEP01(2018)009}{\emph{JHEP} {\bfseries 01}
  (2018) 009} [\href{https://arxiv.org/abs/1709.03397}{{\ttfamily
  1709.03397}}].

\bibitem{Cepedello:2018rfh}
R.~Cepedello, R.~M. Fonseca and M.~Hirsch, \emph{{Systematic classification of
  three-loop realizations of the Weinberg operator}},
  \href{https://doi.org/10.1007/JHEP10(2018)197}{\emph{JHEP} {\bfseries 10}
  (2018) 197} [\href{https://arxiv.org/abs/1807.00629}{{\ttfamily
  1807.00629}}].

\bibitem{Cai:2017jrq}
Y.~Cai, J.~Herrero-García, M.~A. Schmidt, A.~Vicente and R.~R. Volkas,
  \emph{{From the trees to the forest: a review of radiative neutrino mass
  models}}, \href{https://doi.org/10.3389/fphy.2017.00063}{\emph{Front.in
  Phys.} {\bfseries 5} (2017) 63}
  [\href{https://arxiv.org/abs/1706.08524}{{\ttfamily 1706.08524}}].

\bibitem{Babu:2019mfe}
K.~Babu, P.~B. Dev, S.~Jana and A.~Thapa, \emph{{Non-Standard Interactions in
  Radiative Neutrino Mass Models}},
  \href{https://doi.org/10.1007/JHEP03(2020)006}{\emph{JHEP} {\bfseries 03}
  (2020) 006} [\href{https://arxiv.org/abs/1907.09498}{{\ttfamily
  1907.09498}}].

\bibitem{Gargalionis:2019drk}
J.~Gargalionis, I.~Popa-Mateiu and R.~R. Volkas, \emph{{Radiative neutrino mass
  model from a mass dimension-11 $\Delta L =2 $ effective operator}},
  \href{https://doi.org/10.1007/JHEP03(2020)150}{\emph{JHEP} {\bfseries 03}
  (2020) 150} [\href{https://arxiv.org/abs/1912.12386}{{\ttfamily
  1912.12386}}].

\bibitem{Cepedello:2020lul}
R.~Cepedello, M.~Hirsch, P.~Rocha-Morán and A.~Vicente, \emph{{Minimal 3-loop
  neutrino mass models and charged lepton flavor violation}},
  \href{https://arxiv.org/abs/2005.00015}{{\ttfamily 2005.00015}}.

\bibitem{Arbelaez:2020xcg}
C.~Arbeláez, G.~Cottin, J.~C. Helo and M.~Hirsch, \emph{{Long-lived charged
  particles and multilepton signatures from neutrino mass models}},
  \href{https://doi.org/10.1103/PhysRevD.101.095033}{\emph{Phys. Rev. D}
  {\bfseries 101} (2020) 095033}.

\bibitem{Ma:2009gu}
E.~Ma, \emph{{Radiative inverse seesaw mechanism for nonzero neutrino mass}},
  \href{https://doi.org/10.1103/PhysRevD.80.013013}{\emph{Phys. Rev. D}
  {\bfseries 80} (2009) 013013}
  [\href{https://arxiv.org/abs/0904.4450}{{\ttfamily 0904.4450}}].

\bibitem{Ma:2012xj}
E.~Ma and J.~Wudka, \emph{{Vector-Boson-Induced Neutrino Mass}},
  \href{https://doi.org/10.1016/j.physletb.2012.05.008}{\emph{Phys. Lett. B}
  {\bfseries 712} (2012) 391}
  [\href{https://arxiv.org/abs/1202.3098}{{\ttfamily 1202.3098}}].

\bibitem{Kanemura:2010bq}
S.~Kanemura and T.~Ota, \emph{{Neutrino Masses from Loop-induced $d \geq 7$
  Operators}},
  \href{https://doi.org/10.1016/j.physletb.2010.09.064}{\emph{Phys. Lett. B}
  {\bfseries 694} (2011) 233}
  [\href{https://arxiv.org/abs/1009.3845}{{\ttfamily 1009.3845}}].

\bibitem{Krauss:2002px}
L.~M. Krauss, S.~Nasri and M.~Trodden, \emph{{A Model for neutrino masses and
  dark matter}}, \href{https://doi.org/10.1103/PhysRevD.67.085002}{\emph{Phys.
  Rev.} {\bfseries D67} (2003) 085002}
  [\href{https://arxiv.org/abs/hep-ph/0210389}{{\ttfamily hep-ph/0210389}}].

\bibitem{Branco:1988ex}
G.~Branco, W.~Grimus and L.~Lavoura, \emph{{The Seesaw Mechanism in the
  Presence of a Conserved Lepton Number}},
  \href{https://doi.org/10.1016/0550-3213(89)90304-0}{\emph{Nucl. Phys. B}
  {\bfseries 312} (1989) 492}.

\bibitem{Aoki:2009vf}
M.~Aoki, S.~Kanemura and O.~Seto, \emph{{A Model of TeV Scale Physics for
  Neutrino Mass, Dark Matter and Baryon Asymmetry and its Phenomenology}},
  \href{https://doi.org/10.1103/PhysRevD.80.033007}{\emph{Phys. Rev.}
  {\bfseries D80} (2009) 033007}
  [\href{https://arxiv.org/abs/0904.3829}{{\ttfamily 0904.3829}}].

\bibitem{Aoki:2008av}
M.~Aoki, S.~Kanemura and O.~Seto, \emph{{Neutrino mass, Dark Matter and Baryon
  Asymmetry via TeV-Scale Physics without Fine-Tuning}},
  \href{https://doi.org/10.1103/PhysRevLett.102.051805}{\emph{Phys. Rev. Lett.}
  {\bfseries 102} (2009) 051805}
  [\href{https://arxiv.org/abs/0807.0361}{{\ttfamily 0807.0361}}].

\bibitem{Ma:2007yx}
E.~Ma and U.~Sarkar, \emph{{Revelations of the E(6)/U(1)(N) Model: Two-Loop
  Neutrino Mass and Dark Matter}},
  \href{https://doi.org/10.1016/j.physletb.2007.08.019}{\emph{Phys. Lett.}
  {\bfseries B653} (2007) 288}
  [\href{https://arxiv.org/abs/0705.0074}{{\ttfamily 0705.0074}}].

\bibitem{Ma:2006km}
E.~Ma, \emph{{Verifiable radiative seesaw mechanism of neutrino mass and dark
  matter}}, \href{https://doi.org/10.1103/PhysRevD.73.077301}{\emph{Phys. Rev.}
  {\bfseries D73} (2006) 077301}
  [\href{https://arxiv.org/abs/hep-ph/0601225}{{\ttfamily hep-ph/0601225}}].

\bibitem{Jana:2019mgj}
S.~Jana, P.~Vishnu and S.~Saad, \emph{{Minimal Realizations of Dirac Neutrino
  Mass from Generic One-loop and Two-loop Topologies at $d=5$}},
  \href{https://doi.org/10.1088/1475-7516/2020/04/018}{\emph{JCAP} {\bfseries
  04} (2020) 018} [\href{https://arxiv.org/abs/1910.09537}{{\ttfamily
  1910.09537}}].

\bibitem{Cheung:2004xm}
K.~Cheung and O.~Seto, \emph{{Phenomenology of TeV right-handed neutrino and
  the dark matter model}},
  \href{https://doi.org/10.1103/PhysRevD.69.113009}{\emph{Phys. Rev.}
  {\bfseries D69} (2004) 113009}
  [\href{https://arxiv.org/abs/hep-ph/0403003}{{\ttfamily hep-ph/0403003}}].

\bibitem{Cheung:2017kxb}
K.~Cheung and H.~Okada, \emph{{A testable radiative neutrino mass model with
  multi-charged particles}},
  \href{https://doi.org/10.1016/j.physletb.2017.10.010}{\emph{Phys. Lett. B}
  {\bfseries 774} (2017) 446}
  [\href{https://arxiv.org/abs/1708.06111}{{\ttfamily 1708.06111}}].

\bibitem{Cheung:2018itc}
K.~Cheung and H.~Okada, \emph{{Generalized one-loop neutrino mass model with
  charged particles}},
  \href{https://doi.org/10.1103/PhysRevD.97.075027}{\emph{Phys. Rev. D}
  {\bfseries 97} (2018) 075027}
  [\href{https://arxiv.org/abs/1801.00585}{{\ttfamily 1801.00585}}].

\bibitem{Nomura:2018lsx}
T.~Nomura and H.~Okada, \emph{{One-loop neutrino mass model without any
  additional symmetries}},
  \href{https://doi.org/10.1016/j.dark.2019.100359}{\emph{Phys. Dark Univ.}
  {\bfseries 26} (2019) 100359}
  [\href{https://arxiv.org/abs/1808.05476}{{\ttfamily 1808.05476}}].

\bibitem{FileviezPerez:2009ud}
P.~Fileviez~Perez and M.~B. Wise, \emph{{On the Origin of Neutrino Masses}},
  \href{https://doi.org/10.1103/PhysRevD.80.053006}{\emph{Phys. Rev.}
  {\bfseries D80} (2009) 053006}
  [\href{https://arxiv.org/abs/0906.2950}{{\ttfamily 0906.2950}}].

\bibitem{Babu:2002uu}
K.~S. Babu and C.~Macesanu, \emph{{Two loop neutrino mass generation and its
  experimental consequences}},
  \href{https://doi.org/10.1103/PhysRevD.67.073010}{\emph{Phys. Rev.}
  {\bfseries D67} (2003) 073010}
  [\href{https://arxiv.org/abs/hep-ph/0212058}{{\ttfamily hep-ph/0212058}}].

\bibitem{Babu:1989pz}
K.~Babu and E.~Ma, \emph{{Radiative Hierarchy of Majorana Neutrino Masses}},
  \href{https://doi.org/10.1016/0370-2693(89)90983-0}{\emph{Phys. Lett. B}
  {\bfseries 228} (1989) 508}.

\bibitem{Ma:1998dn}
E.~Ma, \emph{{Pathways to naturally small neutrino masses}},
  \href{https://doi.org/10.1103/PhysRevLett.81.1171}{\emph{Phys. Rev. Lett.}
  {\bfseries 81} (1998) 1171}
  [\href{https://arxiv.org/abs/hep-ph/9805219}{{\ttfamily hep-ph/9805219}}].

\bibitem{Babu:1988ki}
K.~S. Babu, \emph{{Model of 'Calculable' Majorana Neutrino Masses}},
  \href{https://doi.org/10.1016/0370-2693(88)91584-5}{\emph{Phys. Lett.}
  {\bfseries B203} (1988) 132}.

\bibitem{Bonnet:2012kz}
F.~Bonnet, M.~Hirsch, T.~Ota and W.~Winter, \emph{{Systematic study of the d=5
  Weinberg operator at one-loop order}},
  \href{https://doi.org/10.1007/JHEP07(2012)153}{\emph{JHEP} {\bfseries 07}
  (2012) 153} [\href{https://arxiv.org/abs/1204.5862}{{\ttfamily 1204.5862}}].

\bibitem{Babu:2009aq}
K.~Babu, S.~Nandi and Z.~Tavartkiladze, \emph{{New Mechanism for Neutrino Mass
  Generation and Triply Charged Higgs Bosons at the LHC}},
  \href{https://doi.org/10.1103/PhysRevD.80.071702}{\emph{Phys. Rev. D}
  {\bfseries 80} (2009) 071702}
  [\href{https://arxiv.org/abs/0905.2710}{{\ttfamily 0905.2710}}].

\bibitem{Anamiati:2018cuq}
G.~Anamiati, O.~Castillo-Felisola, R.~M. Fonseca, J.~Helo and M.~Hirsch,
  \emph{{High-dimensional neutrino masses}},
  \href{https://doi.org/10.1007/JHEP12(2018)066}{\emph{JHEP} {\bfseries 12}
  (2018) 066} [\href{https://arxiv.org/abs/1806.07264}{{\ttfamily
  1806.07264}}].

\bibitem{Gu:2009hu}
P.-H. Gu, H.-J. He, U.~Sarkar and X.-m. Zhang, \emph{{Double Type-II Seesaw,
  Baryon Asymmetry and Dark Matter for Cosmic $e^\pm$ Excesses}},
  \href{https://doi.org/10.1103/PhysRevD.80.053004}{\emph{Phys. Rev. D}
  {\bfseries 80} (2009) 053004}
  [\href{https://arxiv.org/abs/0906.0442}{{\ttfamily 0906.0442}}].

\bibitem{Babu:1999me}
K.~Babu and S.~Nandi, \emph{{Natural fermion mass hierarchy and new signals for
  the Higgs boson}},
  \href{https://doi.org/10.1103/PhysRevD.62.033002}{\emph{Phys. Rev. D}
  {\bfseries 62} (2000) 033002}
  [\href{https://arxiv.org/abs/hep-ph/9907213}{{\ttfamily hep-ph/9907213}}].

\bibitem{Giudice:2008uua}
G.~F. Giudice and O.~Lebedev, \emph{{Higgs-dependent Yukawa couplings}},
  \href{https://doi.org/10.1016/j.physletb.2008.05.062}{\emph{Phys. Lett. B}
  {\bfseries 665} (2008) 79} [\href{https://arxiv.org/abs/0804.1753}{{\ttfamily
  0804.1753}}].

\bibitem{Gogoladze:2008wz}
I.~Gogoladze, N.~Okada and Q.~Shafi, \emph{{NMSSM and Seesaw Physics at LHC}},
  \href{https://doi.org/10.1016/j.physletb.2008.12.068}{\emph{Phys. Lett. B}
  {\bfseries 672} (2009) 235}
  [\href{https://arxiv.org/abs/0809.0703}{{\ttfamily 0809.0703}}].

\bibitem{Chen:2006hn}
M.-C. Chen, A.~de~Gouvea and B.~A. Dobrescu, \emph{{Gauge Trimming of Neutrino
  Masses}}, \href{https://doi.org/10.1103/PhysRevD.75.055009}{\emph{Phys. Rev.
  D} {\bfseries 75} (2007) 055009}
  [\href{https://arxiv.org/abs/hep-ph/0612017}{{\ttfamily hep-ph/0612017}}].

\bibitem{Bonnet:2009ej}
F.~Bonnet, D.~Hernandez, T.~Ota and W.~Winter, \emph{{Neutrino masses from
  higher than d=5 effective operators}},
  \href{https://doi.org/10.1088/1126-6708/2009/10/076}{\emph{JHEP} {\bfseries
  10} (2009) 076} [\href{https://arxiv.org/abs/0907.3143}{{\ttfamily
  0907.3143}}].

\bibitem{Kajiyama:2013rla}
Y.~Kajiyama, H.~Okada and T.~Toma, \emph{{Multicomponent dark matter particles
  in a two-loop neutrino model}},
  \href{https://doi.org/10.1103/PhysRevD.88.015029}{\emph{Phys. Rev. D}
  {\bfseries 88} (2013) 015029}
  [\href{https://arxiv.org/abs/1303.7356}{{\ttfamily 1303.7356}}].

\bibitem{Farzan:2014aca}
Y.~Farzan, \emph{{Two-loop snail diagrams: relating neutrino masses to dark
  matter}}, \href{https://doi.org/10.1007/JHEP05(2015)029}{\emph{JHEP}
  {\bfseries 05} (2015) 029} [\href{https://arxiv.org/abs/1412.6283}{{\ttfamily
  1412.6283}}].

\bibitem{Restrepo:2015sjs}
D.~Restrepo, A.~Rivera, M.~Sánchez and O.~Zapata, \emph{{A model with a viable
  dark matter candidate and massive neutrinos}},
  \href{https://doi.org/10.1016/j.nuclphysbps.2015.10.132}{\emph{Nucl. Part.
  Phys. Proc.} {\bfseries 267-269} (2015) 367}.

\bibitem{Kashiwase:2015pra}
S.~Kashiwase, H.~Okada, Y.~Orikasa and T.~Toma, \emph{{Two Loop Neutrino Model
  with Dark Matter and Leptogenesis}},
  \href{https://doi.org/10.1142/S0217751X16501219}{\emph{Int. J. Mod. Phys. A}
  {\bfseries 31} (2016) 1650121}
  [\href{https://arxiv.org/abs/1505.04665}{{\ttfamily 1505.04665}}].

\bibitem{Ahriche:2016rgf}
A.~Ahriche, K.~L. McDonald, S.~Nasri and I.~Picek, \emph{{A Critical Analysis
  of One-Loop Neutrino Mass Models with Minimal Dark Matter}},
  \href{https://doi.org/10.1016/j.physletb.2016.04.022}{\emph{Phys. Lett. B}
  {\bfseries 757} (2016) 399}
  [\href{https://arxiv.org/abs/1603.01247}{{\ttfamily 1603.01247}}].

\bibitem{Nomura:2016vxr}
T.~Nomura, H.~Okada and Y.~Orikasa, \emph{{Radiative Seesaw Model with
  Degenerate Majorana Dark Matter}},
  \href{https://doi.org/10.1103/PhysRevD.93.113008}{\emph{Phys. Rev. D}
  {\bfseries 93} (2016) 113008}
  [\href{https://arxiv.org/abs/1603.04631}{{\ttfamily 1603.04631}}].

\bibitem{Sierra:2016qfa}
D.~Aristizabal~Sierra, C.~Simoes and D.~Wegman, \emph{{Closing in on minimal
  dark matter and radiative neutrino masses}},
  \href{https://doi.org/10.1007/JHEP06(2016)108}{\emph{JHEP} {\bfseries 06}
  (2016) 108} [\href{https://arxiv.org/abs/1603.04723}{{\ttfamily
  1603.04723}}].

\bibitem{Ahriche:2016ixu}
A.~Ahriche, A.~Manning, K.~L. McDonald and S.~Nasri, \emph{{Scale-Invariant
  Models with One-Loop Neutrino Mass and Dark Matter Candidates}},
  \href{https://doi.org/10.1103/PhysRevD.94.053005}{\emph{Phys. Rev. D}
  {\bfseries 94} (2016) 053005}
  [\href{https://arxiv.org/abs/1604.05995}{{\ttfamily 1604.05995}}].

\bibitem{Guo:2016dzl}
S.-Y. Guo, Z.-L. Han and Y.~Liao, \emph{{Testing the type II radiative seesaw
  model: From dark matter detection to LHC signatures}},
  \href{https://doi.org/10.1103/PhysRevD.94.115014}{\emph{Phys. Rev. D}
  {\bfseries 94} (2016) 115014}
  [\href{https://arxiv.org/abs/1609.01018}{{\ttfamily 1609.01018}}].

\bibitem{Simoes:2017kqb}
C.~Simoes and D.~Wegman, \emph{{Radiative Two-Loop Neutrino Masses with Dark
  Matter}}, \href{https://doi.org/10.1007/JHEP04(2017)148}{\emph{JHEP}
  {\bfseries 04} (2017) 148}
  [\href{https://arxiv.org/abs/1702.04759}{{\ttfamily 1702.04759}}].

\bibitem{Nomura:2017emk}
T.~Nomura and H.~Okada, \emph{{Loop induced type-II seesaw model and GeV dark
  matter with $U(1)_{B-L}$ gauge symmetry}},
  \href{https://doi.org/10.1016/j.physletb.2017.10.033}{\emph{Phys. Lett. B}
  {\bfseries 774} (2017) 575}
  [\href{https://arxiv.org/abs/1704.08581}{{\ttfamily 1704.08581}}].

\bibitem{Yao:2017vtm}
C.-Y. Yao and G.-J. Ding, \emph{{Systematic Study of One-Loop Dirac Neutrino
  Masses and Viable Dark Matter Candidates}},
  \href{https://doi.org/10.1103/PhysRevD.96.095004}{\emph{Phys. Rev. D}
  {\bfseries 96} (2017) 095004}
  [\href{https://arxiv.org/abs/1707.09786}{{\ttfamily 1707.09786}}].

\bibitem{Esch:2018ccs}
S.~Esch, M.~Klasen and C.~E. Yaguna, \emph{{A singlet doublet dark matter model
  with radiative neutrino masses}},
  \href{https://doi.org/10.1007/JHEP10(2018)055}{\emph{JHEP} {\bfseries 10}
  (2018) 055} [\href{https://arxiv.org/abs/1804.03384}{{\ttfamily
  1804.03384}}].

\bibitem{CentellesChulia:2019xky}
S.~Centelles~Chuliá, R.~Cepedello, E.~Peinado and R.~Srivastava,
  \emph{{Systematic classification of two loop $d$ = 4 Dirac neutrino mass
  models and the Diracness-dark matter stability connection}},
  \href{https://doi.org/10.1007/JHEP10(2019)093}{\emph{JHEP} {\bfseries 10}
  (2019) 093} [\href{https://arxiv.org/abs/1907.08630}{{\ttfamily
  1907.08630}}].

\bibitem{Loualidi:2020jlj}
M.~Loualidi and M.~Miskaoui, \emph{{One-loop Type II Seesaw Neutrino Model with
  Stable Dark Matter Candidates}},
  \href{https://arxiv.org/abs/2003.11434}{{\ttfamily 2003.11434}}.

\bibitem{Restrepo:2013aga}
D.~Restrepo, O.~Zapata and C.~E. Yaguna, \emph{{Models with radiative neutrino
  masses and viable dark matter candidates}},
  \href{https://doi.org/10.1007/JHEP11(2013)011}{\emph{JHEP} {\bfseries 11}
  (2013) 011} [\href{https://arxiv.org/abs/1308.3655}{{\ttfamily 1308.3655}}].

\bibitem{delAguila:2013mia}
F.~del Águila and M.~Chala, \emph{{LHC bounds on Lepton Number Violation
  mediated by doubly and singly-charged scalars}},
  \href{https://doi.org/10.1007/JHEP03(2014)027}{\emph{JHEP} {\bfseries 03}
  (2014) 027} [\href{https://arxiv.org/abs/1311.1510}{{\ttfamily 1311.1510}}].

\bibitem{Kanemura:2013vxa}
S.~Kanemura, K.~Yagyu and H.~Yokoya, \emph{{First constraint on the mass of
  doubly-charged Higgs bosons in the same-sign diboson decay scenario at the
  LHC}}, \href{https://doi.org/10.1016/j.physletb.2013.08.054}{\emph{Phys.
  Lett. B} {\bfseries 726} (2013) 316}
  [\href{https://arxiv.org/abs/1305.2383}{{\ttfamily 1305.2383}}].

\bibitem{Du:2018eaw}
Y.~Du, A.~Dunbrack, M.~J. Ramsey-Musolf and J.-H. Yu, \emph{{Type-II Seesaw
  Scalar Triplet Model at a 100 TeV $pp$ Collider: Discovery and Higgs Portal
  Coupling Determination}},
  \href{https://doi.org/10.1007/JHEP01(2019)101}{\emph{JHEP} {\bfseries 01}
  (2019) 101} [\href{https://arxiv.org/abs/1810.09450}{{\ttfamily
  1810.09450}}].

\bibitem{Primulando:2019evb}
R.~Primulando, J.~Julio and P.~Uttayarat, \emph{{Scalar phenomenology in
  type-II seesaw model}},
  \href{https://doi.org/10.1007/JHEP08(2019)024}{\emph{JHEP} {\bfseries 08}
  (2019) 024} [\href{https://arxiv.org/abs/1903.02493}{{\ttfamily
  1903.02493}}].

\bibitem{Dev:2018kpa}
P.~Bhupal~Dev and Y.~Zhang, \emph{{Displaced vertex signatures of doubly
  charged scalars in the type-II seesaw and its left-right extensions}},
  \href{https://doi.org/10.1007/JHEP10(2018)199}{\emph{JHEP} {\bfseries 10}
  (2018) 199} [\href{https://arxiv.org/abs/1808.00943}{{\ttfamily
  1808.00943}}].

\bibitem{Dev:2018sel}
P.~B. Dev, M.~J. Ramsey-Musolf and Y.~Zhang, \emph{{Doubly-Charged Scalars in
  the Type-II Seesaw Mechanism: Fundamental Symmetry Tests and High-Energy
  Searches}}, \href{https://doi.org/10.1103/PhysRevD.98.055013}{\emph{Phys.
  Rev. D} {\bfseries 98} (2018) 055013}
  [\href{https://arxiv.org/abs/1806.08499}{{\ttfamily 1806.08499}}].

\bibitem{Alloul:2013raa}
A.~Alloul, M.~Frank, B.~Fuks and M.~Rausch~de Traubenberg,
  \emph{{Doubly-charged particles at the Large Hadron Collider}},
  \href{https://doi.org/10.1103/PhysRevD.88.075004}{\emph{Phys. Rev. D}
  {\bfseries 88} (2013) 075004}
  [\href{https://arxiv.org/abs/1307.1711}{{\ttfamily 1307.1711}}].

\bibitem{ATLAS:2012hi}
{\scshape ATLAS} collaboration, \emph{{Search for doubly-charged Higgs bosons
  in like-sign dilepton final states at $\sqrt{s}=7$ TeV with the ATLAS
  detector}}, \href{https://doi.org/10.1140/epjc/s10052-012-2244-2}{\emph{Eur.
  Phys. J. C} {\bfseries 72} (2012) 2244}
  [\href{https://arxiv.org/abs/1210.5070}{{\ttfamily 1210.5070}}].

\bibitem{ATLAS:2014kca}
{\scshape ATLAS} collaboration, \emph{{Search for anomalous production of
  prompt same-sign lepton pairs and pair-produced doubly charged Higgs bosons
  with $ \sqrt{s}=8 $ TeV $pp$ collisions using the ATLAS detector}},
  \href{https://doi.org/10.1007/JHEP03(2015)041}{\emph{JHEP} {\bfseries 03}
  (2015) 041} [\href{https://arxiv.org/abs/1412.0237}{{\ttfamily 1412.0237}}].

\bibitem{Aaboud:2018qcu}
{\scshape ATLAS} collaboration, \emph{{Search for doubly charged scalar bosons
  decaying into same-sign $W$ boson pairs with the ATLAS detector}},
  \href{https://doi.org/10.1140/epjc/s10052-018-6500-y}{\emph{Eur. Phys. J. C}
  {\bfseries 79} (2019) 58} [\href{https://arxiv.org/abs/1808.01899}{{\ttfamily
  1808.01899}}].

\bibitem{Chatrchyan:2012ya}
{\scshape CMS} collaboration, \emph{{A Search for a Doubly-Charged Higgs Boson
  in $pp$ Collisions at $\sqrt{s}=7$ TeV}},
  \href{https://doi.org/10.1140/epjc/s10052-012-2189-5}{\emph{Eur. Phys. J. C}
  {\bfseries 72} (2012) 2189}
  [\href{https://arxiv.org/abs/1207.2666}{{\ttfamily 1207.2666}}].

\bibitem{CMS:2017pet}
{\scshape CMS} collaboration, \emph{{A search for doubly-charged Higgs boson
  production in three and four lepton final states at
  $\sqrt{s}=13~\mathrm{TeV}$}}, .

\bibitem{CMS:2016cpz}
{\scshape CMS} collaboration, \emph{{Search for a doubly-charged Higgs boson
  with $\sqrt{s}=8~\mathrm{TeV}$ $pp$ collisions at the CMS experiment}}, .

\bibitem{Aad:2019pfm}
{\scshape ATLAS} collaboration, \emph{{Search for Magnetic Monopoles and Stable
  High-Electric-Charge Objects in 13 Tev Proton-Proton Collisions with the
  ATLAS Detector}},
  \href{https://doi.org/10.1103/PhysRevLett.124.031802}{\emph{Phys. Rev. Lett.}
  {\bfseries 124} (2020) 031802}
  [\href{https://arxiv.org/abs/1905.10130}{{\ttfamily 1905.10130}}].

\bibitem{Aaboud:2018kbe}
{\scshape ATLAS} collaboration, \emph{{Search for heavy long-lived multicharged
  particles in proton-proton collisions at $\sqrt{s}$ = 13 TeV using the ATLAS
  detector}}, \href{https://doi.org/10.1103/PhysRevD.99.052003}{\emph{Phys.
  Rev.} {\bfseries D99} (2019) 052003}
  [\href{https://arxiv.org/abs/1812.03673}{{\ttfamily 1812.03673}}].

\bibitem{PhysRevD.50.2335}
M.~Drees, R.~M. Godbole, M.~Nowakowski and S.~D. Rindani,
  \emph{\ensuremath{\gamma}\ensuremath{\gamma} processes at high energy pp
  colliders}, \href{https://doi.org/10.1103/PhysRevD.50.2335}{\emph{Phys. Rev.
  D} {\bfseries 50} (1994) 2335}.

\bibitem{PhysRevD.76.075013}
T.~Han, B.~Mukhopadhyaya, Z.~Si and K.~Wang, \emph{Pair production of doubly
  charged scalars: Neutrino mass constraints and signals at the cern lhc},
  \href{https://doi.org/10.1103/PhysRevD.76.075013}{\emph{Phys. Rev. D}
  {\bfseries 76} (2007) 075013}.

\bibitem{Babu:2016rcr}
K.~S. Babu and S.~Jana, \emph{{Probing Doubly Charged Higgs Bosons at the LHC
  through Photon Initiated Processes}},
  \href{https://doi.org/10.1103/PhysRevD.95.055020}{\emph{Phys. Rev.}
  {\bfseries D95} (2017) 055020}
  [\href{https://arxiv.org/abs/1612.09224}{{\ttfamily 1612.09224}}].

\bibitem{Agarwalla:2018xpc}
S.~K. Agarwalla, K.~Ghosh, N.~Kumar and A.~Patra, \emph{{Same-sign multilepton
  signatures of an SU(2)$_{R}$ quintuplet at the LHC}},
  \href{https://doi.org/10.1007/JHEP01(2019)080}{\emph{JHEP} {\bfseries 01}
  (2019) 080} [\href{https://arxiv.org/abs/1808.02904}{{\ttfamily
  1808.02904}}].

\bibitem{Ghosh:2017jbw}
K.~Ghosh, S.~Jana and S.~Nandi, \emph{{Neutrino Mass Generation at TeV Scale
  and New Physics Signatures from Charged Higgs at the LHC for Photon Initiated
  Processes}}, \href{https://doi.org/10.1007/JHEP03(2018)180}{\emph{JHEP}
  {\bfseries 03} (2018) 180}
  [\href{https://arxiv.org/abs/1705.01121}{{\ttfamily 1705.01121}}].

\bibitem{Baines:2018ltl}
S.~Baines, N.~E. Mavromatos, V.~A. Mitsou, J.~L. Pinfold and A.~Santra,
  \emph{{Monopole production via photon fusion and Drell–Yan processes:
  MadGraph implementation and perturbativity via velocity-dependent coupling
  and magnetic moment as novel features}},
  \href{https://doi.org/10.1140/epjc/s10052-018-6440-6,
  10.1140/epjc/s10052-019-6678-7}{\emph{Eur. Phys. J.} {\bfseries C78} (2018)
  966} [\href{https://arxiv.org/abs/1808.08942}{{\ttfamily 1808.08942}}].

\bibitem{Kurochkin:2006jr}
Y.~Kurochkin, I.~Satsunkevich, D.~Shoukavy, N.~Rusakovich and Y.~Kulchitsky,
  \emph{{On production of magnetic monopoles via gamma gamma fusion at high
  energy p p collisions}},
  \href{https://doi.org/10.1142/S0217732306022237}{\emph{Mod. Phys. Lett. A}
  {\bfseries 21} (2006) 2873}.

\bibitem{Acharya:2019vtb}
{\scshape MoEDAL} collaboration, \emph{{Magnetic Monopole Search with the Full
  MoEDAL Trapping Detector in 13 TeV pp Collisions Interpreted in Photon-Fusion
  and Drell-Yan Production}},
  \href{https://doi.org/10.1103/PhysRevLett.123.021802}{\emph{Phys. Rev. Lett.}
  {\bfseries 123} (2019) 021802}
  [\href{https://arxiv.org/abs/1903.08491}{{\ttfamily 1903.08491}}].

\bibitem{_nan_2012}
S.~C. {\.{I}}nan, \emph{Direct graviton production via photon-photon fusion at
  the {CERN}-{LHC}},
  \href{https://doi.org/10.1088/0256-307x/29/3/031301}{\emph{Chinese Physics
  Letters} {\bfseries 29} (2012) 031301}.

\bibitem{Danielsson:2016nyy}
U.~Danielsson, R.~Enberg, G.~Ingelman and T.~Mandal, \emph{{Heavy photophilic
  scalar at the LHC from a varying electromagnetic coupling}},
  \href{https://doi.org/10.1016/j.nuclphysb.2017.04.003}{\emph{Nucl. Phys. B}
  {\bfseries 919} (2017) 569}
  [\href{https://arxiv.org/abs/1601.00624}{{\ttfamily 1601.00624}}].

\bibitem{Bethe:1930ku}
H.~Bethe, \emph{{Theory of the Passage of Fast Corpuscular Rays Through
  Matter}}, \href{https://doi.org/10.1002/andp.19303970303}{\emph{Annalen
  Phys.} {\bfseries 5} (1930) 325}.

\bibitem{Aad:2013pqd}
{\scshape ATLAS} collaboration, \emph{{Search for long-lived, multi-charged
  particles in pp collisions at $\sqrt{s}$=7 TeV using the ATLAS detector}},
  \href{https://doi.org/10.1016/j.physletb.2013.04.036}{\emph{Phys. Lett. B}
  {\bfseries 722} (2013) 305}
  [\href{https://arxiv.org/abs/1301.5272}{{\ttfamily 1301.5272}}].

\bibitem{Khachatryan:2016sfv}
{\scshape CMS} collaboration, \emph{{Search for long-lived charged particles in
  proton-proton collisions at $\sqrt s=$ 13 TeV}},
  \href{https://doi.org/10.1103/PhysRevD.94.112004}{\emph{Phys. Rev. D}
  {\bfseries 94} (2016) 112004}
  [\href{https://arxiv.org/abs/1609.08382}{{\ttfamily 1609.08382}}].

\bibitem{Veeraraghavan:2013rqa}
V.~Veeraraghavan, \emph{{Search for multiply charged Heavy Stable Charged
  Particles in data collected with the CMS detector.}}, Ph.D. thesis, Florida
  State U., 2013.
\newblock 10.2172/1128814.

\bibitem{Aaboud:2018zeb}
{\scshape ATLAS} collaboration, \emph{{Search for supersymmetry in events with
  four or more leptons in $\sqrt{s}=13$ TeV $pp$ collisions with ATLAS}},
  \href{https://doi.org/10.1103/PhysRevD.98.032009}{\emph{Phys. Rev. D}
  {\bfseries 98} (2018) 032009}
  [\href{https://arxiv.org/abs/1804.03602}{{\ttfamily 1804.03602}}].

\bibitem{Aad:2014iza}
{\scshape ATLAS} collaboration, \emph{{Search for supersymmetry in events with
  four or more leptons in $\sqrt{s}$ = 8 TeV pp collisions with the ATLAS
  detector}}, \href{https://doi.org/10.1103/PhysRevD.90.052001}{\emph{Phys.
  Rev. D} {\bfseries 90} (2014) 052001}
  [\href{https://arxiv.org/abs/1405.5086}{{\ttfamily 1405.5086}}].

\bibitem{ATLAS:2012kr}
{\scshape ATLAS} collaboration, \emph{{Search for R-parity-violating
  supersymmetry in events with four or more leptons in $\sqrt{s}=7$ TeV $pp$
  collisions with the ATLAS detector}},
  \href{https://doi.org/10.1007/JHEP12(2012)124}{\emph{JHEP} {\bfseries 12}
  (2012) 124} [\href{https://arxiv.org/abs/1210.4457}{{\ttfamily 1210.4457}}].

\bibitem{Aaboud:2017qph}
{\scshape ATLAS} collaboration, \emph{{Search for doubly charged Higgs boson
  production in multi-lepton final states with the ATLAS detector using
  proton--proton collisions at $\sqrt{s}=13\,\text {TeV}$}},
  \href{https://doi.org/10.1140/epjc/s10052-018-5661-z}{\emph{Eur. Phys. J. C}
  {\bfseries 78} (2018) 199}
  [\href{https://arxiv.org/abs/1710.09748}{{\ttfamily 1710.09748}}].

\bibitem{Aad:2015dha}
{\scshape ATLAS} collaboration, \emph{{Search for heavy lepton resonances
  decaying to a $Z$ boson and a lepton in $pp$ collisions at $\sqrt{s}=8$ TeV
  with the ATLAS detector}},
  \href{https://doi.org/10.1007/JHEP09(2015)108}{\emph{JHEP} {\bfseries 09}
  (2015) 108} [\href{https://arxiv.org/abs/1506.01291}{{\ttfamily
  1506.01291}}].

\bibitem{Sirunyan:2017lae}
{\scshape CMS} collaboration, \emph{{Search for electroweak production of
  charginos and neutralinos in multilepton final states in proton-proton
  collisions at $\sqrt{s}=$ 13 TeV}},
  \href{https://doi.org/10.1007/JHEP03(2018)166}{\emph{JHEP} {\bfseries 03}
  (2018) 166} [\href{https://arxiv.org/abs/1709.05406}{{\ttfamily
  1709.05406}}].

\bibitem{Khachatryan:2016iqn}
{\scshape CMS} collaboration, \emph{{Searches for $R$-parity-violating
  supersymmetry in $pp $collisions at $\sqrt(s) =$ 8 TeV in final states with
  0-4 leptons}}, \href{https://doi.org/10.1103/PhysRevD.94.112009}{\emph{Phys.
  Rev. D} {\bfseries 94} (2016) 112009}
  [\href{https://arxiv.org/abs/1606.08076}{{\ttfamily 1606.08076}}].

\bibitem{Chatrchyan:2014aea}
{\scshape CMS} collaboration, \emph{{Search for anomalous production of events
  with three or more leptons in $pp$ collisions at $\sqrt(s) =$ 8 TeV}},
  \href{https://doi.org/10.1103/PhysRevD.90.032006}{\emph{Phys. Rev. D}
  {\bfseries 90} (2014) 032006}
  [\href{https://arxiv.org/abs/1404.5801}{{\ttfamily 1404.5801}}].

\bibitem{Chatrchyan:2013xsw}
{\scshape CMS} collaboration, \emph{{Search for Top Squarks in
  $R$-Parity-Violating Supersymmetry using Three or More Leptons and B-Tagged
  Jets}}, \href{https://doi.org/10.1103/PhysRevLett.111.221801}{\emph{Phys.
  Rev. Lett.} {\bfseries 111} (2013) 221801}
  [\href{https://arxiv.org/abs/1306.6643}{{\ttfamily 1306.6643}}].

\bibitem{Chatrchyan:2012mea}
{\scshape CMS} collaboration, \emph{{Search for anomalous production of
  multilepton events in $pp$ collisions at $\sqrt{s}=7$ TeV}},
  \href{https://doi.org/10.1007/JHEP06(2012)169}{\emph{JHEP} {\bfseries 06}
  (2012) 169} [\href{https://arxiv.org/abs/1204.5341}{{\ttfamily 1204.5341}}].

\bibitem{Maki:1962mu}
Z.~Maki, M.~Nakagawa and S.~Sakata, \emph{{Remarks on the unified model of
  elementary particles}}, \href{https://doi.org/10.1143/PTP.28.870}{\emph{Prog.
  Theor. Phys.} {\bfseries 28} (1962) 870}.

\bibitem{Pontecorvo:1967fh}
B.~Pontecorvo, \emph{{Neutrino Experiments and the Problem of Conservation of
  Leptonic Charge}}, {\emph{Sov. Phys. JETP} {\bfseries 26} (1968) 984}.

\bibitem{Casas:2006hf}
J.~Casas, A.~Ibarra and F.~Jimenez-Alburquerque, \emph{{Hints on the
  high-energy seesaw mechanism from the low-energy neutrino spectrum}},
  \href{https://doi.org/10.1088/1126-6708/2007/04/064}{\emph{JHEP} {\bfseries
  04} (2007) 064} [\href{https://arxiv.org/abs/hep-ph/0612289}{{\ttfamily
  hep-ph/0612289}}].

\bibitem{Casas:2001sr}
J.~Casas and A.~Ibarra, \emph{{Oscillating neutrinos and $\mu \to e, \gamma$}},
  \href{https://doi.org/10.1016/S0550-3213(01)00475-8}{\emph{Nucl. Phys. B}
  {\bfseries 618} (2001) 171}
  [\href{https://arxiv.org/abs/hep-ph/0103065}{{\ttfamily hep-ph/0103065}}].

\bibitem{Cordero-Carrion:2019qtu}
I.~Cordero-Carrión, M.~Hirsch and A.~Vicente, \emph{{General parametrization
  of Majorana neutrino mass models}},
  \href{https://doi.org/10.1103/PhysRevD.101.075032}{\emph{Phys. Rev. D}
  {\bfseries 101} (2020) 075032}
  [\href{https://arxiv.org/abs/1912.08858}{{\ttfamily 1912.08858}}].

\bibitem{Lesgourgues:2014zoa}
J.~Lesgourgues and S.~Pastor, \emph{{Neutrino cosmology and Planck}},
  \href{https://doi.org/10.1088/1367-2630/16/6/065002}{\emph{New J. Phys.}
  {\bfseries 16} (2014) 065002}
  [\href{https://arxiv.org/abs/1404.1740}{{\ttfamily 1404.1740}}].

\bibitem{Capozzi:2017ipn}
F.~Capozzi, E.~Di~Valentino, E.~Lisi, A.~Marrone, A.~Melchiorri and A.~Palazzo,
  \emph{{Global constraints on absolute neutrino masses and their ordering}},
  \href{https://doi.org/10.1103/PhysRevD.95.096014}{\emph{Phys. Rev.}
  {\bfseries D95} (2017) 096014}
  [\href{https://arxiv.org/abs/1703.04471}{{\ttfamily 1703.04471}}].

\bibitem{Cirelli:2005uq}
M.~Cirelli, N.~Fornengo and A.~Strumia, \emph{{Minimal dark matter}},
  \href{https://doi.org/10.1016/j.nuclphysb.2006.07.012}{\emph{Nucl. Phys.}
  {\bfseries B753} (2006) 178}
  [\href{https://arxiv.org/abs/hep-ph/0512090}{{\ttfamily hep-ph/0512090}}].

\bibitem{Davier:2010nc}
M.~Davier, A.~Hoecker, B.~Malaescu and Z.~Zhang, \emph{{Reevaluation of the
  Hadronic Contributions to the Muon g-2 and to alpha(MZ)}},
  \href{https://doi.org/10.1140/epjc/s10052-012-1874-8}{\emph{Eur. Phys. J. C}
  {\bfseries 71} (2011) 1515}
  [\href{https://arxiv.org/abs/1010.4180}{{\ttfamily 1010.4180}}].

\bibitem{Aoyama:2012wk}
T.~Aoyama, M.~Hayakawa, T.~Kinoshita and M.~Nio, \emph{{Complete Tenth-Order
  QED Contribution to the Muon g-2}},
  \href{https://doi.org/10.1103/PhysRevLett.109.111808}{\emph{Phys. Rev. Lett.}
  {\bfseries 109} (2012) 111808}
  [\href{https://arxiv.org/abs/1205.5370}{{\ttfamily 1205.5370}}].

\bibitem{Keshavarzi:2018mgv}
A.~Keshavarzi, D.~Nomura and T.~Teubner, \emph{{Muon $g-2$ and $\alpha(M_Z^2)$:
  a new data-based analysis}},
  \href{https://doi.org/10.1103/PhysRevD.97.114025}{\emph{Phys. Rev. D}
  {\bfseries 97} (2018) 114025}
  [\href{https://arxiv.org/abs/1802.02995}{{\ttfamily 1802.02995}}].

\bibitem{Hagiwara:2011af}
K.~Hagiwara, R.~Liao, A.~D. Martin, D.~Nomura and T.~Teubner,
  \emph{{$(g-2)\_mu$ and $\alpha(M\_Z^2)$ re-evaluated using new precise
  data}}, \href{https://doi.org/10.1088/0954-3899/38/8/085003}{\emph{J. Phys.
  G} {\bfseries 38} (2011) 085003}
  [\href{https://arxiv.org/abs/1105.3149}{{\ttfamily 1105.3149}}].

\bibitem{Adam:2013mnn}
{\scshape MEG} collaboration, \emph{{New constraint on the existence of the
  $\mu^+ \to e^+\gamma$ decay}},
  \href{https://doi.org/10.1103/PhysRevLett.110.201801}{\emph{Phys. Rev. Lett.}
  {\bfseries 110} (2013) 201801}
  [\href{https://arxiv.org/abs/1303.0754}{{\ttfamily 1303.0754}}].

\bibitem{TheMEG:2016wtm}
{\scshape MEG} collaboration, \emph{{Search for the lepton flavour violating
  decay $\mu ^+ \rightarrow \mathrm {e}^+ \gamma $ with the full dataset of the
  MEG experiment}},
  \href{https://doi.org/10.1140/epjc/s10052-016-4271-x}{\emph{Eur. Phys. J. C}
  {\bfseries 76} (2016) 434}
  [\href{https://arxiv.org/abs/1605.05081}{{\ttfamily 1605.05081}}].

\bibitem{Lindner:2016bgg}
M.~Lindner, M.~Platscher and F.~S. Queiroz, \emph{{A Call for New Physics : The
  Muon Anomalous Magnetic Moment and Lepton Flavor Violation}},
  \href{https://doi.org/10.1016/j.physrep.2017.12.001}{\emph{Phys. Rept.}
  {\bfseries 731} (2018) 1} [\href{https://arxiv.org/abs/1610.06587}{{\ttfamily
  1610.06587}}].

\bibitem{Meucci:2019jog}
{\scshape MEG} collaboration, \emph{{Status of charged lepton flavour violation
  search with MEG II experiment}},  in \emph{{18th Incontri di Fisica delle
  Alte Energie}}, 12, 2019, \href{https://arxiv.org/abs/1912.08656}{{\ttfamily
  1912.08656}}.

\bibitem{Ball:2014uwa}
{\scshape NNPDF} collaboration, \emph{{Parton distributions for the LHC Run
  II}}, \href{https://doi.org/10.1007/JHEP04(2015)040}{\emph{JHEP} {\bfseries
  04} (2015) 040} [\href{https://arxiv.org/abs/1410.8849}{{\ttfamily
  1410.8849}}].

\bibitem{Ball:2013hta}
{\scshape NNPDF} collaboration, \emph{{Parton distributions with QED
  corrections}},
  \href{https://doi.org/10.1016/j.nuclphysb.2013.10.010}{\emph{Nucl. Phys. B}
  {\bfseries 877} (2013) 290}
  [\href{https://arxiv.org/abs/1308.0598}{{\ttfamily 1308.0598}}].

\bibitem{Martin:2004dh}
A.~Martin, R.~Roberts, W.~Stirling and R.~Thorne, \emph{{Parton distributions
  incorporating QED contributions}},
  \href{https://doi.org/10.1140/epjc/s2004-02088-7}{\emph{Eur. Phys. J. C}
  {\bfseries 39} (2005) 155}
  [\href{https://arxiv.org/abs/hep-ph/0411040}{{\ttfamily hep-ph/0411040}}].

\bibitem{Schmidt:2015zda}
C.~Schmidt, J.~Pumplin, D.~Stump and C.~Yuan, \emph{{CT14QED parton
  distribution functions from isolated photon production in deep inelastic
  scattering}}, \href{https://doi.org/10.1103/PhysRevD.93.114015}{\emph{Phys.
  Rev. D} {\bfseries 93} (2016) 114015}
  [\href{https://arxiv.org/abs/1509.02905}{{\ttfamily 1509.02905}}].

\bibitem{Alwall:2014hca}
J.~Alwall, R.~Frederix, S.~Frixione, V.~Hirschi, F.~Maltoni, O.~Mattelaer
  et~al., \emph{{The automated computation of tree-level and next-to-leading
  order differential cross sections, and their matching to parton shower
  simulations}}, \href{https://doi.org/10.1007/JHEP07(2014)079}{\emph{JHEP}
  {\bfseries 07} (2014) 079} [\href{https://arxiv.org/abs/1405.0301}{{\ttfamily
  1405.0301}}].

\bibitem{Conte:2012fm}
E.~Conte, B.~Fuks and G.~Serret, \emph{{MadAnalysis 5, A User-Friendly
  Framework for Collider Phenomenology}},
  \href{https://doi.org/10.1016/j.cpc.2012.09.009}{\emph{Comput. Phys. Commun.}
  {\bfseries 184} (2013) 222}
  [\href{https://arxiv.org/abs/1206.1599}{{\ttfamily 1206.1599}}].

\bibitem{ATLAS:2017iqw}
{\scshape ATLAS} collaboration, \emph{{Search for doubly-charged Higgs boson
  production in multi-lepton final states with the ATLAS detector using
  proton-proton collisions at $\sqrt{s}=13\,\mathrm{TeV}$}}, .

\bibitem{Ucchielli:2017qad}
{\scshape ATLAS} collaboration, \emph{{Search for doubly-charged Higgs boson in
  multi-lepton final states at $\sqrt{s}$=13 TeV with the ATLAS detector}},
  \href{https://doi.org/10.22323/1.314.0722}{\emph{PoS} {\bfseries EPS-HEP2017}
  (2017) 722}.

\bibitem{Nuti:2014eaa}
{\scshape ATLAS} collaboration, \emph{{Doubly-charged Higgs searches by
  ATLAS}}, \href{https://doi.org/10.22323/1.209.0012}{\emph{PoS} {\bfseries
  Charged2014} (2014) 012}.

\bibitem{Zee:1985id}
A.~Zee, \emph{{Quantum Numbers of Majorana Neutrino Masses}},
  \href{https://doi.org/10.1016/0550-3213(86)90475-X}{\emph{Nucl. Phys. B}
  {\bfseries 264} (1986) 99}.

\bibitem{Nebot:2007bc}
M.~Nebot, J.~F. Oliver, D.~Palao and A.~Santamaria, \emph{{Prospects for the
  Zee-Babu Model at the CERN LHC and low energy experiments}},
  \href{https://doi.org/10.1103/PhysRevD.77.093013}{\emph{Phys. Rev. D}
  {\bfseries 77} (2008) 093013}
  [\href{https://arxiv.org/abs/0711.0483}{{\ttfamily 0711.0483}}].

\bibitem{Gunion:1989in}
J.~Gunion, J.~Grifols, A.~Mendez, B.~Kayser and F.~I. Olness, \emph{{Higgs
  Bosons in Left-Right Symmetric Models}},
  \href{https://doi.org/10.1103/PhysRevD.40.1546}{\emph{Phys. Rev. D}
  {\bfseries 40} (1989) 1546}.

\bibitem{Pati:1974yy}
J.~C. Pati and A.~Salam, \emph{{Lepton Number as the Fourth Color}},
  \href{https://doi.org/10.1103/PhysRevD.10.275}{\emph{Phys. Rev. D} {\bfseries
  10} (1974) 275}.

\bibitem{Mohapatra:1974hk}
R.~N. Mohapatra and J.~C. Pati, \emph{{Left-Right Gauge Symmetry and an
  Isoconjugate Model of CP Violation}},
  \href{https://doi.org/10.1103/PhysRevD.11.566}{\emph{Phys. Rev. D} {\bfseries
  11} (1975) 566}.

\bibitem{Senjanovic:1975rk}
G.~Senjanovic and R.~N. Mohapatra, \emph{{Exact Left-Right Symmetry and
  Spontaneous Violation of Parity}},
  \href{https://doi.org/10.1103/PhysRevD.12.1502}{\emph{Phys. Rev. D}
  {\bfseries 12} (1975) 1502}.

\bibitem{Gunion:1989ci}
J.~Gunion, R.~Vega and J.~Wudka, \emph{{Higgs triplets in the standard model}},
  \href{https://doi.org/10.1103/PhysRevD.42.1673}{\emph{Phys. Rev. D}
  {\bfseries 42} (1990) 1673}.

\bibitem{ArkaniHamed:2002qx}
N.~Arkani-Hamed, A.~Cohen, E.~Katz, A.~Nelson, T.~Gregoire and J.~G. Wacker,
  \emph{{The Minimal moose for a little Higgs}},
  \href{https://doi.org/10.1088/1126-6708/2002/08/021}{\emph{JHEP} {\bfseries
  08} (2002) 021} [\href{https://arxiv.org/abs/hep-ph/0206020}{{\ttfamily
  hep-ph/0206020}}].

\bibitem{Muhlleitner:2003me}
M.~Muhlleitner and M.~Spira, \emph{{A Note on doubly charged Higgs pair
  production at hadron colliders}},
  \href{https://doi.org/10.1103/PhysRevD.68.117701}{\emph{Phys. Rev. D}
  {\bfseries 68} (2003) 117701}
  [\href{https://arxiv.org/abs/hep-ph/0305288}{{\ttfamily hep-ph/0305288}}].

\bibitem{Perez:2008zc}
P.~Fileviez~Perez, T.~Han, G.-Y. Huang, T.~Li and K.~Wang, \emph{{Testing a
  Neutrino Mass Generation Mechanism at the LHC}},
  \href{https://doi.org/10.1103/PhysRevD.78.071301}{\emph{Phys. Rev. D}
  {\bfseries 78} (2008) 071301}
  [\href{https://arxiv.org/abs/0803.3450}{{\ttfamily 0803.3450}}].

\bibitem{Georgi:1985nv}
H.~Georgi and M.~Machacek, \emph{{DOUBLY CHARGED HIGGS BOSONS}},
  \href{https://doi.org/10.1016/0550-3213(85)90325-6}{\emph{Nucl. Phys. B}
  {\bfseries 262} (1985) 463}.

\end{thebibliography}\endgroup
\end{document}